\documentclass[11pt]{article}
\usepackage{amsmath}
\usepackage{amsfonts}
\usepackage{amssymb}
\usepackage{epsfig}
\usepackage{times}
\begin{document}

%%%%%%%%%%%%%%%%%%%%%%%%%%%%%%%%%%%%%%%%%%%%%%%%%%%%%%%%%%%%%%%%%%%%%%%%%%%%%%

\begin{titlepage}

%%%%%%%%%%%%%%%%%%%%%%%%%%%%%%%%%%%%%%%%%%%%%%%%%%%%%%%%%%%%%%%%%%%%%%%%%%%%%%

January 2006 (revised March 2006)\hfill hep-ph/0512236

\vskip 5cm

\centerline{\bf INCLUSIVE HADRONIC DISTRIBUTIONS INSIDE ONE JET
AT HIGH ENERGY COLLIDERS}

\smallskip

\centerline{\bf AT ``MODIFIED LEADING LOGARITHMIC APPROXIMATION''}

\smallskip

\centerline{\bf OF QUANTUM CHROMODYNAMICS}

\vskip .75 cm

\centerline{R. Perez-Ramos
\footnote{E-mail: perez@lpthe.jussieu.fr}
\& B. Machet
\footnote{E-mail: machet@lpthe.jussieu.fr}
}

\baselineskip=15pt

\smallskip
\centerline{\em Laboratoire de Physique Th\'eorique et Hautes Energies
\footnote{LPTHE, tour 24-25, 5\raise 3pt \hbox{\tiny \`eme} \'etage,
Universit\'e P. et M. Curie, BP 126, 4 place Jussieu,
F-75252 Paris Cedex 05 (France)}}
\centerline{\em Unit\'e Mixte de Recherche UMR 7589}
\centerline{\em Universit\'e Pierre et Marie Curie-Paris6; CNRS;
Universit\'e Denis Diderot-Paris7}

\vskip 1.5cm

{\bf Abstract}: After demonstrating their general expressions valid at all $x$,
double differential 1-particle inclusive distributions 
inside a quark and a gluon jet produced in a hard process, together with the
inclusive $k_\perp$ distributions, are calculated at small $x$ in the
Modified Leading Logarithmic Approximation (MLLA), as functions of the
transverse momentum $k_\perp$ of the outgoing hadron.
Results are compared with
the Double Logarithmic Approximation (DLA) and a naive DLA-inspired
evaluation; sizable corrections are exhibited,  which, associated with the
requirement to stay in a perturbative regime, set the limits of
the interval where our calculations can be trusted.
We give predictions for the LHC and Tevatron colliders.

\vskip 1 cm

{\em Keywords: perturbative Quantum Chromodynamics,
jets, high-energy colliders}

\vfill

\null\hfil\epsffile{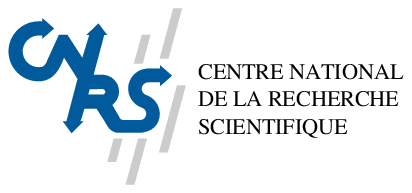}

%%%%%%%%%%%%%%%%%%%%%%%%%%%%%%%%%%%%%%%%%%%%%%%%%%%%%%%%%%%%%%%%%%%%%%%%%%%%%%

\end{titlepage}

%%%%%%%%%%%%%%%%%%%%%%%%%%%%%%%%%%%%%%%%%%%%%%%%%%%%%%%%%%%%%%%%%%%%%%%%%%%%%%

\baselineskip=15pt

%%%%%%%%%%%%%%%%%%%%%%%%%
\section{INTRODUCTION}
%%%%%%%%%%%%%%%%%%%%%%%%%

\vskip .4cm

In high energy collisions, perturbative Quantum Chromodynamics (pQCD)
successfully predicts inclusive energy spectra of particles in jets.
They have been determined within the
Modified Leading Logarithmic Approximation (MLLA) ~\cite{EvEq}~\cite{FW}
as  functions of the logarithm of the energy ($\ln(1/x)$) and the result
is in nice agreement with the data of -- $e^+e^-$ and hadronic --
 colliders and of deep inelastic scattering (DIS) (see for example
\cite{OPAL1} \cite{CdF} \cite{DIS}).
Though theoretical predictions have been derived for small $x$
(energy fraction of one parton inside the jet, $x\ll 1$)
\footnote{as the exact solution of the MLLA evolution equations}
, the agreement turns out to hold even for $x\sim 1$. 
The shape of the inclusive spectrum can even be successfully described by
setting the infrared transverse momentum  cutoff $Q_0$ as low as the
intrinsic QCD scale $\Lambda_{QCD}$ (this is the so-called
``limiting spectrum'').

\bigskip

This work concerns the production of two hadrons inside  a high energy jet
(quark or gluon); they hadronize out of two partons 
at the end of a cascading process that we calculate in pQCD;
considering this transition as a ``soft'' process is
the essence of the ``Local Parton Hadron Duality'' (LPHD) hypothesis
\cite{EvEq} \cite{DKTM} \cite{KO}, that experimental data have,
up to now, not put in jeopardy.

\bigskip

More specifically, we study, in the MLLA scheme of resummation,
the double differential inclusive 1-particle distribution
and the inclusive $k_\perp$ distribution as functions of  the transverse
momentum of the emitted hadrons; they have up to now only been investigated
in DLA (Double Logarithmic Approximation) \cite{EvEq}.
After giving  general expressions valid at all $x$, we are concerned in
the rest of the paper with the small $x$ region (the range of which is
extensively discussed) where explicit analytical formul{\ae} can be obtained;
  we furthermore consider the limit $Q_0
\approx \Lambda_{QCD}$, which leads to tractable results.
We deal with jets of small aperture; as far as hadronic colliders are
concerned, this has in particular the advantage to avoid interferences
between ingoing and outgoing states.

\bigskip

The paper is organized as follows:

\bigskip

$\bullet$\quad The description of the process, the
notations and conventions are presented in section \ref{section:descri}.
We  set there the general formula of the inclusive 2-particle
differential cross section for the production of two
hadrons $h_1$ and $h_2$  at angle $\Theta$ within a jet of opening angle
$\Theta_0$,  carrying respectively  the fractions $x_1$ and $x_2$ of
the jet energy $E$;  the axis of the jet is identified
with the direction of the energy flow.

\medskip

$\bullet$\quad In section \ref{section:EA}, we determine the 
 double differential inclusive 1-particle distribution
$\frac{d^2N}{d\ln\left(1/x_1\right)\,d\ln\Theta}$ for the hadron
$h_1$ emitted with the energy
fraction $x_1$ of the jet energy $E$, at an angle $\Theta$ with respect to
the jet axis.
This expression is valid for all $x$; it however only simplifies for $x \ll
1$, where an analytical expression can be obtained; this concerns the
rest of the paper. 

\medskip

$\bullet$\quad In section \ref{section:lowEA}, we go to the small $x$
region and determine
$\frac{d^2N}{d\ln\left(1/x_1\right)\,d\ln\Theta},\ x_1\ll 1$ 
both for a gluon jet and for a quark jet.
It is plotted as a function of $\ln k_\perp$ (or $\ln\Theta$)
for different values of $\ell_1 = \ln(1/x_1)$; the role of the
opening angle $\Theta_0$ of the jet is also considered; we compare in
particular the MLLA calculation  with a naive approach,
inspired by DLA calculations, in which  furthermore the evolution of the
starting jet from  $\Theta_0$, its initial aperture, to the angle $\Theta$
between the two outgoing hadrons  is not taken into account.

The MLLA expressions of the average gluon and quark color currents $<C>_g$ and $<C>_q$
involve potentially large corrections with respect to their expressions at
leading order; the larger the (small) $x$ domain extends, the larger they
are; keeping then under control  sets the bound
$\ell \equiv \ln\frac{1}{x}\geq 2.5$.

\medskip

$\bullet$\quad In section \ref{section:ktdist}, we study the inclusive 
$k_\perp$ distribution $\frac{dN}{d\ln k_\perp}$, which is
 the  integral of
$\frac{d^2N}{dx_1\,d\ln\Theta}$ with respect to $x_1$;
It is shown in particular how MLLA corrections ensure its positivity.
The domain of validity of our predictions is discussed; it is a $k_\perp$ interval,
 limited by the necessity of staying in the perturbative regime
 and the range of applicability of our small $x$ approximation;
it increases with the jet hardness.
The  case of mixed gluon and quark jets is evoked.

\medskip

$\bullet$\quad A conclusion briefly summarizes the results of this work and
comments on its extensions under preparation.

\medskip
\medskip

Five appendices complete this work;

\medskip

$\bullet$\quad  Appendix \ref{section:exactsol} is dedicated to the 
MLLA evolution equation for the partonic fragmentation functions
$D_g^{g\ or \ q}$ and their exact solutions \cite{Perez}\cite{Perez2}.
  They are
plotted, together with their derivatives with respect to $\ln(1/x)$
and $\ln k_\perp$.
This eases the understanding of the figures in the core of the paper and
shows the consistency of our calculations.

\medskip

$\bullet$ \quad Appendix \ref{section:leadingxF}
  presents the explicit expressions at leading order for
the average color currents of partons $<C>_{A_0}$.

\medskip

$\bullet$ \quad Appendix \ref{section:udeltau}
 completes section \ref{section:lowEA} and appendix  \ref{section:leadingxF}
 by providing explicit {formul\ae} necessary to evaluate
the MLLA corrections $\delta\!<C>_{A_0}$ to the average color currents;

\medskip

$\bullet$ \quad While the core of the paper mainly give results for LHC,
 Appendix \ref{section:LEP} provides an overview at LEP and Tevatron energies.
It is shown  how,
considering too large values of $x$ ($\ln\frac{1}{x} < 2$) endanger the
positivity of $\frac{d^2N}{d\ell\,d\ln k_\perp}$ at low $k_\perp$.
Curves are also given  for $\frac{dN}{d\ln k_\perp}$;
the range of applicability of our approximation is discussed in relation
with the core of the paper.

\medskip

$\bullet$ \quad in Appendix \ref{section:DLA}, we compare the DLA and MLLA
approximations for the spectrum, the
double differential 1-particle inclusive distribution, and the inclusive
$k_\perp$ distribution.

\vskip .7 cm

%%%%%%%%%%%%%%%%%%%%%%%%%%%%%%%%%%%%%%%%%%%%%%%%%%%%%%%%%%%%%%%%%%%%%%%%%%%%%%
\section{THE PROCESS UNDER CONSIDERATION}
\label{section:descri}
%%%%%%%%%%%%%%%%%%%%%%%%%%%%%%%%%%%%%%%%%%%%%%%%%%%%%%%%%%%%%%%%%%%%%%%%%%%%%%

\vskip .4cm

It is depicted in Fig.~1 below.  In a hard collision, a parton $A_0$
is produced, which can be a quark or a gluon
\footnote{
in $p-p$ or $p-\bar p$ collisions, two partons collide which
can create $A_0$ either as a quark or as a gluon; in the deep inelastic
scattering (DIS) and in $e^+ e^-$ colliders,
a vector boson ($\gamma$ or $Z$) decays
into a quark-antiquark pair, and $A_0$ is a quark (or an antiquark);
}
.
$A_0$, by a succession of partonic emissions  (quarks, gluons), produces
a jet of opening angle $\Theta_0$, which, in particular, contains
the parton $A$; $A$ splits  into $B$ and $C$, which hadronize respectively
into the two hadrons $h_1$  and $h_2$ (and other hadrons).
$\Theta$ is the angle between $B$ and $C$.

Because the virtualities of $B$ and $C$ are much smaller than that
of $A$ \cite{DDT},
$\Theta$ can be considered to be close to the angle between $h_1$ and $h_2$
\cite{DDT}\cite{DD}; angular ordering is also a necessary condition for
this property to hold.

\bigskip

\vbox{
\begin{center}
\epsfig{file=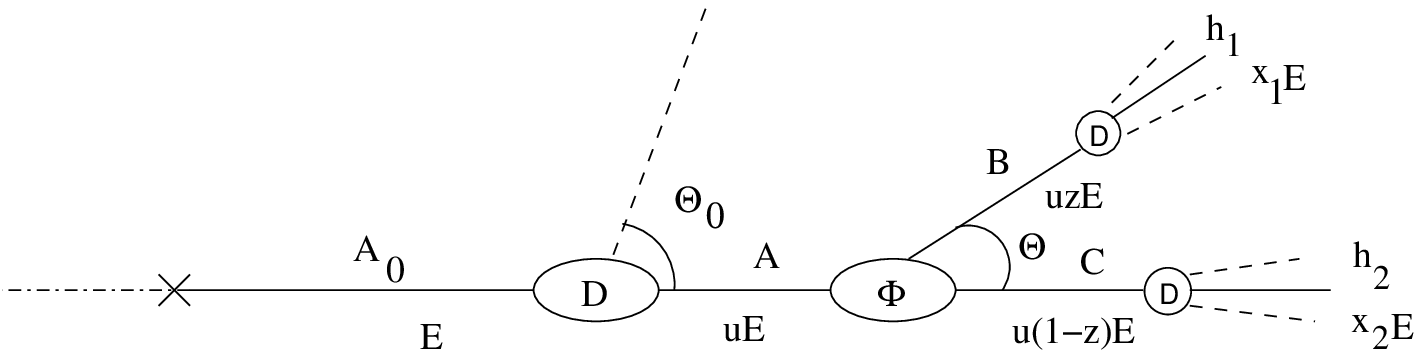, height=4.5truecm,width=11.9truecm}
\vskip .4cm
{\em Fig.~1: process under consideration: two hadrons $h_1$ and $h_2$ inside
one jet.}
\end{center}
}

\bigskip

$A_0$ carries the energy $E$.
 With a probability $D_{A_0}^A$, it
gives rise to the (virtual) parton $A$, which carries the fraction
$u$ of the energy $E$;
$\Phi_A^{BC}(z)$ is the splitting function of $A$ into $B$ and $C$,
carrying respectively the fractions $uz$ and $u(1-z)$ of $E$;
$h_1$ carries the fraction $x_1$ of $E$; $h_2$ carries the fraction $x_2$
of $E$; $D_B^{h_1}\left(\displaystyle\frac{x_1}{uz},uzE\Theta,Q_0\right)$
and $D_C^{h_2} \left(\displaystyle\frac{x_2}{u(1-z)},u(1-z)E\Theta,Q_0\right)$
are their respective energy distributions.

One has $\Theta \leq\Theta_0$.
On the other hand, since $k_{\perp} \geq Q_0$ ($Q_0$  is the collinear
cutoff), the emission angle
must satisfy $\Theta \geq \Theta_{min}=Q_0/(xE)$, $x$ being the fraction of the
energy $E$ carried away by this particle (see also subsection
\ref{subsection:notations} below).

The following expression for the inclusive double differential 2-particle
 cross section has been demonstrated in \cite{DDT} \cite{DD}:
\begin{eqnarray}
&&\frac{d\sigma}{d\Omega_{jet}\,
dx_1\,dx_2\,d\ln\left({\sin^2\displaystyle\frac{\Theta}{2}}\right)\,\displaystyle\frac{d\varphi}{2\pi}}
= 
\left(\frac{d\sigma}{d\Omega_{jet}}\right)_0\; \sum_{A,B,C} 
 \int \frac{du}{u^2} \int dz \Bigg[ \frac{1}{z(1-z)}
 \frac{\alpha_s(k_{\perp}^2)}{4\pi}\cr
 &&\Phi_A^{BC}\left(z\right)
 D_{A_0}^{A}\left(u,E\Theta_0,uE
 \Theta\right)
D_{B}^{h_1}\left(\frac{x_1}{uz},
uzE\Theta,Q_0\right)
 D_{C}^{h_2}\left(\frac{x_2}{u\left(1-z\right)},u
 (1-z)E\Theta,Q_0\right)\Bigg],\cr
&&
\label{eq:basic}
\end{eqnarray}
where $\left(\displaystyle\frac{d\sigma}{d\Omega_{jet}}\right)_0$ is the Born cross section
for the production of $A_0$, $\Omega_{jet}$ is the solid angle of the jet
 and $\varphi$ is the azimuthal angle between $B$ and $C$.

$\alpha_s(q^2)$ is the QCD running coupling constant:
\begin{equation}
\alpha_s(q^2) = \frac{4\pi}{4N_c\;\beta\ln\displaystyle
\frac{q^2}{\Lambda_{QCD}^2}},
\label{eq:alphas}
\end{equation}
where $\Lambda_{QCD} \approx$ a few hundred MeV is the intrinsic scale
of QCD and
\begin{equation}
\beta=\frac{1}{4N_c}\left(\frac{11}{3} N_c - \frac{4}{3}T_R\right) \\
\label{eq:beta}
\end{equation}
is the first term in the perturbative expansion of the $\beta$-function,
$N_c$ is the number of colors, $T_R = n_f/2$, where $n_f$ is the number
of light quark flavors ($n_f=3$); it is convenient to scale all relevant
parameters in units of $4N_c$.

In (\ref{eq:basic}), the integrations over $u$ and $z$ are performed
from $0$ to $1$;
the appropriate step functions ensuring $uz \geq x_1$, $u(1-z) \geq x_2$
(positivity of energy) are included in $D_B^{h_1}$ and $D_C^{h_2}$.

\vskip .7 cm

\subsection{Notations and variables}
\label{subsection:notations}
%%%%%%%%%%%%%%%%%%%%%%%%%%%%%%%%%%%%%%

\vskip .4cm

The notations and conventions, that are used above and throughout the paper
 are the following. For any given particle with 4-momentum $(k_0,\vec k)$,
transverse momentum $k_\perp \geq Q_0$ ($k_\perp$ is the modulus of the
trivector $\vec k_\perp$),
carrying the fraction $x=k_0/E$ of the jet energy $E$, one defines

\begin{equation}
   \ell = \ln\frac{E}{k_0}=\ln(1/x), \quad y = \ln \frac{k_{\perp}}{Q_0}.
\end{equation}
$Q_0$ is the infrared cutoff parameter (minimal transverse momentum).

If the radiated parton is emitted with an angle $\vartheta$ with
respect to the direction of the jet, one has
\begin{equation}
k_{\perp} = |\vec k|\sin{\vartheta}
\approx k_0\sin{\vartheta}.
\label{eq:appro}
\end{equation}
The r.h.s. of (\ref{eq:appro}) uses $|\vec k| \approx k_0$, resulting from
the property that the virtuality $k^2$ of the emitted parton is negligible in
the logarithmic approximation.
For collinear emissions ($\vartheta \ll 1$), $k_{\perp}\sim |\vec k|\vartheta
\approx k_0\vartheta$.

One also defines the variable $Y_\vartheta$
\begin{equation}
Y_\vartheta =\ell + y = \ln\left(E\frac{k_\perp}{k_0}\frac{1}{Q_0}\right)
\approx \ln \frac{E\vartheta}{Q_0};
\label{eq:defY}
\end{equation}
to the opening angle $\Theta_0$ of the jet corresponds
\begin{equation}
Y_{\Theta_0} = \ln\frac{E\Theta_0}{Q_0};
\label{eq:Y0}
\end{equation}
$E\Theta_0$ measures the ``hardness'' of the jet.
Since $\vartheta < \Theta_0$, one has the
condition, valid for any emitted soft parton off its ``parent''
\begin{equation}
Y_\vartheta < Y_{\Theta_0}.
\label{eq:condY}
\end{equation}

The partonic fragmentation  function $D_a^b(x_b,Q,q)$
represents the probability of finding
the parton $b$ having the fraction $x_b$ of the energy of $a$
inside the dressed parton $a$;  the virtuality (or transverse momentum)
$k_a^2$ of $a$ can go up to $|Q^2|$,  that of $b$ can go down to $|q^2|$.

\vskip .7 cm

\subsection{The jet axis}
\label{subsection:axis}
%%%%%%%%%%%%%%%%%%%%%%%%%%%

\vskip .4cm

The two quantities studied in the following paragraphs (double
differential 1-particle inclusive distribution and inclusive
 $k_\perp$ distribution)
refer to the direction (axis) of the jet, with respect to which the angles
are measured.
We identify it with the direction of the energy flow.

The double differential 1-particle inclusive distribution
$\frac{d^2N}{dx_1 d\ln\Theta}$ is accordingly defined
by summing the inclusive double differential 2-particle cross section
 over all $h_2$ hadrons and integrating it over
their energy fraction $x_2$ {\em with a weight which is the energy
($x_2$) itself};
it measures the angular distribution of
an outgoing hadron $h_1$ with energy fraction $x_1$ of the jet energy,
produced at an angle $\Theta$ with respect to the direction of
the energy flow.

Once the axis has been fixed, a second (unweighted) integration with
respect to the energy of the other hadron ($x_1$) leads to the inclusive
 $k_\perp$ distribution $\frac{dN}{d\ln k_\perp}$.

\vskip .7 cm

%%%%%%%%%%%%%%%%%%%%%%%%%%%%%%%%%%%%%%%%%%%%%%%%%%%%%%%%%%%%%%%%%%%%%%%%%%%%%%
\section{DOUBLE DIFFERENTIAL 1-PARTICLE INCLUSIVE DISTRIBUTION 
$\boldsymbol{\displaystyle\frac{d^2N}{dx_1\;d\ln\Theta}}$}
\label{section:EA}
%%%%%%%%%%%%%%%%%%%%%%%%%%%%%%%%%%%%%%%%%%%%%%%%%%%%%%%%%%%%%%%%%%%%%%%%%%%%%%

\vskip .4cm

After integrating trivially over the azimuthal angle
 (at this approximation the cross-section does not depend on it),
and going to small $\Theta$, the positive quantity $\frac{d^2N}{dx_1\,d\ln
\Theta}$ reads

\begin{equation}
\frac{d^2N}{dx_1\,
d\ln \Theta}
=
\sum_{h_2}\int _{0}^{1} dx_2\ {x_2}\;
\frac{d\sigma}
{d\Omega_{jet}\,
dx_1\,dx_2\,
d\ln\Theta}\,
\frac{1}{\left(\displaystyle\frac{d\sigma}{d\Omega_{jet}}\right)_0}.
\label{eq:eac}
\end{equation}

We use the energy conservation sum rule \cite{Politzer}
\begin{equation}
\sum_{h}\int _{0}^{1} dx\,x D_C^{h} (x,\ldots) =1
\label{eq:ensr}
\end{equation}
expressing that all partons $h_2$ within a dressed parton ($C$) carry
the total momentum of $C$,
then make the change of variable $v=\frac{x}{u(1-z)}$
where $u(1-z)$ is the upper kinematic limit for $x_2$, to get
\begin{equation}
\sum_{h_2}\int _{0}^{u(1-z)}
dx_2\,x_2 D_C^{h_2}
\left(\frac{x_2}{u(1-z)},u(1-z) E\Theta,Q_0\right)
=u^2(1-z)^2,
\label{eq:sumrule}
\end{equation}
and finally obtain the desired quantity;
\begin{equation}
  \frac{d^2N}{dx_1\,d\ln\Theta}=
\sum_{A,B}
\int du \int dz\ 
\frac{1-z}{z} \frac{\alpha_s\left(k_{\perp}^2\right)}
{2\pi}
  \Phi_A^B(z)
  D_{A_0}^A\left(u,E\Theta_0,uE\Theta\right)
  D_{B}^{h_1}\left(\frac{x_1}{uz},uzE \Theta,Q_0\right)
\label{eq:2};
\end{equation}
the summation index $C$ has been suppressed since knowing $A$ and $B$
fixes $C$.

We can transform (\ref{eq:2}) by using the following trick:
\begin{equation}
\int du\,\int \frac{dz}{z}(1-z)=\int du\,\int\frac{dz}{z}
-\int d(uz)\int\frac{du}{u},
\label{eq:trick}
\end{equation}
and (\ref{eq:2}) becomes
\begin{equation}
\begin{split}
  \frac{d^2N}{dx_1\,d\ln\Theta}
= 
&\sum_{A}\int
du\,
  D_{A_0}^{A}(u,E\Theta_0,uE\Theta)
\sum_{B}
  \int
\frac{dz}{z}
  \frac{\alpha_s\left(k_{\perp}^2\right)}
{2\pi}
\Phi_A^B (z)
  D_{B}^{h_1}\left(\frac{x_1}{uz},uzE\Theta,Q_0\right)\\
&- \sum_{B}\int
d(uz)D_{B}^{h_1}\left (\frac{x_1}{uz},uzE\Theta,Q_0\right)
  \sum_{A}\int
\frac{du}{u}\frac{\alpha_s (k_{\perp}^2)}{4\pi}
  \Phi_A^B\left(\frac{uz}u\right)
  D_{A_0}^A(u,E\Theta_0,uE\Theta).
\end{split}
\label{eq:inter}
\end{equation} 
We then make use of the two complementary DGLAP (see also the beginning of
section \ref{section:lowEA}) evolution equations
~\cite{DGLAP} 
which contain the Sudakov form factors  $d_A$ and $d_B$ of the partons
$A$ and $B$ respectively:
\begin{equation}
d_A^{-1}(k_{A}^2)\frac{d}{d\ln k_A^2}
  \left[d_A(k_{A}^2)
    D_A^{h_1}\left(\frac{x_1}{u},uE\Theta,Q_0\right)\right]
    =
\frac{\alpha_s(k_{\perp}^2)}{4\pi}
\sum_{B}\int
  \frac{dz}{z}
  \Phi_A^{B}\left(z\right) D_{B}^{h_1}\left(\frac{x_1}{uz},uzE\Theta,Q_0\right),
\label{eq:dglap1}
\end{equation}
\begin{equation}
 d_B(k_{B}^2)\frac{d}{d\ln k_B^2}\left[d_B^{-1}
 (k_{B}^2)
 D_{A_0}^B(w,E\Theta_0,wE\Theta)\right]
=
- \frac{\alpha_s(k_{\perp}^2)}{4\pi}
\sum_{A} \int
 \frac{du}{u}
\Phi_A^B \left(\frac{w}{u}\right) D_{A_0}^A(u,E\Theta_0,uE\Theta);
\label{eq:dglap2}
\end{equation}
the variable $uz$  occurring in (\ref{eq:trick}) has been introduced;
in (\ref{eq:dglap1}) and (\ref{eq:dglap2}),
$(uE\Theta)^2$ refers respectively to the virtualities $k_A^2$ and $k_B^2$
of $A$ and $B$.
Using (\ref{eq:dglap1}) and (\ref{eq:dglap2}),
(\ref{eq:inter}) transforms into 
\begin{equation}
\begin{split}
  \frac{d^2N}{\displaystyle dx_1\,d\ln\Theta}
&=\sum_{A}\int
du
  \,D_{A_0}^A(u,E\Theta_0,uE\Theta)
d_A^{-1}(k_{A}^2)\frac{d}{d\ln k_A^2}\left[d_A
(k_{A}^2)
  D_A^{h_1}\left(\frac{x_1}{u},uE\Theta,Q_0\right)\right]\\
  &+\sum_B\int
dw\,D_B^{h_1}\left(\frac{x_1} {w},wE\Theta,Q_0\right)
  d_B(k_{B}^2)\frac{d}{d\ln k_B^2}\left[d_B^{-1}
  (k_{B}^2) D_{A_0}^B\left(w,E\Theta_0,wE\Theta\right)\right].
\end{split}
\label{eq:dist}  
\end{equation}
$D_A^{h_1}$ depends  on the virtuality of $A$ through the variable
\cite{EvEq}
$\Delta \xi = \xi(k_A^2) -\xi(Q_0^2) =
\displaystyle\frac{1}{4N_c\beta}\ln\left(
\displaystyle\frac{\ln(k_A^2/\Lambda_{QCD}^2)}{\ln(Q_0^2/\Lambda_{QCD}^2)}\right)$
and elementary kinematic considerations \cite{DDT} lead to
$k_A^2 \sim \left(uE \Theta\right)^2$.

By renaming $B\rightarrow A$ and  $w\rightarrow u$,
(\ref{eq:dist}) finally becomes
\begin{equation}
\begin{split}
 \frac{d^2N}{dx_1\,d\ln{\Theta}}
&=
\sum_A\int
du
 \left[D_{A_0}^A (u,E\Theta_0,uE\Theta)d_A^{-1}(k_{A}^2)
\frac{d} {d\ln \Theta}
\left[d_A(k_{A}^2)
D_A^{h_1} \left(\frac{x_1}{u},uE\Theta,Q_0\right)\right]\right.\\
&\left.+D_A^{h_1}\left(\frac{x_1}{u},uE\Theta,Q_0\right)
d_A(k_{A}^2)
\frac{d}{d\ln \Theta}
 \left[d_A^{-1}(k_{A}^2)D_{A_0}^A\left(u,E\Theta_0,uE\Theta\right)
 \right]\right]\\
&=\sum_A\frac{d}{d\ln \Theta}\left[\int
du
 D_{A_0}^{A}\left(u,E\Theta_0,uE\Theta\right)D_A^{h_1}
\left( \frac{x_1}{u},uE\Theta,Q_0\right)\right],
\end{split}
\label{eq:dist1}
\end{equation}
and one gets
\begin{equation}
\frac{d^2N}{dx_1\,d\ln{\Theta}}=
\frac{d}{d\ln\Theta}F_{A_0}^{h_1}\left(x_1,
\Theta,E,\Theta_0\right)
\label{eq:DD}
\end{equation}
with
\begin{equation}
 F_{A_0}^{h_1}\left(x_1,\Theta,E,\Theta_0\right)
\equiv
\sum_{A}\int
du
 D_{A_0}^A\left(u,E\Theta_0,uE\Theta\right)D_A^{h_1}\left(\frac{x_1}
 {u},uE\Theta,Q_0\right);
%}
\label{eq:F}
\end{equation}

\medskip

$F$ defined in (\ref{eq:F}) is the inclusive double differential
distribution in $x_1$ and $\Theta$  with respect to the energy flux
(the energy fraction of the hadron $h_1$ within the registered
energy flux) and is represented by the convolution of the two functions
$D_{A_0}^A$ and $D_A^h$.

The general formula (\ref{eq:DD}) is valid for all $x_1$; its 
analytical expression in the small $x_1$ region will be written in
the next section.

\vskip .7 cm

%%%%%%%%%%%%%%%%%%%%%%%%%%%%%%%%%%%%%%%%%%%%%%%%%%%%%%%%%%%%%
\section{SOFT APPROXIMATION (SMALL-$\boldsymbol x_1$) FOR
$\boldsymbol{\displaystyle\frac{d^2N}{d\ell_1\;d\ln k_\perp}}$}
\label{section:lowEA}
%%%%%%%%%%%%%%%%%%%%%%%%%%%%%%%%%%%%%%%%%%%%%%%%%%%%%%%%%%%%%

\vskip .4cm

At $\ell_1$ fixed, since  $y_1=\ln (k_{\perp}/Q_0)$ and
$Y = \ln(E\Theta/Q_0) = \ell_1 + y_1$, $dy_1 = d\ln k_\perp = d\ln\Theta$
and we write hereafter $\frac{d^2N}{d\ell_1\;d\ln k_\perp}$ or
 $\frac{d^2N}{d\ell_1\;d y_1}$ instead of
$\frac{d^2N}{d\ell_1\;d\ln\Theta}$.

Since the $u$-integral (\ref{eq:F}) is dominated by $u={\cal O}(1)$
\footnote{$D_A^{h_1}\left(\frac{x_1}{u},uE\Theta,Q_0\right) \approx
(u/x_1)\times$ (slowly varying function) -- see (\ref{eq:Dlowx2}) --
and the most singular possible behavior
of $D_{A_0}^A(u, E\Theta_0, uE\Theta,Q_0)$, which could enhance the
contribution of small $u$, is $\sim 1/u$; however, the integrand then
behaves like Const. $\times$ (slowly\ varying\ function) and the
contribution of small
$u$ to the integral is still negligible.}
, the DGLAP
\cite{EvEq} partonic distributions  $D_{A_0}^A(u, \ldots)$  are to be used
and, since, on the other hand,  we restrict to small $x_1$, 
$x_1/u \ll 1$ and the MLLA inclusive $D_A^{h_1}((x_1/u),\ldots)$ are
requested. The latter will be taken as the exact solution (see
\cite{Perez}) of the (MLLA) evolution equations 
that we briefly sketch out, for the sake of completeness,
in appendix \ref{section:exactsol}.
MLLA evolution equations accounts for the constraints of angular
ordering (like DLA but unlike DGLAP equations) and of energy-momentum
conservation (unlike DLA).

For soft hadrons, the behavior of the function
$D_{A}^{h_1}(x_1,E\Theta,Q_0)$ at $x_1\ll 1$  is \cite{EvEq}
\begin{equation}
 D_A^{h_1}(x_1,E\Theta,Q_0)\approx\frac{1}{x_1}\rho_A^{h_1}\left(\ln\frac{1}{x_1},
 \ln\frac{E\Theta}{Q_0}\equiv Y_\Theta\right),
\label{eq:Dlowx}
\end{equation}
where $\rho_A^{h_1}$ is a slowly varying function of two logarithmic
variables that describes the
``hump-backed'' plateau.

For $D_A^{h_1}\left(\displaystyle\frac{x_1}{u},uE\Theta,Q_0\right)$ occurring in
(\ref{eq:F}), this yields
\begin{equation}
D_A^{h_1}\left(\frac{x_1}{u},uE\Theta,Q_0\right) \approx
\frac{u}{x_1}\rho_A^{h_1}\left(\ln\frac{u}{x_1},\ln u + Y_\Theta\right).
\label{eq:Dlowx2}
\end{equation}
Because of (\ref{eq:defY}), one has
\begin{equation}
\rho_A^{h}(\ell, Y_\Theta) = \rho_A^{h_1}(\ell, \ell + y) = \tilde
D_A^{h}(\ell, y),
\label{eq:rhoD}
\end{equation}
and, in what follows, we shall always consider the functions
\begin{equation}
xD_A(x,E\Theta,Q_0)=\tilde D_A(\ell,y).
\label{eq:Dtilde}
\end{equation}

The expansion of
$\rho_A^{h_1}\left(\ln\displaystyle\frac{u}{x_1},\ln u + Y_\Theta\right)$
around $u=1$ ($\ln u=\ln 1$)  reads
\begin{eqnarray}
\frac{x_1}{u}D_A^{h_1}\left(\frac{x_1}{u},uE\Theta,Q_0\right)
&=& \rho_A^{h_1}(\ell_1 + \ln u,Y_\Theta + \ln u)
= \rho_A^{h_1}(\ell_1 + \ln u, y_1 + \ell_1 + \ln u)\cr
&=& \tilde D_A^{h_1}(\ell_1 + \ln u, y_1)
= \tilde D_A^{h_1}(\ell_1,y_1) + \ln u \; \frac{d}{d\ell_1}\tilde
D_A^{h_1}(\ell_1, y_1) + \ldots,
\label{eq:rhodev}
\end{eqnarray}
such that

\vbox{
\begin{eqnarray}
x_1 F_{A_0}^{h_1}(x_1,\Theta,E,\Theta_0)
&\approx&
\sum_A \int du\;
{u} D_{A_0}^A(u,E\Theta_0,uE\Theta)
\left(\tilde D_A^{h_1}(\ell_1,y_1) + \ln u\;
\frac{d\tilde D_A^{h_1}(\ell_1,y_1)}{d\ell_1}
\right)\cr
&=& \sum_A \left[\int du\, u D_{A_0}^A(u,E\Theta_0,uE\Theta)\right]
 \tilde D_A^{h_1}(\ell_1,y_1)\cr
&&  + \sum_A \left[\int du\, u\ln u D_{A_0}^A(u,E\Theta_0,uE\Theta)\right]
\frac{d\tilde D_A^{h_1}(\ell_1,y_1)}{d\ell_1};
\label{eq:Fdev}
\end{eqnarray}
}

the second line in (\ref{eq:Fdev}) is the ${\cal O}(1)$
 main contribution; the third line, which accounts for the derivatives,
including the variation of $\alpha_s$, makes up corrections of relative order
 ${\cal O}(\sqrt{\alpha_s})$ with respect to the leading terms
(see also (\ref{eq:correc})), which have never been considered before;
since, in the last
line of (\ref{eq:Fdev}), $u\leq 1 \Rightarrow \ln u \leq 0$ and
$\frac{d \tilde D_{A}^{h_1}}{d\ell_1}$ is positive
(see appendix \ref{subsection:deriv}), the corresponding correction is
negative.
A detailed discussion of all corrections  is made in subsections
\ref{subsection:colcur} and \ref{subsection:correcs}

It is important for further calculations that (\ref{eq:F}) has now factorized.

While (\ref{eq:F}) (\ref{eq:Fdev}) involve (inclusive)
{\em hadronic} fragmentation functions $\tilde D_A^{h_1} = \tilde
D_g^{h_1}$ or $\tilde D_q^{h_1}$,
the MLLA {\em partonic} functions $\tilde D_A^b(\ell, y)$ satisfy
the evolution equations (\ref{eq:eveqincl}) with
exact solution (\ref{eq:ifD}), demonstrated in
\cite{Perez} and recalled in appendix \ref{section:exactsol}.
The link between the latter
($\tilde D_g^g$, $\tilde D_q^g$,  $\tilde D_g^q$,  $\tilde D_q^q$) and the
former goes as follows.
 At small $x$, since quarks are secondary products of gluons, for a given
``parent'', the number
of emitted quarks is a universal function of the number of emitted gluons:
the upper indices of emitted partons are thus correlated, and
we can  replace  in (\ref{eq:Fdev}) the inclusive fragmentation functions by
the partonic ones,  go to the functions
$\tilde D_A(\ell, y)$, where the upper index (which we will omit)
 is indifferently $g$ or $q$,
and rewrite
\begin{equation}
x_1 F_{A_0}^{h_1}(x_1,\Theta,E,\Theta_0)
\approx
\sum_A \Big( <u>^A_{A_0} + \delta \! <u>^A_{A_0}
\psi_{A,\ell_1}(\ell_1, y_1)
\Big)
\tilde D_A(\ell_1,y_1),
\label{eq:xF2}
\end{equation}
with
\footnote{In (\ref{eq:udef}), $u$ is integrated form $0$ to $1$, while,
kinematically, it cannot get lower than $x_1$; since we are
working at small $x_1$, this approximation is reasonable.}
\begin{eqnarray}
<u>^A_{A_0}&=&\int_{0}^{1}du\,uD_{A_0}^A\left(u,E\Theta_0,uE\Theta\right)
\approx \int_{0}^{1}du\,uD_{A_0}^A\left(u,E\Theta_0,E\Theta\right),\cr
\delta \! <u>^A_{A_0}&=&
\int_{0}^{1}du\,(u\ln u)D_{A_0}^A\left(u,E\Theta_0,uE\Theta\right)
\approx \int_{0}^{1}du(\,u\ln u)D_{A_0}^A\left(u,E\Theta_0,E\Theta\right),
\label{eq:udef}
\end{eqnarray}
and
\begin{equation}
\psi_{A,\ell_1}(\ell_1,y_1) = \frac{1}{\tilde D_A(\ell_1, y_1)}
\frac{d \tilde D_A(\ell_1, y_1)}{d\ell_1}.
\label{eq:psidef}
\end{equation}

Thus, for a gluon jet
\begin{eqnarray}
x_1 F_{g}^{h_1}\left(x_1,\Theta,E,\Theta_0\right)
&\approx&
<u>^g_{g} \tilde D_g(\ell_1,y_1)
+ <u>^q_{g} \tilde D_q(\ell_1,y_1)\cr
 &+& \delta \! <u>^g_{g}\psi_{g,\ell_1}(\ell_1,y_1) 
\tilde D_g(\ell_1,y_1)
\cr
&+&  \delta \! <u>^q_{g}\psi_{q,\ell_1}(\ell_1,y_1)
\tilde D_q(\ell_1,y_1),\cr
&&
\label{eq:Fg}
\end{eqnarray}
and for a quark jet
\begin{eqnarray}
x_1 F_{q}^{h_1}\left(x_1,\Theta,E,\Theta_0\right)
&\approx&
<u>^g_{q}\tilde D_g(\ell_1,y_1)
+ <u>^q_{q} \tilde D_q(\ell_1,y_1)\cr
&+& \delta \! <u>^g_{q}\psi_{g,\ell_1}(\ell_1,y_1)
 \tilde D_g(\ell_1,y_1)\cr
&+& \delta \! <u>^q_{q}\psi_{q,\ell_1}(\ell_1,y_1)
\tilde D_q(\ell_1,y_1).\cr
&&
\label{eq:Fq}
\end{eqnarray}
It turns out (see \cite{EvEq}) that the MLLA corrections to
the formul\ae
\begin{equation}
\tilde D_q^g \approx \frac{C_F}{N_c}\tilde D_g^g, \quad
\tilde D_q^q \approx \frac{C_F}{N_c} \tilde D_g^q,
\label{eq:DgDq}
\end{equation}
do not modify the results and we use (\ref{eq:DgDq}) in the
following.
We rewrite accordingly (\ref{eq:Fg}) and (\ref{eq:Fq})

\vbox{
\begin{eqnarray}
x_1 F_{g}^{h_1}\left(x_1,\Theta,E,\Theta_0\right)
&\approx&
\frac{<C>_g^0 + \delta \! <C>_g}{N_c}\;\tilde D_g(\ell_1,y_1)
\equiv \frac{<C>_g}{N_c}\;\tilde D_g(\ell_1,y_1),\cr
x_1 F_{q}^{h_1}\left(x_1,\Theta,E,\Theta_0\right)
&\approx&
\frac{<C>_q^0 + \delta \! <C>_q}{N_c}\;\tilde D_g(\ell_1,y_1)
\equiv \frac{<C>_q}{N_c}\;\tilde D_g(\ell_1,y_1),
\label{eq:Fgq}
\end{eqnarray}
}

with 
\begin{eqnarray}
<C>_{g}^0&=&<u>^g_{g} N_c+<u>^q_{g} C_F,\cr
&&\cr
<C>_{q}^0&=&<u>^g_{q} N_c+<u>^q_{q} C_F,
\label{eq:C}
\end{eqnarray}
and where we have called
\begin{eqnarray}
\delta \!<C>_{g}&=&N_c\;\delta \!<u>^g_{g}\psi_{g,\ell_1}(\ell_1,y_1)
 +C_F\;\delta \!<u>^q_{g}\psi_{q,\ell_1}(\ell_1,y_1),\cr
\delta \!<C>_{q}&=&N_c\;\delta \!<u>^g_{q}\psi_{g,\ell_1}(\ell_1,y_1)
+C_F\;\delta \!<u>^q_{q}\psi_{q,\ell_1}(\ell_1,y_1).
\label{eq:deltaC}
\end{eqnarray}
$<C>_{A_0}$ is the average color current of partons caught by the
calorimeter.

Plugging (\ref{eq:Fgq}) into (\ref{eq:DD}) yields the general formula
\begin{equation}
\left(\frac{d^2N}{d\ell_1\, d\ln k_\perp}\right)_{q,g}=
\frac{d}{d y_1}\left[\frac{<C>_{q,g}}{N_c}\;\tilde
D_g(\ell_1,y_1)\right]
\label{eq:gen}
\end{equation}
\medskip

The first line of (\ref{eq:Fg}) and (\ref{eq:Fq}) are the leading terms,
the second and third lines are  corrections.
Their relative order  is easily determined by the following
relations (see (\ref{eq:gamma0}) for the definition of $\gamma_0$)
\begin{eqnarray}
&& \frac{d^2N}{d\ell_1\,d\ln k_\perp} = \frac{<C>_{q,g}}{N_c}
\frac{d}{dy_1}\tilde D_g(\ell_1,y_1) 
+ \frac{1}{N_c} \tilde D_g(\ell_1,y_1) \frac{d}{dy_1}<C>_{q,g},\cr
&& \frac{d}{dy_1}\tilde D_g(\ell_1,y_1) = {\cal O}(\gamma_0)
= {\cal O}(\sqrt{\alpha_s}),\cr
&& \frac{d}{dy_1}<C>_{q,g} = {\cal O}(\gamma_0^2) = {\cal
O}(\alpha_s);
\label{eq:correc}
\end{eqnarray}
 The different contributions are discussed in subsections
\ref{subsection:colcur} and \ref{subsection:correcs} below.

\medskip

$\bullet$\ $\frac{d\tilde D_g(\ell,y)}{d\ln k_\perp}
\equiv \frac{d\tilde D_g(\ell,y)}{dy}$ (see the beginning of this section)
occurring in (\ref{eq:gen}) is plotted
in Fig.~12 and 13 of appendix \ref{section:exactsol},
and $\frac{d \tilde D_g(\ell, y)}{d\ell}$ occurring in
(\ref{eq:xF2}) (\ref{eq:psidef}) is plotted in
Figs.~14 and 15.

\medskip

$\bullet$\ The expressions for the leading terms of
$x_1 F_{A_0}^{h_1}\left(x_1,\Theta,E,\Theta_0\right)$ 
together with the ones of  $<C>_{g}^0$ and $<C>_{q}^0$ are given in
appendix \ref{section:leadingxF}.

\medskip

$\bullet$\ The calculations of $\delta \! <C>_g$ and $\delta \! <C>_q$ are detailed
in appendix \ref{section:udeltau}, where the explicit analytical expressions
for the $<u>$'s and $\delta <u>$'s are also given.

\medskip

We call ``naive'' the approach'' in which
one disregards the evolution of the jet between $\Theta_0$ and
$\Theta$; this amounts to taking to zero the derivative of $<C>_{q,g}$ in
(\ref{eq:gen});
(\ref{eq:Cquark}), (\ref{eq:Cgluon}), (\ref{eq:c1c2c3}) then yield
\begin{equation}
<C>_g^{naive} = N_c,\quad <C>_q^{naive}= C_F.
\label{eq:Cnaive}
\end{equation}

\vskip .7cm

\subsection{The average color current $\boldsymbol{<C>_{A_0}}$}
\label{subsection:colcur}
%%%%%%%%%%%%%%%%%%%%%%%%%%%%%%%%%%%%%%%%%%%%%%%%%%%%%%%%%%%%%%%%

\vskip .4cm

On Fig.~2 below, we plot, for $Y_{\Theta_0}=7.5$,
  $<C>_q^0$, $<C>_q^0 + \delta\!<C>_q$, $<C>_g^0$, $<C>_g^0 + \delta\!<C>_g$
as functions of $y$, for $\ell = 2.5$ on the left and $\ell = 3.5$ on the
right. Since $\Theta \leq \Theta_{0}$, the curves
stop at $y$ such that $y + \ell = Y_{\Theta_{0}}$; they reach then their
respective asymptotic values $N_c$ for $<C>_g$ and $C_F$ for $<C>_q$, at
which $\delta\! <C>_q$ and $\delta\! <C>_g$ also vanish (see also
the naive approach (\ref{eq:Cnaive})).
These corrections also
vanish at $y=0$ because they are proportional to the logarithmic
derivative $(1/\tilde D(\ell,y))(d\tilde D(\ell,y)/d\ell)$
(see (\ref{eq:deltaC}))
which both vanish, for $q$ and $g$,
at $y=0$ (see appendix \ref{section:exactsol}, and Figs.~16-17); there, the
values of $<C>_g$ and $<C>_q$ can be determined from
(\ref{eq:Cquark})(\ref{eq:Cgluon}).

The curves corresponding to LEP and Tevatron working conditions,
$Y_{\Theta_0}=5.2$, are shown in appendix \ref{section:LEP}.

\bigskip

\vbox{
\begin{center}
\epsfig{file=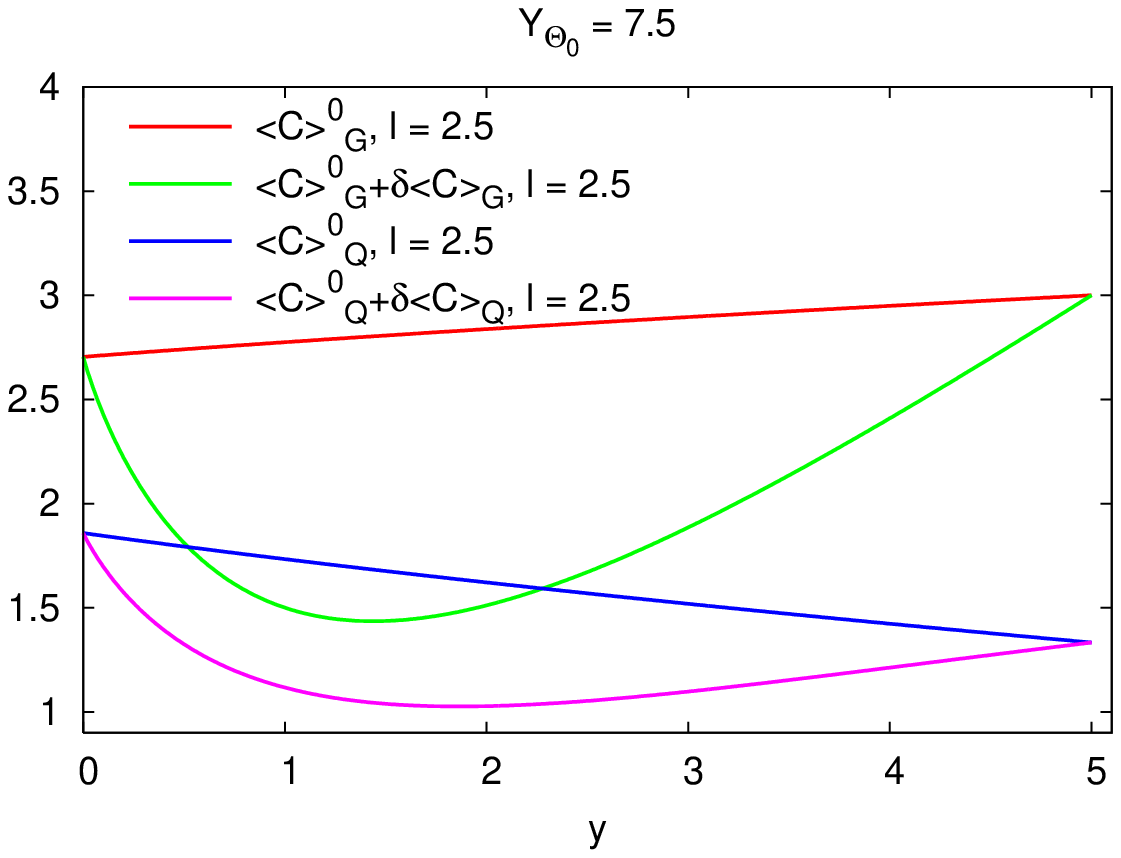, height=5truecm,width=7.5truecm}
\hfill
\epsfig{file=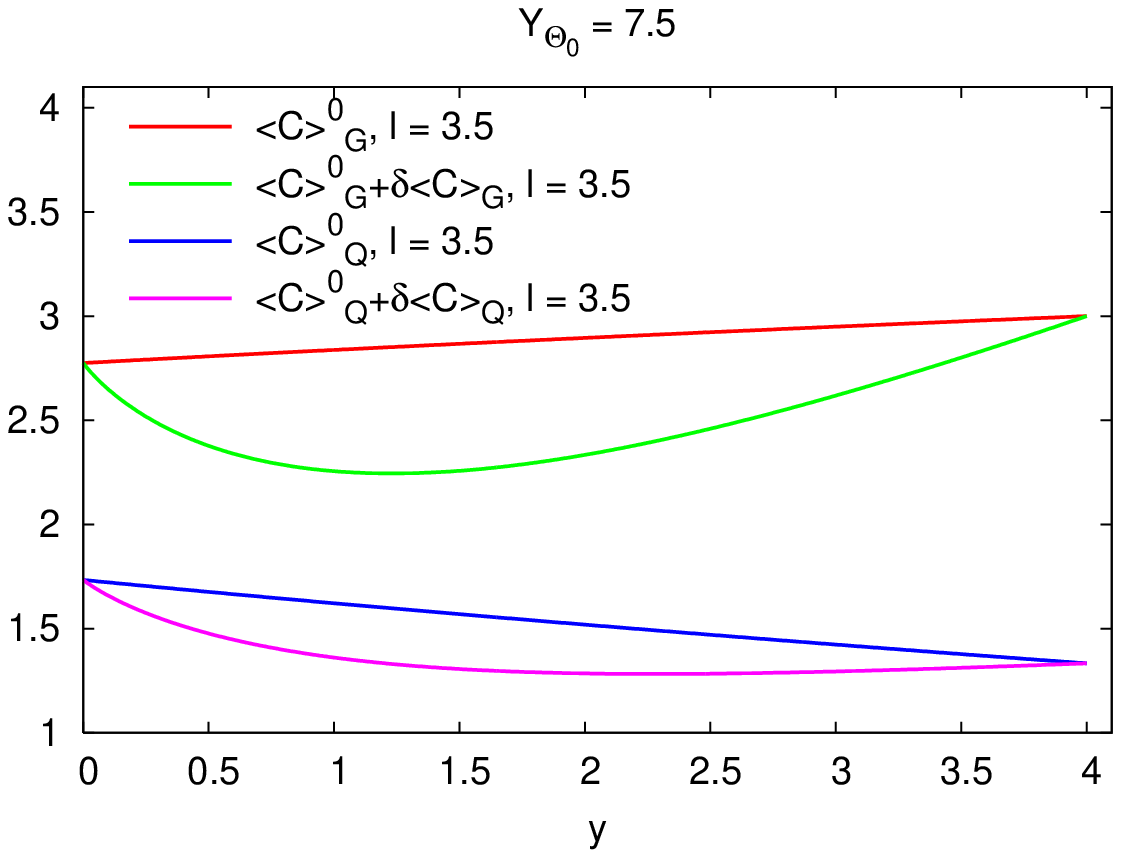, height=5truecm,width=7.5truecm}
\end{center}

\centerline{\em Fig.~2: $<C>_{A_0}^0$ and $<C>_{A_0}^0 + \delta\!<C>_{A_0}$ for quark and gluon
jets, as functions of $y$,}

\centerline{\em for $Y_{\Theta_0}=7.5$, $\ell=2.5$ on the left and
$\ell=3.5$ on the right.}
}

\bigskip

Two types of MLLA corrections arise in our calculation, which are
easily visualized on Fig.~2:

$\ast$\ through the expansion (\ref{eq:rhodev}) around $u=1$,
the average color current $<C>^0_{A_0}$ gets modified by $\delta\!<C>_{A_0}
\leq 0$
of relative order ${\cal O}(\sqrt{\alpha_s})$; it is represented on Fig.~2
by the vertical difference between the straight lines ($<C>_{A_0}^0$) and
the curved ones ($<C>^0_{A_0} +\delta\!<C>_{A_0}$);

$\ast$\ the derivative of $<C>^0_{A_0}$ with respect to $y$ is itself of
relative order ${\cal O}(\sqrt{\alpha_s})$ with respect to that of $\tilde
D_g$; it is the slopes of the straight lines in Fig.~2.

The $y$ derivatives of $<C>_{A_0}^0 +
\delta\!<C>_{A_0}$  differ from the ones of the leading $<C>^0_{A_0}$;
this effect combines the two types of MLLA corrections mentioned above:
the derivation of $<C>$ with respect to $y$ and the existence of
$\delta\!<C>$.

For  $Y_{\Theta_0}=7.5$, the $\delta\!<C>$ correction can represent
$50\%$ of $<C>_{g}$ at  $\ell=2.5$ and $y\approx 1.5$; for
higher values of $\ell$ (smaller $x$), as can be seen on the right figure,
its importance decreases; it is remarkable that, when $\delta\!<C>$ is
large, the corrections to $\frac{d<C>}{dy}$ with respect to
$\frac{d<C>^0}{dy}$ become small, {\em and vice-versa}: at both extremities
of the curves for the color current, the $\delta\!<C>$ corrections vanish,
but their slopes are very different from the ones of the
 straight lines corresponding to $<C>^0$.

So, all corrections that we have uncovered are potentially
large, even $\frac{d\delta\!<C>}{dy}$, which is the $y$ derivative of a
MLLA corrections. This raises the question of the validity of our
calculations. Several conditions need to be fulfilled at the same time:

$\ast$\ one must stay in the perturbative regime, which needs $y_1\geq 1$
($k_\perp > 2.72 \Lambda_{QCD} \approx .7\,$GeV;
this condition excludes in particular the zone of very large increase of
$\frac{d^2N}{d\ell_1\,d\ln k_\perp}$ when $y_1 \to 0$ (this property is
linked to the divergence of the running
coupling constant of QCD $\alpha_s(k_\perp^2) \to \infty$
when $k_\perp \to \Lambda_{QCD}$).

$\ast$\ $x$ must be small, that is $\ell$ large enough, since this is the
limit at which we have obtained analytical results; we see on Fig.~2 that
it cannot go reasonably below $\ell = 2.5$;
this lower threshold turns out to be of the same order magnitude as the one found
in the forthcoming study of 2-particle correlations
inside one jet in the MLLA approximation \cite{Perez2};

$\ast$\ (MLLA) corrections to the leading behavior must stay under control
(be small ``enough''); if one only looks at the size of the $\delta\!<C>$
corrections at $Y_{\Theta_0}=7.5$, it would be very tempting to
exclude $y\in [.5, 2.5]$; however this is
without taking into account the $y$ derivatives of $<C>$, which also play an
important role, as stressed above;
our attitude, which will be confirmed or not by experimental results, is to
only globally constrain the overall size of all corrections by setting $x$
small enough.

\bigskip

Would the corrections become
excessively large, the expansion (\ref{eq:rhodev}) should be pushed one
step further, which corresponds to next-to-MLLA (NMLLA) corrections; this
should then be associated with NMLLA evolution equations for the inclusive
spectrum, which lies out of the scope of the present work. 

\medskip

Though $\delta\!<C>$ can be large, specially at small
values of $\ell$, the positivity of $<C>^0 + \delta\! <C>$ is always
preserved on the whole allowed range of $y$.

\medskip

The difference between the naive and MLLA calculations lies in 
neglecting or not  the evolution of the jet between $\Theta_0$ and
$\Theta$, or, in practice, in considering or not the average
color current $<C>_{A_0}$ as a constant.

\vskip .5 cm

We present below our results for a gluon and for a quark jet.
We choose two values  $Y_{\Theta_0} = 7.5$,  which can be associated with
the LHC environment
\footnote{Sharing equally the $14$ TeV of available center of mass energy
between the six constituent partons of the two colliding nucleons yields
$E\approx 2.3$ TeV by colliding parton, one considers a jet
opening angle of $\Theta \approx .25$ and $Q_0\approx\Lambda_{QCD}
\approx 250$\,MeV; this gives $Y=\ln\frac{E\Theta}{Q_0}\approx 7.7$.
\label{footnote:LHC}}
, and the unrealistic $Y_{\Theta_0}= 10$ (see appendix \ref{section:LEP}
for  $Y_{\Theta_0}=5.2$ and $5.6$, corresponding to the LEP and Tevatron
 working conditions).
For each value of $Y_{\Theta_0}$ we  make the plots for two values
of $\ell_1$, and compare one of them with the naive approach.

In the rest of the paper we always consider the limiting case
$Q_0 \to \Lambda_{QCD} \Leftrightarrow \lambda \approx 0$,
\begin{equation}
\lambda = \ln\frac{Q_0}{\Lambda_{QCD}}.
\label{eq:lambda}
\end{equation}

The curves stop at their kinematic limit $y_{1\,max}$ such that
$y_{1\,max} + \ell_1 = Y_{\Theta_0}$.

\vskip .7 cm

\subsection{$\boldsymbol{\displaystyle\frac{d^2N}{d\ell_1\,d\ln{ k_\perp}}}$
at small $\boldsymbol x_1$: gluon jet}
%%%%%%%%%%%%%%%%%%%%%%%%%%%%%%%%%%%%%%%%%%%%%%%%%%%%%%%%%%%%%%%%%%%%%%%%%%%%%%

\vskip .4cm

On Fig.~3 below is plotted the double differential distribution
$\frac{d^2N}{d\ell_1\,d\ln{ k_\perp}}$ of a parton
inside a gluon jet as a function of $y_1$ for different values of $\ell_1$
(fixed).

\vbox{
\begin{center}
\epsfig{file=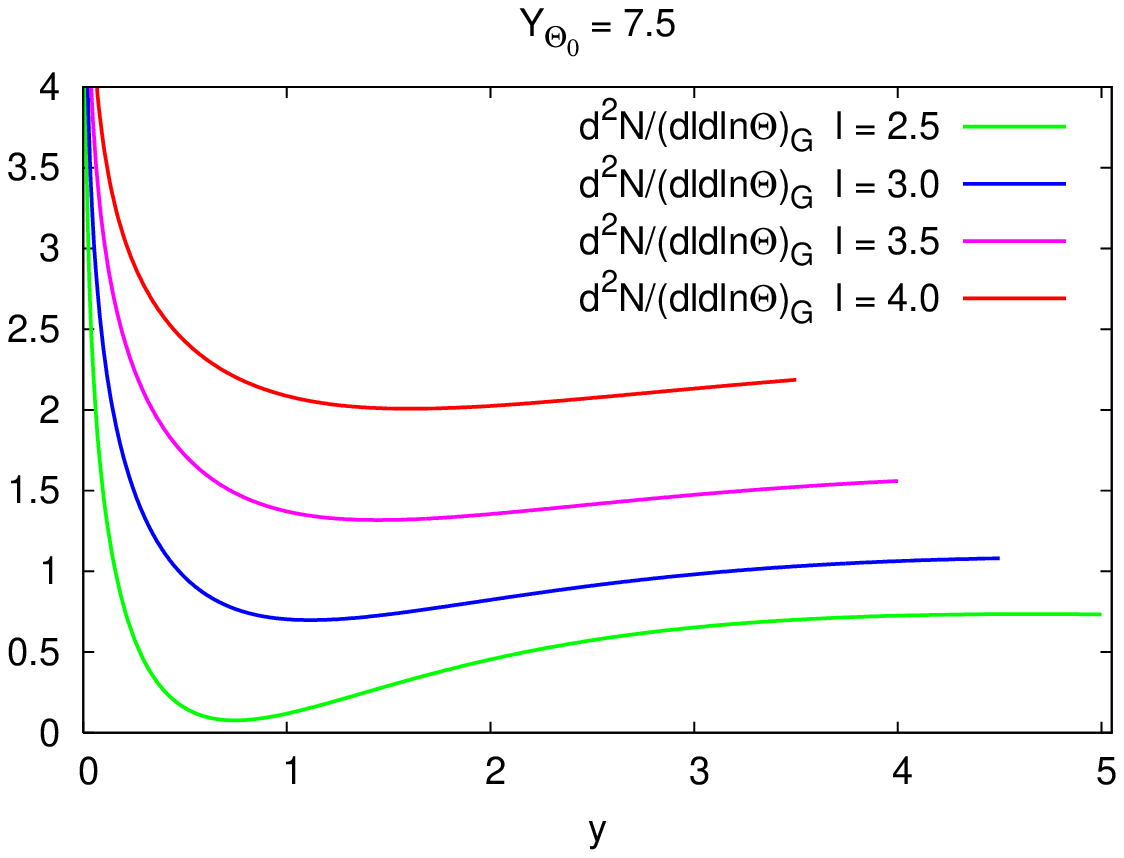, height=5truecm,width=7.5truecm}
\hfill
\epsfig{file=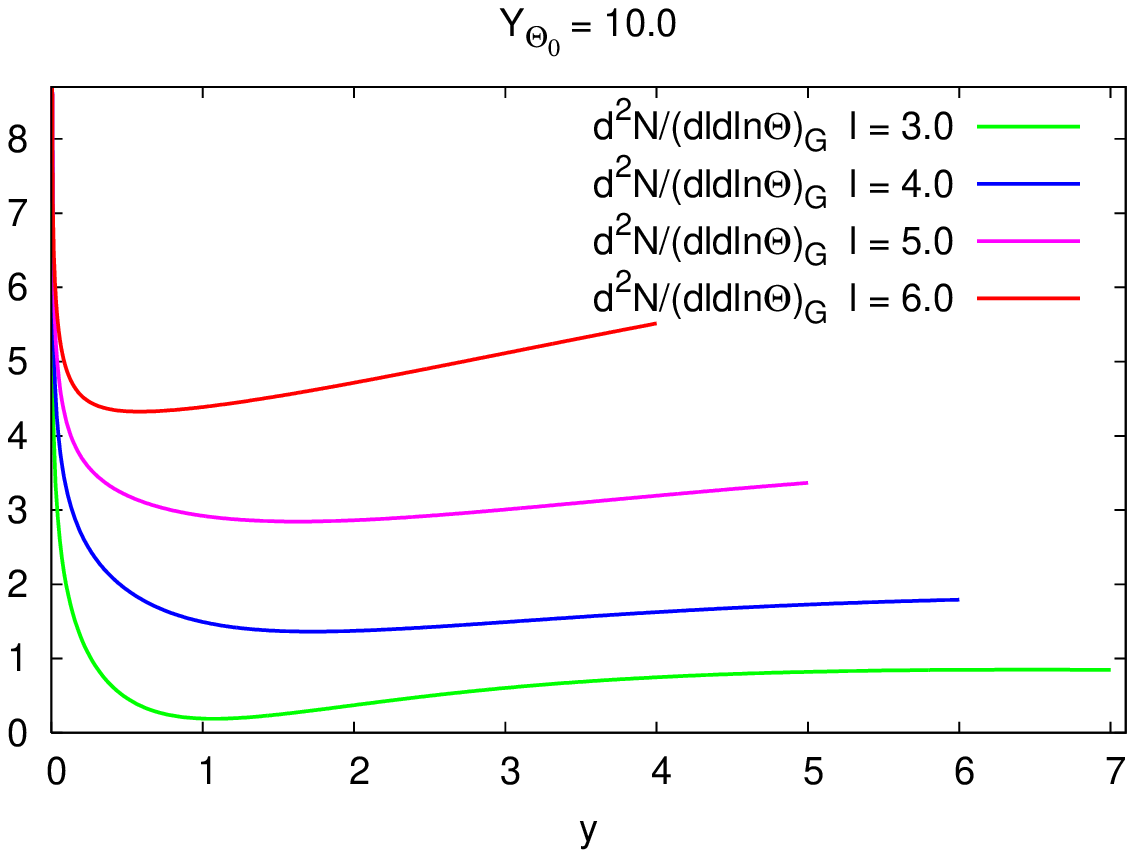, height=5truecm,width=7.5truecm}
\end{center}

\centerline{\em Fig.~3: $\frac{d^2N}{d\ell_1\;d\ln k_\perp}$ for a gluon jet.}
}

\bigskip
\bigskip

On Fig.~4 are compared, for a given value of  $\ell_1$,
the two following cases:

\medskip

$\ast$ the first corresponds to the full formul{\ae} (\ref{eq:Fgq})
(\ref{eq:gen});

$\ast$ the second corresponds to the naive approach (see the definition
above (\ref{eq:Cnaive}))
\begin{equation}
 \left(\frac{d^2N}{d\ell_1\,d\ln{ k_\perp}}\right)^{naive}_{g}
=\frac{d}{dy_1}\tilde D_g(\ell_1,y_1);
\label{eq:EAgnaive}
\end{equation}
$\displaystyle\frac{d\tilde D_g(\ell_1,y_1)}{dy_1}$ is given in (\ref{eq:derivy}).

\vbox{
\begin{center}
\epsfig{file=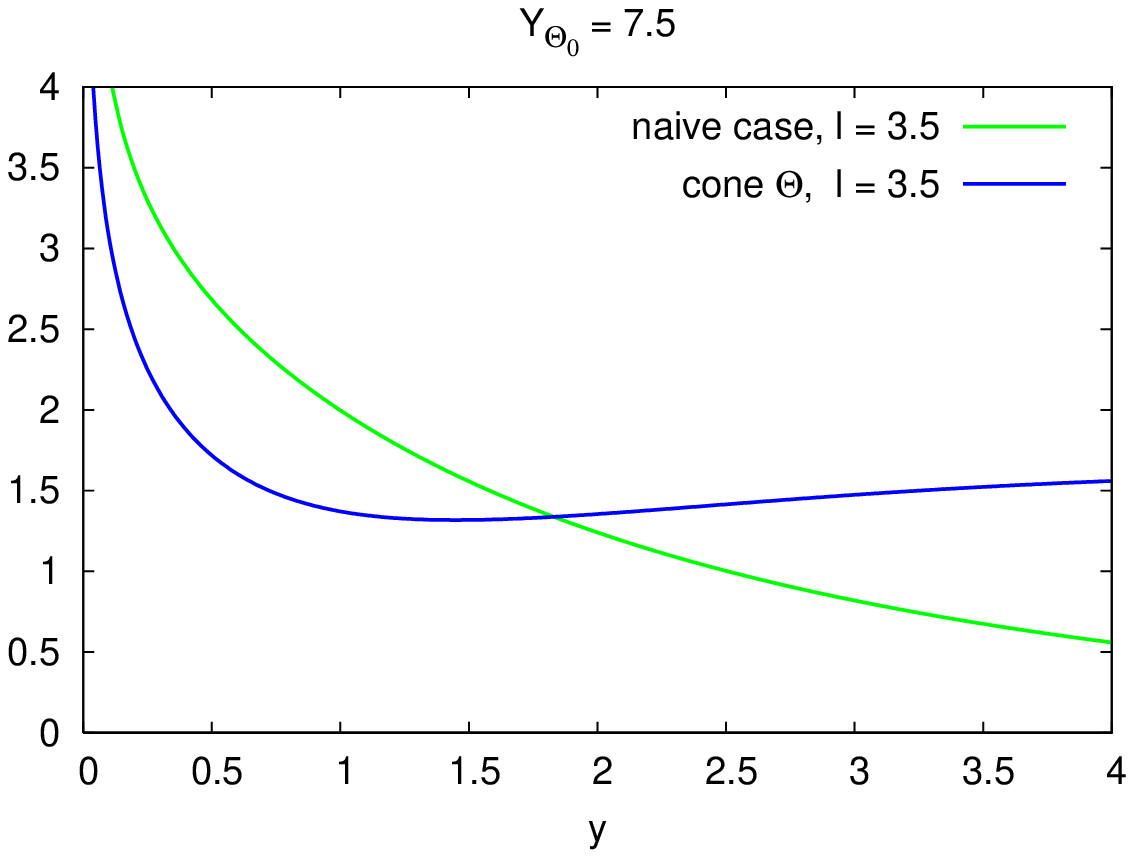, height=5truecm,width=7.5truecm}
\hfill
\epsfig{file=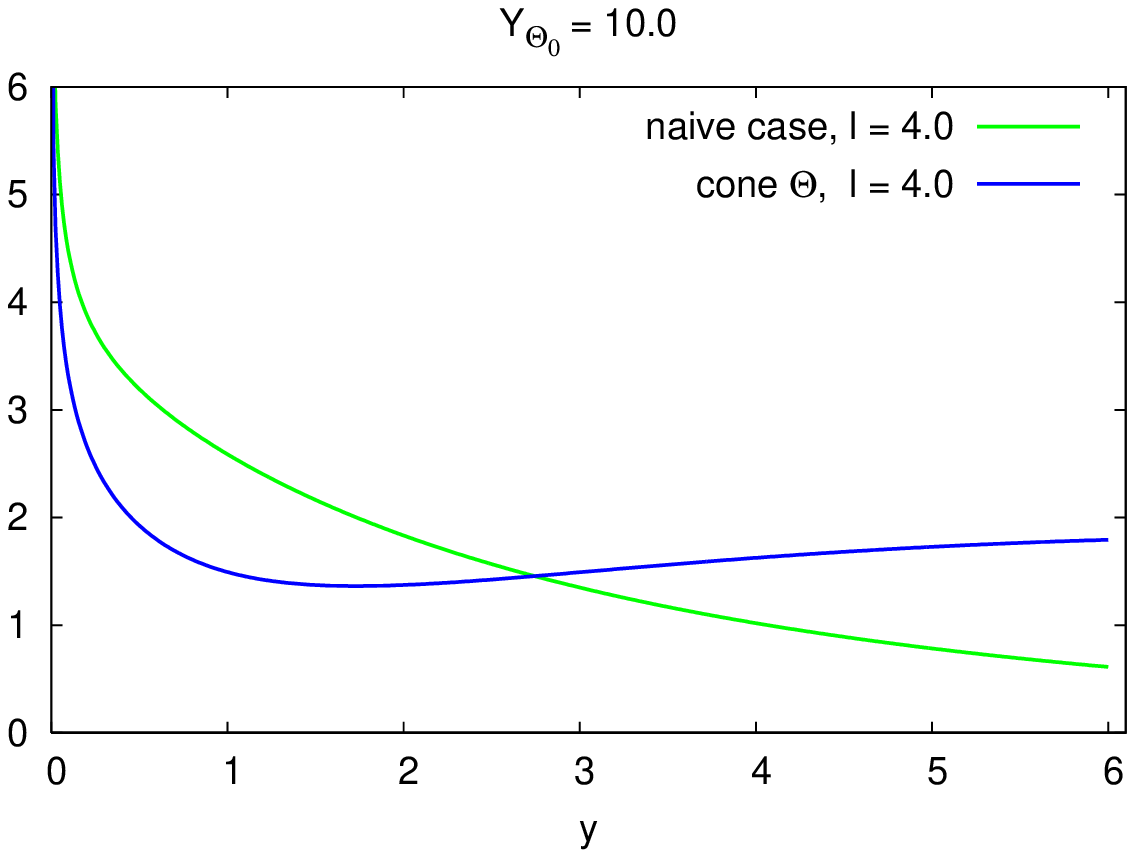, height=5truecm,width=7.5truecm}
\end{center}

\centerline{\em Fig.~4: $\frac{d^2N}{d\ell_1\;d\ln k_\perp}$ for a gluon jet
at fixed $\ell_1$,  MLLA and naive approach.}
}

\bigskip
\bigskip

The raise of the distribution at large $k_\perp$ is due to the positive
corrections already mentioned in the beginning of this section, which
arise from the evolution of the jet between $\Theta$ and $\Theta_0$.

\vskip .7 cm

\subsection{$\boldsymbol{\displaystyle\frac{d^2N}{d\ell_1\,d\ln{ k_\perp}}}$
at small $\boldsymbol x_1$: quark jet}
%%%%%%%%%%%%%%%%%%%%%%%%%%%%%%%%%%%%%%%%%%%%%%%%%%%%%%%%%%%%%%%%%%%%%%%%%%%%%

\vskip .4cm

On Fig.~5 is plotted
the double differential distribution
$\frac{d^2N}{d\ell_1\,d\ln{ k_\perp}}$ of a parton
inside a quark jet as a function of $y_1$ for different values of $\ell_1$
(fixed).

\vbox{
\begin{center}
\epsfig{file=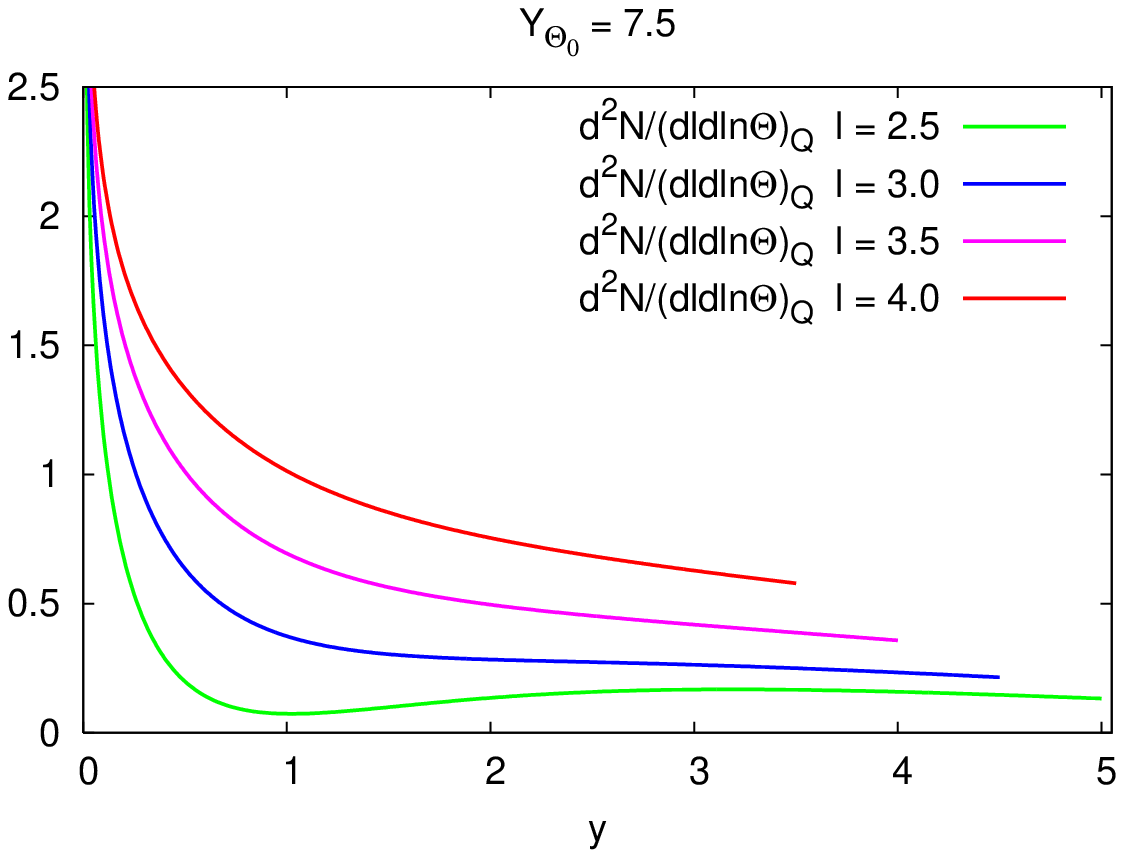, height=5truecm,width=7.5truecm}
\hfill
\epsfig{file=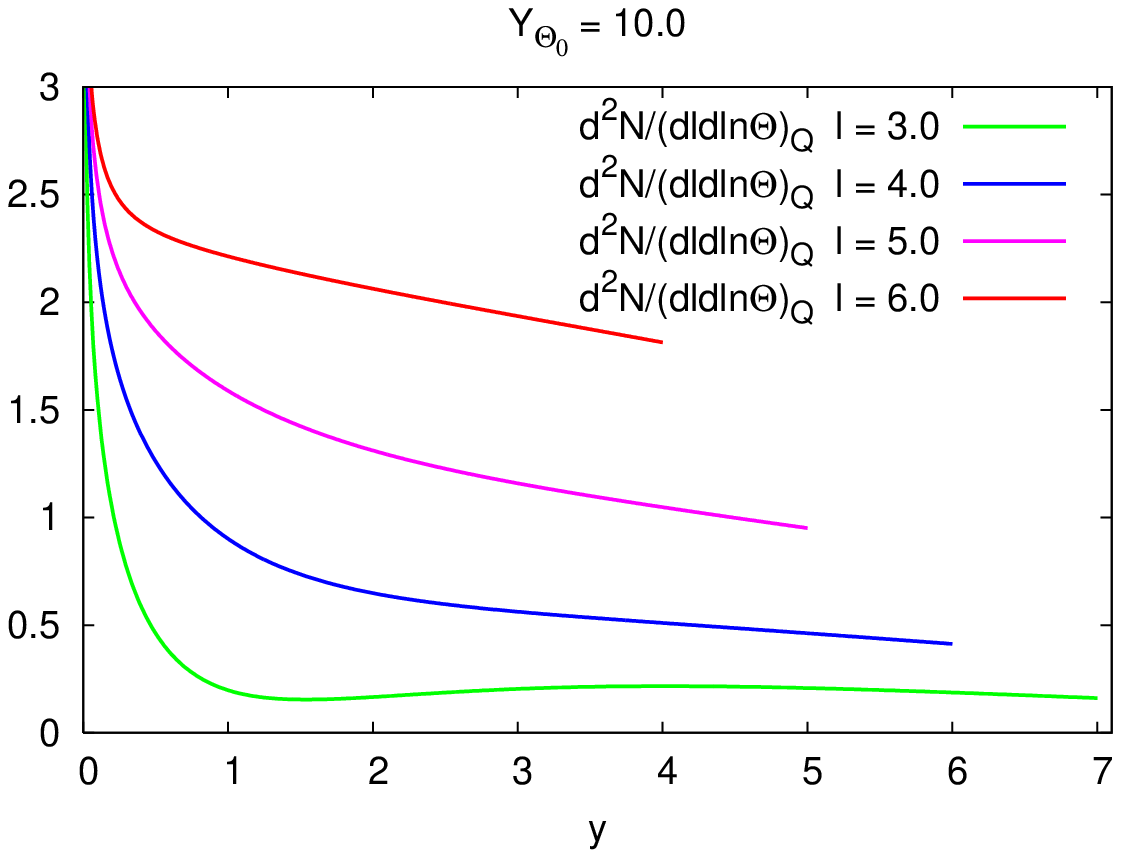, height=5truecm,width=7.5truecm}
\end{center}

\centerline{\em Fig.~5: $\frac{d^2N}{d\ell_1\;d\ln k_\perp}$ for a quark jet.}

}

\bigskip
\bigskip

On Fig.~6 are compared, for a given $\ell_1$ fixed,
the full formul{\ae} (\ref{eq:Fgq})
(\ref{eq:gen})  and the naive approach
\begin{equation}
 \left(\frac{d^2N}{d\ell_1\,d\ln k_\perp}\right)_{q}^{naive}
=\frac{C_F}{N_c}\frac{d}{dy_1} \tilde D_g(\ell_1,y_1).
\label{eq:EAqnaive}
\end{equation}

\vbox{
\begin{center}
\epsfig{file=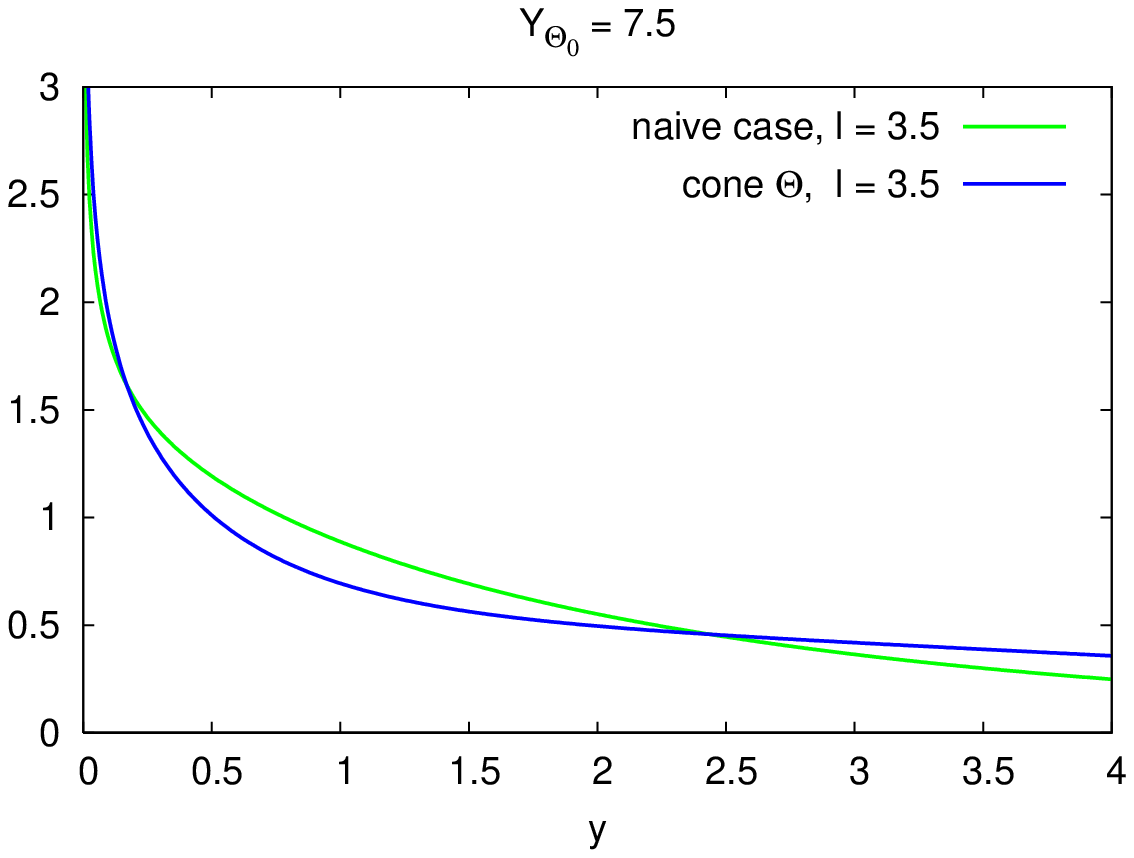, height=5truecm,width=7.5truecm}
\hfill
\epsfig{file=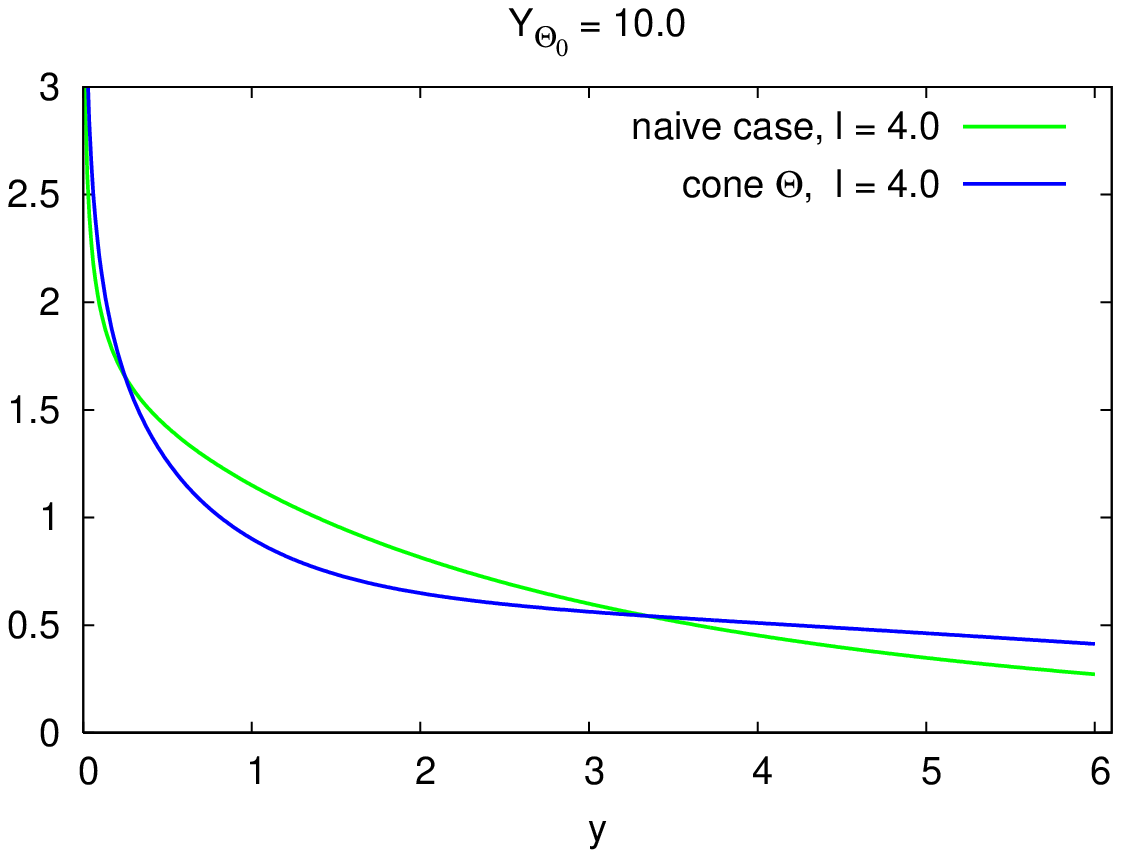, height=5truecm,width=7.5truecm}
\end{center}

\centerline{\em Fig.~6: $\frac{d^2N}{d\ell_1\;d\ln k_\perp}$ for a quark jet
at fixed $\ell_1$,  MLLA and naive approach.}

}

\bigskip
\bigskip

We note, like for gluon jets, at large $y$, a (smaller) increase of the
distribution,
due to taking into account the jet evolution between $\Theta$ and
$\Theta_0$.

\vskip .7 cm

\subsection{Comments}
\label{subsection:correcs}
%%%%%%%%%%%%%%%%%%%%%%%%%%%%%

\vskip .4cm

The gluon distribution is always larger than the quark distribution; this
can also be traced in Fig.~2  which
measures in particular the ratio of the color currents $<C>_g/<C>_q$.

\medskip

The curves for  $\frac{d^2N}{d\ell_1\,d\ln{ k_\perp}}$
have been drawn for $\ell_1 \equiv \ln (1/x_1) \geq 2.5$; going below this
threshold exposes to excessively large MLLA corrections.

\medskip

The signs of the two types of MLLA corrections pointed at in subsection
\ref{subsection:colcur} vary with $y$: $\delta\!<C>$
always brings a negative correction to $<C>^0$, and to
$\frac{d^2N}{d\ell_1\,d\ln{ k_\perp}}$;  for $y \geq 1.5$,
the slope of $<C>$ is always larger that the one of
$<C>^0$, while for $y \leq 1.5$ it is the opposite. 
 It is accordingly not surprising
that, on Figs.~4 and 6, the relative positions of the curves corresponding to
the MLLA calculation and  to a naive calculation change with the value of $y$.
At large $y$, one gets a growing behavior
 of $\frac{d^2N}{d\ell_1\,d\ln{ k_\perp}}$ for gluon jets (Fig.~4),
and a slowly decreasing one for quark jets (Fig.~6), which could not have
been anticipated {\em a priori}.

\medskip

We study in appendix \ref{subsection:doubleDLA}, how  MLLA results
compare with DLA \cite{DLA} \cite{DLA1},
in which the running of $\alpha_s$ has been ``factored out''.

\vskip .7 cm

%%%%%%%%%%%%%%%%%%%%%%%%%%%%%%%%%%%%%%%%%%%%%%%%%%%%%%%%
\section{INCLUSIVE $\boldsymbol{k_\perp}$ DISTRIBUTION
$\boldsymbol{\displaystyle\frac{dN}{d\ln k_\perp}}$}
\label{section:ktdist}
%%%%%%%%%%%%%%%%%%%%%%%%%%%%%%%%%%%%%%%%%%%%%%%%%%%%%%%%

\vskip .4cm

Another quantity of interest is the  inclusive $k_\perp$ distribution which
is defined by
\begin{equation}
\left(\frac{dN}{d\ln k_\perp}\right)_{g\ or\ q} = \int dx_1
\left(\frac{d^2N}{dx_1\, d\ln k_\perp}\right)_{g\ or\ q}
\equiv \int_{\ell_{min}}^{Y_{\Theta_0}-y} d\ell_1
\left(\frac{d^2N}{d\ell_1\, d\ln k_\perp}\right)_{g\ or\ q};
\label{eq:ktdist}
\end{equation}
it measures the transverse momentum distribution of one particle with
respect to the direction of the energy flow (jet axis).

We have introduced in (\ref{eq:ktdist}) a lower bound of integration
$\ell_{min}$ because our calculations are valid for small $x_1$, that is for
large $\ell_1$. In a first step we take $\ell_{min}=0$, then vary it to
study the sensitivity of the calculation to the region of large $x_1$.

We plot below the inclusive $k_\perp$ distributions for gluon and quark
jets, for the same two values $Y_{\Theta_0}=7.5$ and
$Y_{\Theta_0}=10$ as above, and compare them, on the same graphs, with the
``naive calculations'' of the same quantity.

\vskip .7 cm

\subsection{Gluon jet; $\boldsymbol{\ell_{min}=0}$}
%%%%%%%%%%%%%%%%%%%%%%%%%%%%%%%%%%%%%%%%%%%%%%%%%%%

\vskip .4cm

\vbox{
\begin{center}
\epsfig{file=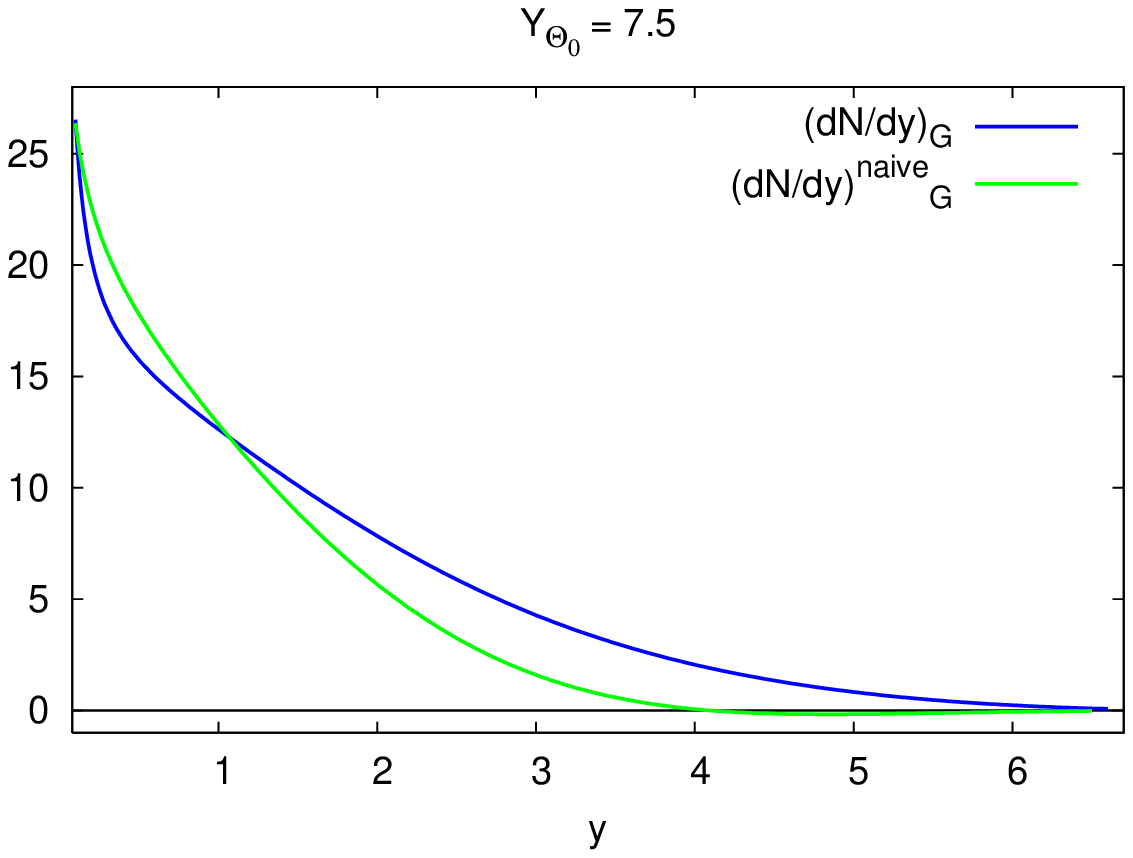, height=5truecm,width=7.5truecm}
\hfill
\epsfig{file=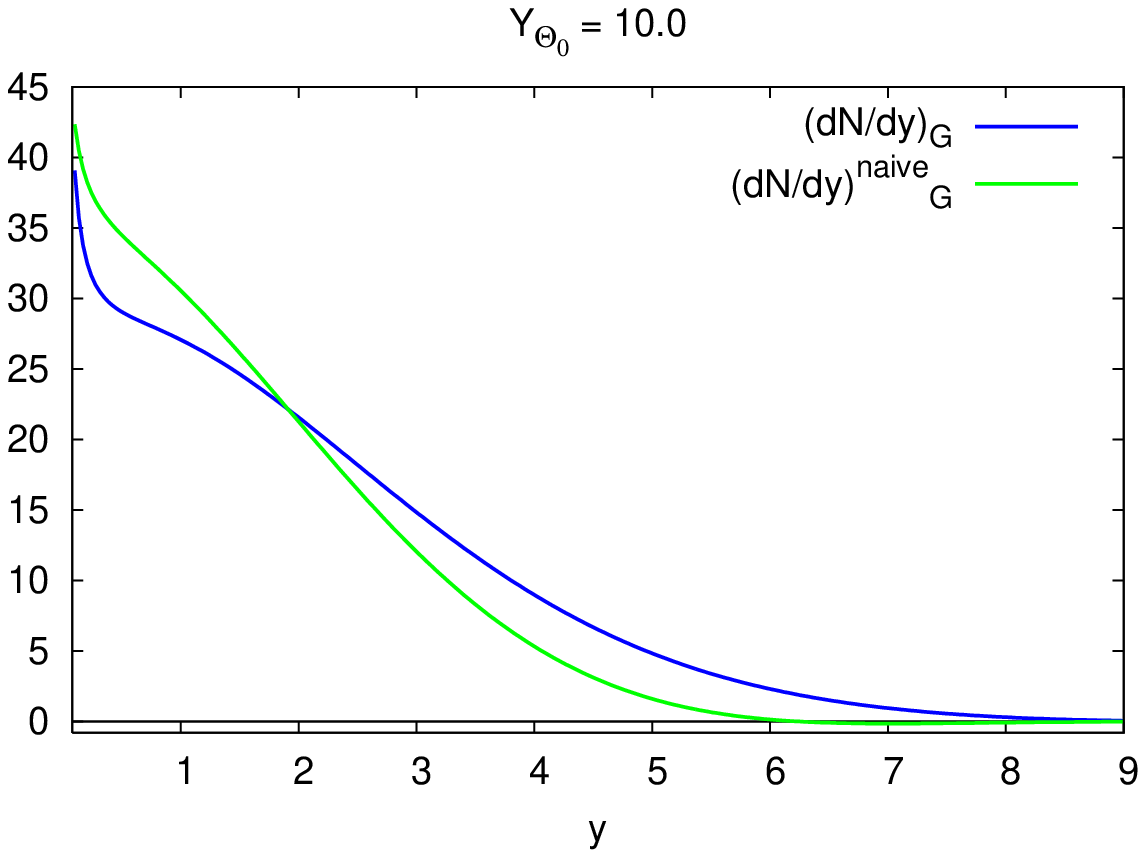, height=5truecm,width=7.5truecm}
\end{center}

\centerline{\em Fig.~7:  $\frac{d{N}} {d\ln k_\perp}$  for a gluon jet,
MLLA and naive approach,}

\centerline{\em for $\ell_{min=0}$,
$Y_{\Theta_0} =7.5$ and $Y_{\Theta_0}=10$.}
}

\vskip .3cm

\vbox{
\begin{center}
\epsfig{file=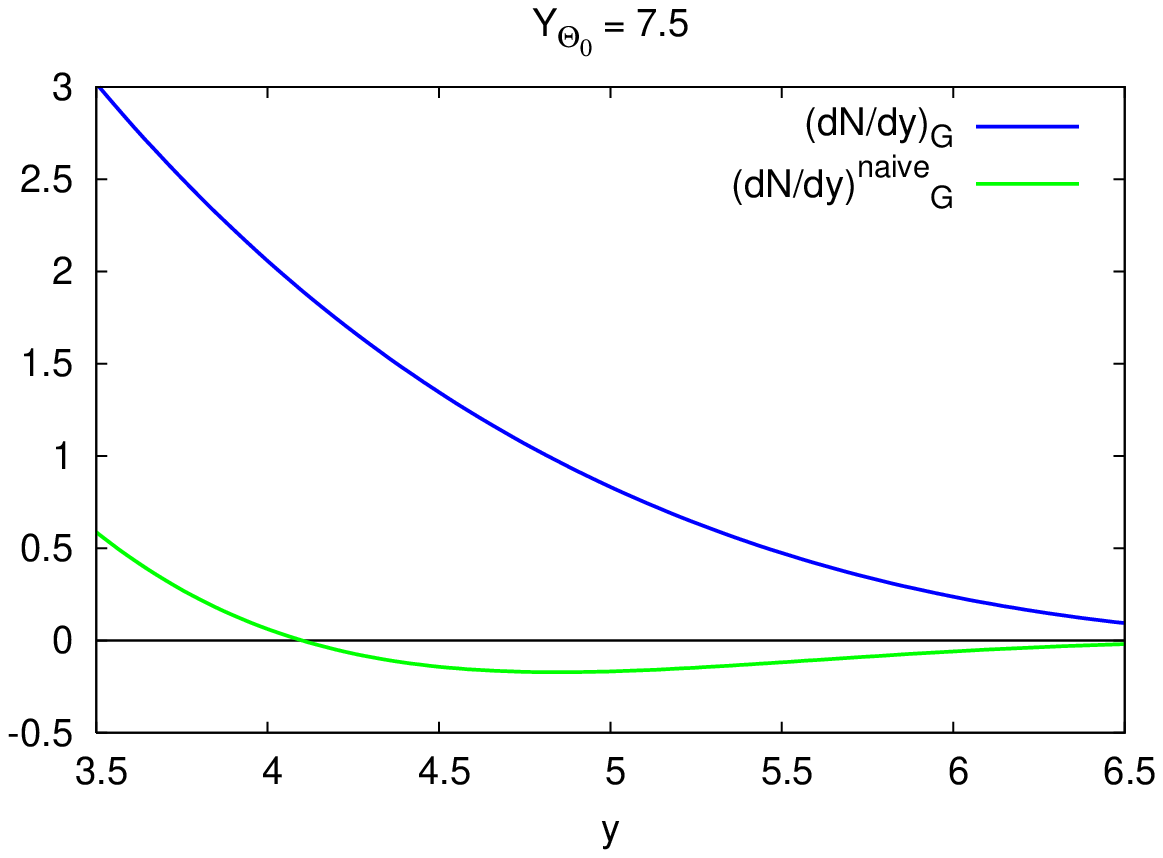, height=5truecm,width=7.5truecm}
\hfill
\epsfig{file=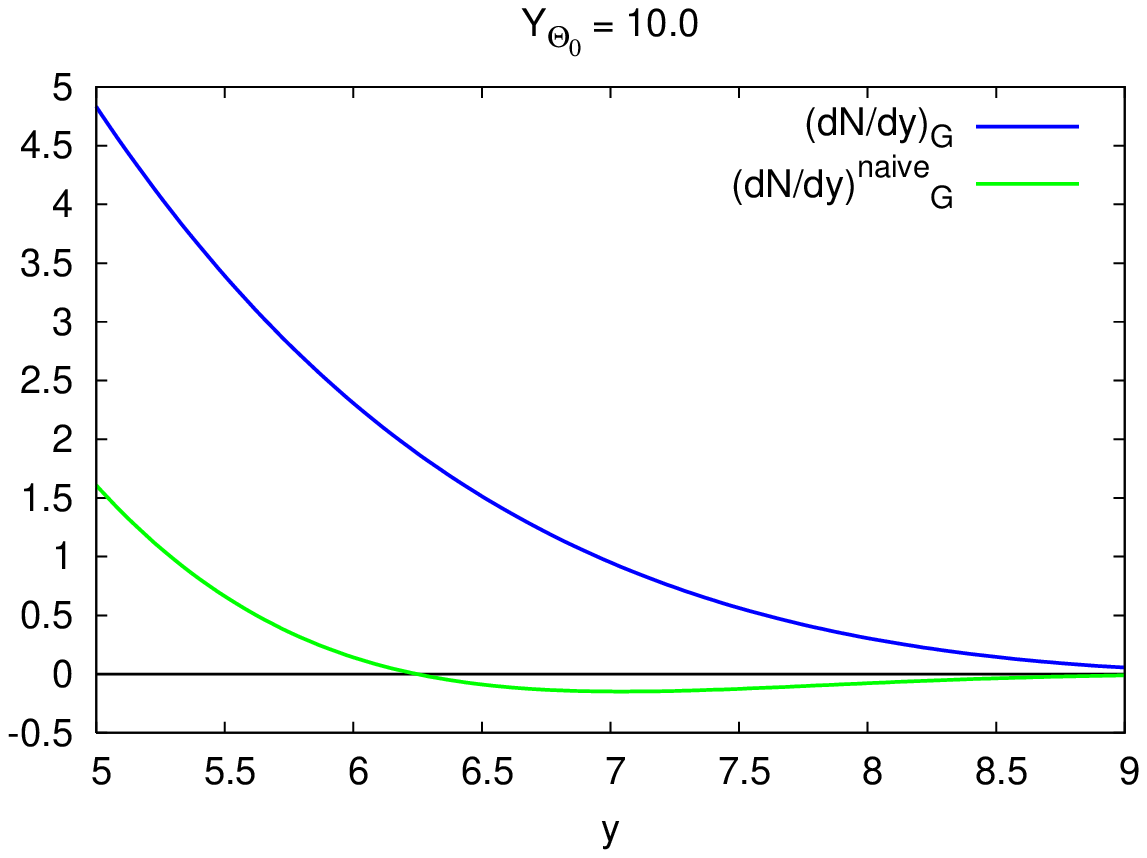, height=5truecm,width=7.5truecm}
\end{center}

\centerline{\em Fig.~8: enlargements of Fig.~7 at large $k_\perp$}
}

\vskip .7 cm

\subsection{Quark jet; $\boldsymbol{\ell_{min}=0}$}
%%%%%%%%%%%%%%%%%%%%%%%%%%%%%%%%%%%%%%%%%%%%%%%%%%%

\vskip .4cm

%\vbox{
\begin{center}
\epsfig{file=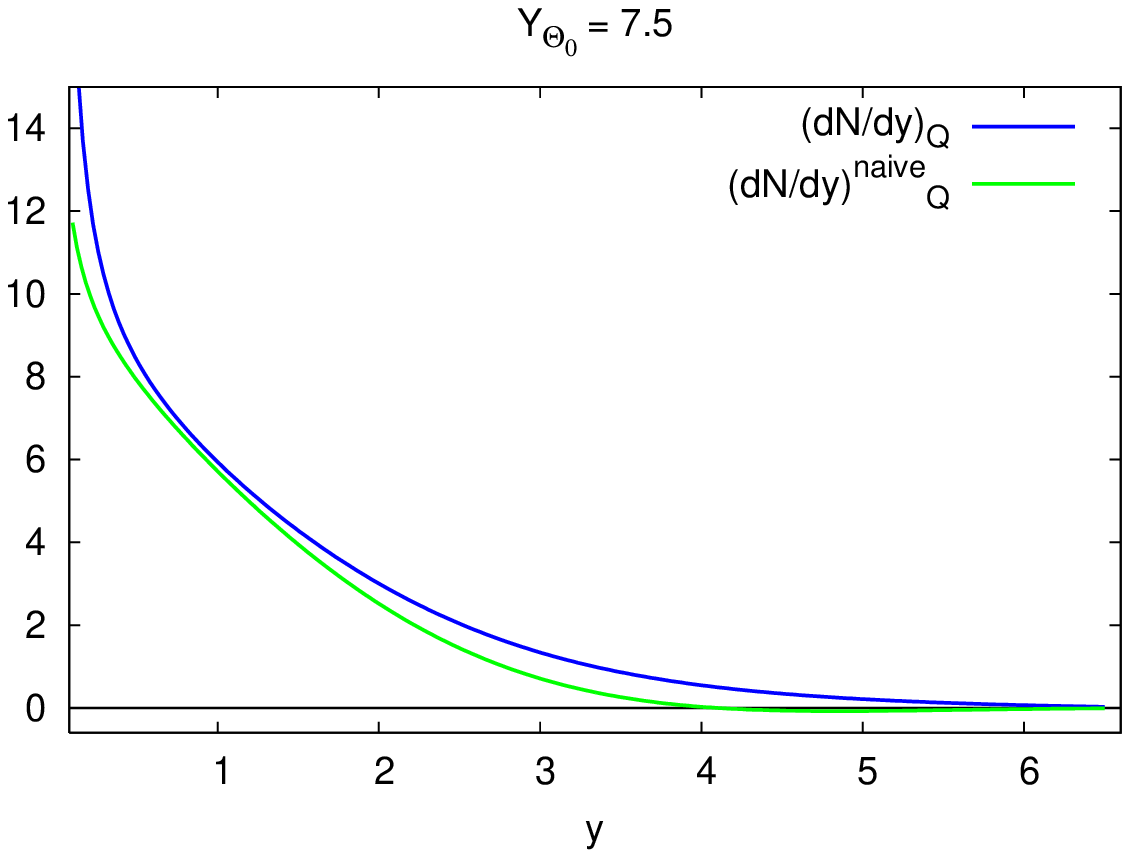, height=5truecm,width=7.5truecm}
\hfill
\epsfig{file=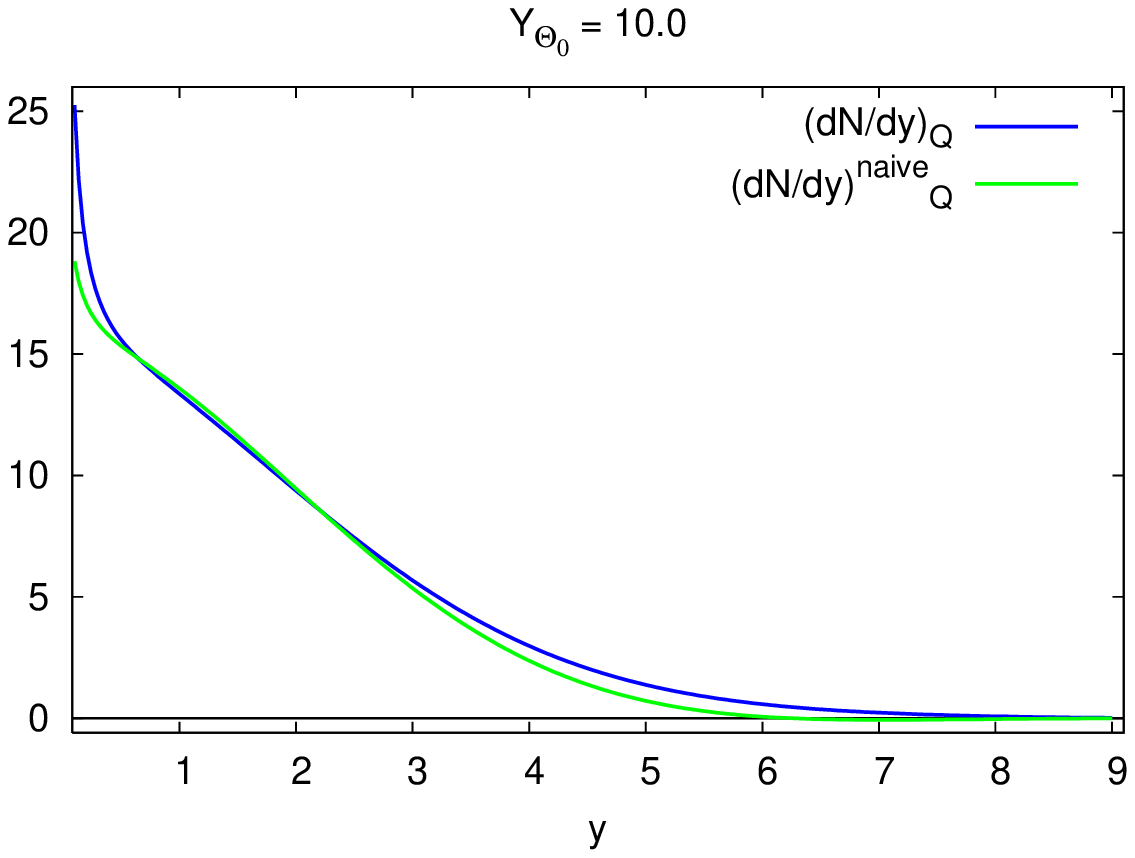, height=5truecm,width=7.5truecm}
\end{center}

\centerline{\em Fig.~9:  $\frac{d{N}} {d\ln k_\perp}$ for a quark jet,
MLLA and naive approach,}

\centerline{\em  for $\ell_{min=0}$,
$Y_{\Theta_0} =7.5$ and $Y_{\Theta_0}=10$.}
%}

\vskip .3cm

%\vbox{
\begin{center}
\epsfig{file=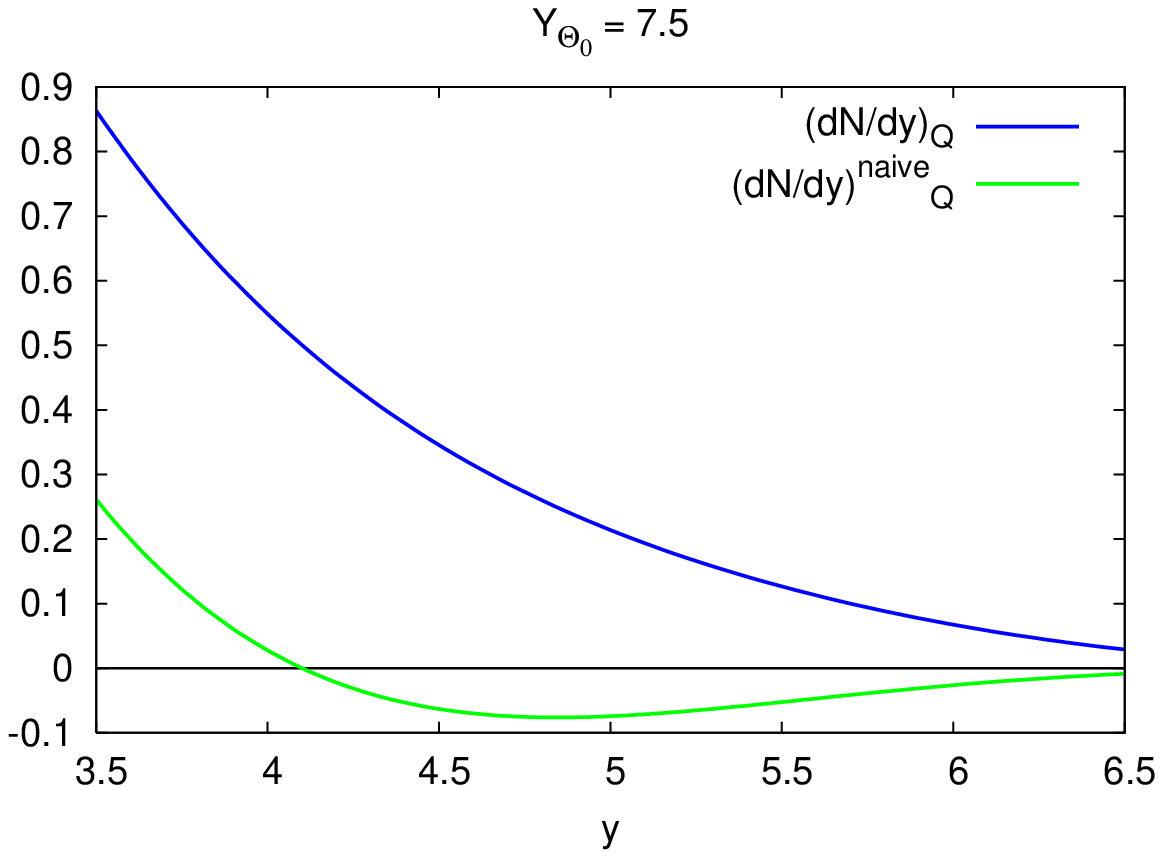, height=5truecm,width=7.5truecm}
\hfill
\epsfig{file=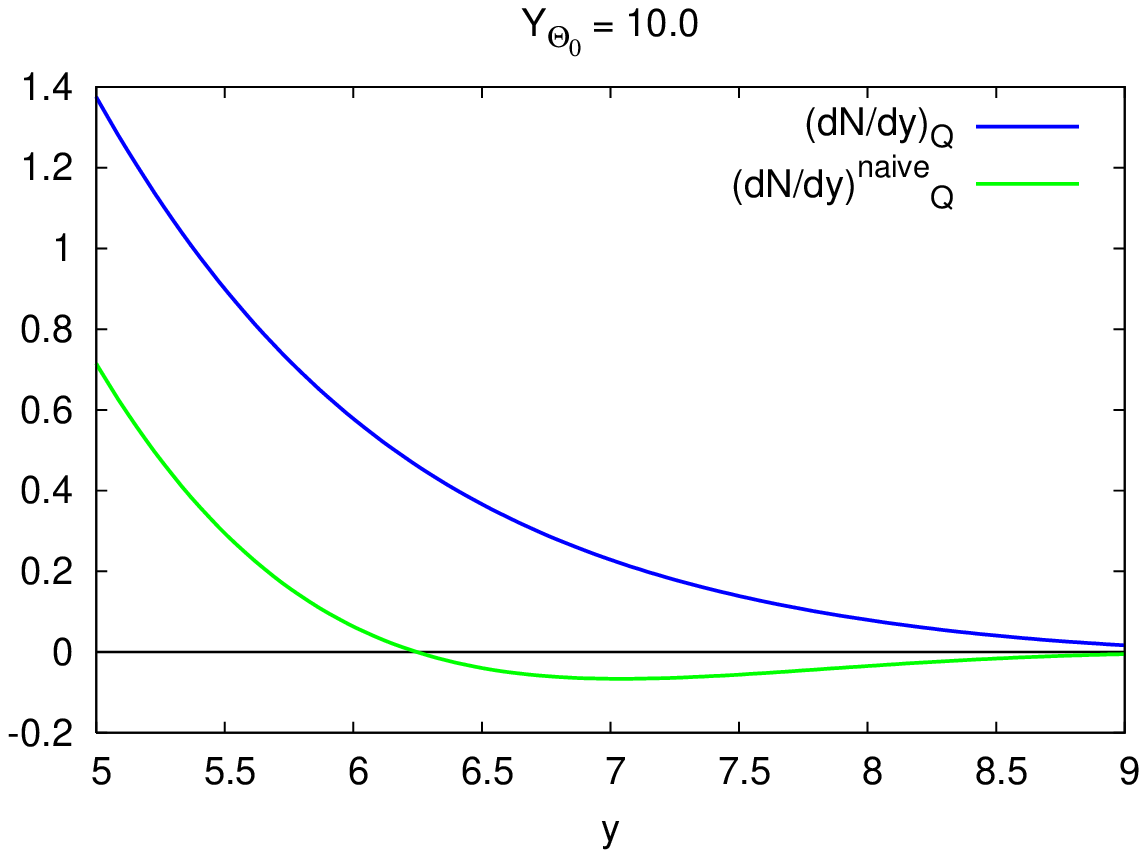, height=5truecm,width=7.5truecm}
\end{center}

\centerline{\em Fig.~10: enlargements of Fig.~9 at large $k_\perp$}
%}

\vskip .7 cm

\subsection{Role of the lower limit of integration $\boldsymbol{\ell_{min}}$}
%%%%%%%%%%%%%%%%%%%%%%%%%%%%%%%%%%%%%%%%%%%%%%%%%%%%%%%%%%%%%%%%%%%%%%%%%%%%%

\vskip .4cm

To get an estimate of the sensitivity of the calculation of
$\frac{dN}{d\ln\,k_\perp}$ to the lower bound of integration
in (\ref{eq:ktdist}), we plot in Fig.~11 below the two results obtained at
$Y_{\Theta_0}=7.5$ for $\ell_{min}=2$ and $\ell_{min}=0$, for a gluon jet
(left) and a quark jet (right).

\bigskip

\vbox{
\begin{center}
\epsfig{file=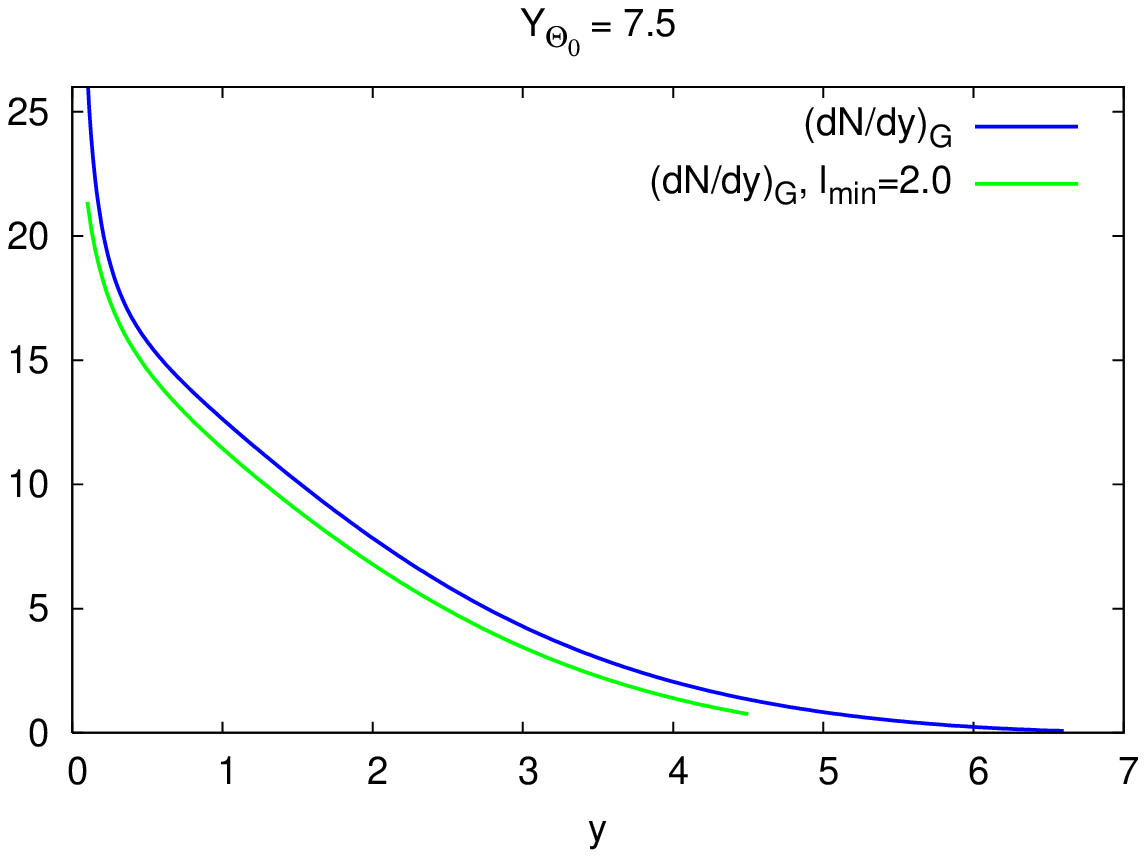, height=5truecm,width=7.5truecm}
\hfill
\epsfig{file=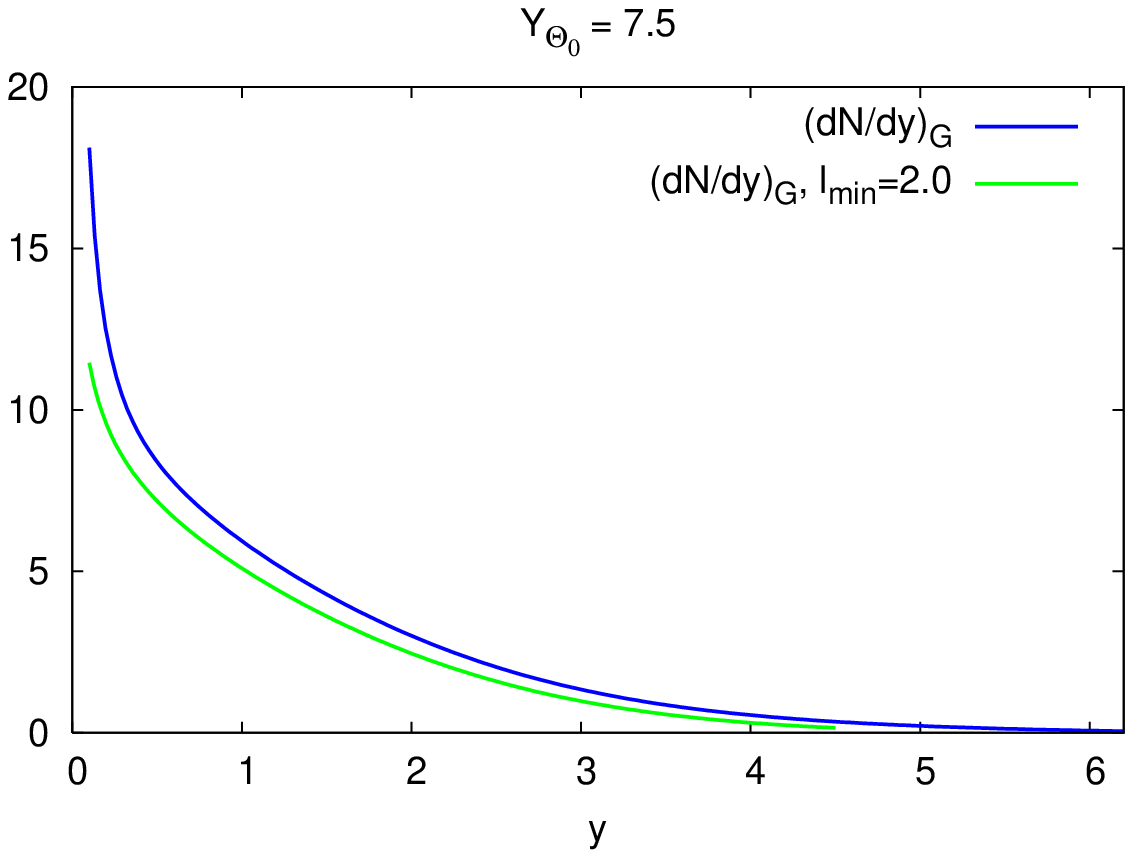, height=5truecm,width=7.5truecm}
\end{center}

\centerline{\em Fig.~11: $\frac{dN}{d\ln\,k_\perp}$ with $\ell_{min}=2$ and
$\ell_{min}=0$}
\centerline{\em for gluon (left) and quark (right) jet.}
}

\bigskip
\bigskip

The shapes of the corresponding distributions are identical; they only
differ by a vertical shift which is small in the perturbative region
$y \geq 1$ (restricting the domain of integration -- increasing
$\ell_{min}$ -- results as expected in a decrease of $\frac{dN}{d\ln
k_\perp}$).
This shows that, though our calculation is only valid at
small $x_1$, the sensitivity of the final result to this parameter is small.

\vskip .7 cm

\subsection{Discussion}
\label{subsection:discuss}
%%%%%%%%%%%%%%%%%%%%%%%%%%%%

\vskip .4cm

MLLA corrections are seen on Fig.~8 and Fig.~10 to cure the problems
of positivity which occur in the naive approach.

The range of $\ell_1$ integration in the definition (\ref{eq:ktdist}) of
$\frac{dN}{d\ln k_\perp}$ should be such that, at least,
 its upper bound corresponds to $x_1$ small enough; we have seen in the
discussion of MLLA corrections to
the color current in subsection \ref{subsection:colcur} that one should
reasonably consider $\ell_1 \geq 2.5$; at fixed $Y_{\Theta_0}$ this yields
the upper bound $y_1 \leq Y_{\Theta_0}-2.5$, that is, at LHC $y_1 \leq 5$.

On the other side, the perturbative regime we suppose to start at $y_1 \geq
1$. These  mark the limits of the interval where our
calculation can be trusted $1 \leq y_1 \leq 5$ at LHC.
For $y_1 <1$ non-perturbative corrections will dominate, and for
$y_1 > Y_{\Theta_0} - \ell_1^{min} \approx Y_{\Theta_0} -2.5$, the
integration defining $\frac{dN}{d\ln k_\perp}$ ranges over values of $x_1$
which lie outside our small $x$ approximation and for which the MLLA
corrections become accordingly out of control.

On the curves of Figs.~7 and 9 at $Y_{\Theta_0}=10$, the small $y$ region
exhibits a bump which comes from the competition between two phenomena: the
divergence of $\alpha_s(k_\perp^2)$ when $k_\perp \to Q_0$ and coherence
effects which deplete multiple production at very small momentum.
The separation of these two effects is still more visible at
$Y_{\Theta_0}=15$, which is studied in appendix \ref{subsection:ktDLA},
where a comparison with DLA calculations is performed.
At smaller $Y_{\Theta_0}$, the divergence of $\alpha_s$ wins
over coherence effects and the bump disappears.

The curves corresponding to the LEP and Tevatron working conditions are
given in appendix \ref{section:LEP}.

\vskip .7 cm

\subsubsection{Mixed quark and gluon jets}
\label{subsection:mixed}
%%%%%%%%%%%%%%%%%%%%%%%%%%%%%%%%%%%%%%%%%%%

\vskip .4cm

In many experiments, the nature of the jet (quark or gluon) is not
determined, and one simply detects outgoing hadrons, which can originate
from  either type; one then introduces a ``mixing'' parameter $\omega$,
which is to be determined experimentally, such that, for example if one
deals with the inclusive $k_\perp$ distribution
\begin{equation}
\left(\frac{dN}{d\ln k_{\perp}}\right)_{mixed}=
\omega\left(\frac{dN}{d\ln k_{\perp}}\right)_{g}+\left(1-\omega\right)
\left(\frac{dN}{d\ln k_{\perp}}\right)_{q}.
\end{equation}
It is in this framework that forthcoming data from the LHC will be
compared with our theoretical predictions; since outgoing
charged hadrons are detected, one introduces the phenomenological
parameter ${\cal K}^{ch}$ \cite{EvEq}\cite{KO}
 normalizing  partonic distributions to the ones
of charged hadrons
\begin{equation}
\Bigg(\frac {dN}{d\ln k_\perp}\Bigg)^{ch} = {\cal K}^{ch}\Bigg(\frac
{dN}{d\ln k_\perp}\Bigg)_{mixed}.
\label{eq:K}
\end{equation}

\vskip .7cm

%%%%%%%%%%%%%%%%%%%%%%%%
\section{CONCLUSION}
%%%%%%%%%%%%%%%%%%%%%%%%

\vskip .4cm

After deducing a general formula, valid for all $x$, for the double
differential 2-particle inclusive cross section for  jet production in a
hard collision process,
the exact solutions of the MLLA evolution equations 
\cite{Perez} have been used to perform a small $x$ calculation of the
double differential 1-particle inclusive distributions and of the inclusive
$k_\perp$ distributions for quark and gluon jets.

Sizable differences with the naive approach in which one forgets the
jet evolution between its  opening angle $\Theta_0$ and the emission angle
$\Theta$  have been found; their role is emphasized to recover, in
particular, the positivity of the distributions.

MLLA corrections increase with  $x$ and decrease when the transverse
momentum $k_\perp$ of the outgoing hadrons gets larger;
that they stay ``within control''
requires in practice that the small $x$ region should not be extended
beyond $\ell < 2.5$; it is remarkable that
similar bounds arise in the study of 2-particle correlations
\cite{Perez2}. At fixed $Y_{\Theta_0}$, the lower bound for $\ell$
translates into an upper bound for $y$; this fixes in particular the upper
limit of confidence for our calculation of $\frac{dN}{d\ln k_\perp}$; above
this threshold, though $k_\perp$ is larger (more
``perturbative''), the small $x$ approximation is no longer valid.

The ``divergent'' behavior of the MLLA distributions for $y \to 0$ forbids
extending the confidence
domain of MLLA  lower that $y \geq 1$, keeping away from the singularity
of $\alpha_s(k_\perp^2)$ when $k_\perp \to \Lambda_{QCD}$.

The two (competing) effects of coherence (damping of multiple production at
small momentum) and  divergence of
$\alpha_s(k_\perp^2)$ at small $k_\perp$ for the inclusive $k_\perp$
distribution have been exhibited.

MLLA and DLA calculations have been compared; in ``modified'' MLLA
calculations, we have furthermore factored out the $\alpha_s$ dependence to
ease the comparison with DLA.

While the goal of this work is a comparison of our theoretical predictions with
forthcoming data from LHC and  Tevatron, we have also given results for
LEP.  LHC energies will provide a larger trustable
 domain of comparison with theoretical predictions at small $x$.

Further developments of this work aim at getting rid of the limit $Q_0
\approx \Lambda_{QCD}$ and extending the calculations to a larger range of
values of $x$; then, because of the lack of analytical expressions,
the general formul{\ae}
(\ref{eq:DD}) and (\ref{eq:F}) should be numerically investigated,
which will also  provide a deeper insight into the connection
between DGLAP and MLLA evolution equations \cite{Perez-Salam}.

\vskip .4 cm

{\em \underline{Acknowledgments}: It is a pleasure to thank M. Cacciari,
Yu.L. Dokshitzer and G.P. Salam for many stimulating discussions, and
for expert help in numerical calculations.
R. P-R. wants to specially  thank Y.L. Dokshitzer for his guidance and
encouragements.}

%}

%%%%%%%%%%%%%%%%%%%%%%%%%%%%%%%%%%%%%%%%%%%%%%%%%%%%%%%%%%%%%%%%%%%%%%%%%%%%
%%%%%%%%%%%%%%%%%%%%%%%%%%%%%%%%%%%%%%%%%%%%%%%%%%%%%%%%%%%%%%%%%%%%%%%%%%%%
\newpage

\appendix

{\bf\Large APPENDIX}

\vskip .75 cm

%%%%%%%%%%%%%%%%%%%%%%%%%%%%%%%%%%%%%%%%%%%%%%%%%%%%%%%%%%%%%%%%%%%%%%%%%%%%
\section{EXACT SOLUTION OF THE  MLLA EVOLUTION EQUATION FOR THE
FRAGMENTATION FUNCTIONS; THE SPECTRUM AND ITS DERIVATIVES }
\label{section:exactsol}
%%%%%%%%%%%%%%%%%%%%%%%%%%%%%%%%%%%%%%%%%%%%%%%%%%%%%%%%%%%%%%%%%%%%%%%%%%%%

\vskip .5cm

\subsection{MLLA evolution equation for a gluon jet}
%%%%%%%%%%%%%%%%%%%%%%%%%%%%%%%%%%%%%%%%%%%%%%%%%%%%%%

\vskip .5cm

Because of (\ref{eq:DgDq}),
we will only write the evolution equations for  gluonic fragmentation
functions $D_g^b$.

The partonic structure functions $D_a^b$ satisfy an evolution equation which is
best written when expressed in terms of the variables $\ell$ and $y$ and
the functions $\tilde D_a^b$ defined by \cite{EvEq}
(see also (\ref{eq:Dlowx}) (\ref{eq:rhoD})):
\begin{equation}
x_b D_a^b(x_b,k_a,q) = \tilde D_a^b(\ell_b,y_b).
\label{eq:D2}
\end{equation}

The parton content $\tilde D_g$ of a gluon is shown in \cite{Perez} to
satisfy the evolution equation ($Y$ and $y$ are linked by (\ref{eq:defY}))
\begin{equation}
\tilde D_g(\ell,y) = \delta(\ell)+ \int_0^{y} dy' \int_0^{\ell} d\ell'
\gamma_0^2
(\ell'+y')\left[1  -a\delta(\ell'\!-\!\ell) \right] \tilde D_g(\ell',y'),
\label{eq:eveqincl}
\end{equation}
where the anomalous dimension $\gamma_0(y)$ is given by ($\lambda$ is defined
in (\ref{eq:lambda}))
\begin{equation}
  \gamma_0^2(y)=4N_c\frac{\alpha_s(k_{\perp}^2)}{2\pi} \approx
  \frac1{\beta(y+ \lambda)}.
\label{eq:gamma0}
\end{equation}
(see the beginning of section \ref{section:descri}
for $\beta$, $T_R$, $C_F$, $\alpha_s$, $N_c$) and
\begin{equation}
a=\frac1{4N_c}\left[\frac{11}3 N_c + \frac{4}{3}T_R 
\left(1-\frac{2\,C_F}{N_c}\right) \right]; \quad  C_F= 4/3\ for\ SU(3)_c.
\label{eq:adef}
\end{equation}
The (single logarithmic) subtraction term proportional to $a$
in (\ref{eq:eveqincl}) accounts
for {\em gluon $\to$ quark}\/ transitions in parton cascades as well
as for energy conservation -- the so-called
``hard corrections'' to parton cascading --.

No superscript has been written in the structure
functions $D_g$ because the same equation is valid indifferently for
$D_g^g$ and $D_g^q$ (see section \ref{section:lowEA}).
One considers that the same evolution equations govern
the (inclusive) hadronic distributions $D_g^h$ (Local Hadron Parton Duality).

\vskip .75 cm

\subsection{Exact solution of the MLLA evolution equation for particle spectra}
%%%%%%%%%%%%%%%%%%%%%%%%%%%%%%%%%%%%%%%%%%%%%%%%%%%%%%%%%%%%%%%%%%%%%%%%%%%%%%%
 
\vskip .5cm

The exact solution of the evolution equation (\ref{eq:eveqincl}), which
includes constraints of energy conservation and
the running of $\alpha_s$, is
demonstrated in ~\cite{Perez} to be given by the following Mellin's
representation
\begin{equation}
\begin{split}
 \tilde D_g\left(\ell,y,\lambda\right)
&= \left(\ell + y +\lambda\right) 
\int\frac{d\omega}{2\pi i}\int\frac{d\nu}{2\pi i}\, e^{\omega\ell+\nu y}\\
 &\int_{0}^{\infty}\frac{ds}{\nu+s}\left(\frac{\omega\left(\nu+s\right)}
 {\left(\omega+s\right)\nu}\right)^{1/(\beta(\omega-\nu))}
 \left(\frac{\nu}{\nu+s}\right)^{a/\beta}\,e^{-\lambda s}.
\end{split}
\label{eq:red4}
\end{equation}
From (\ref{eq:red4}) and taking the 
high energy limit $\ell+y \equiv Y\gg\lambda$ 
\footnote{$Y\gg\lambda \Leftrightarrow E\Theta \gg Q_0^2/\Lambda_{QCD}$
is not strictly equivalent to $Q_0 \to \Lambda_{QCD}$ (limiting spectrum).
\label{footnote:cutoff}}
one gets ~\cite{EvEq}\cite{KO} the explicit formula  

\begin{equation}
 \tilde D_g(\ell,y) = \frac{\ell + y}{\beta B(B+1)} \int \frac{d\omega}{2\pi i}
 \> e^{-\omega y}\> \Phi\big(A+1,B+2,\omega(\ell+y)\big),
\label{eq:confrep}
\end{equation}
where $\Phi$ is the confluent hypergeometric function the integral 
representation of which reads ~\cite{GR} ~\cite{SDP} 
\begin{eqnarray}
 &&\Phi(A+1,B+2,\omega Y) = \Gamma(B+2)\,(\omega Y)^{-B-1}
 \int \frac{dt}{(2\pi i)} \frac{t^{-B}}{t(t-1)}
 \left(\frac{t}{t-1}\right)^A  e^{\omega Y t};\cr
&& \cr
&& \text{with}\quad A = \frac{1}{\beta \omega},\quad
B=\displaystyle{\frac{a}{\beta}},\quad \Gamma(n)=\int_0^\infty d\chi \,
\chi^{n-1}e^{-\chi}.
\label{eq:hyperg}
\end{eqnarray}
Exchanging the $t$ and $\omega$ integrations of
(\ref{eq:confrep}) (\ref{eq:hyperg}) and going from  $t$ to
the new variable $\displaystyle \alpha = \frac{1}{2} \ln\frac{t}{t-1}$,
(\ref{eq:confrep}) becomes
\begin{equation}
\tilde  D_g(\ell,y) = 2\frac{\Gamma(B)}{\beta}
\Re\left( \int_0^\frac{\pi}{2}
  \frac{d\tau}{\pi}\, e^{-B\alpha}\  {\cal F}_B(\tau,y,\ell)\right),
\label{eq:ifD}
\end{equation}
where the integration is performed with respect to $\tau$ defined by
$\displaystyle \alpha = \frac{1}{2}\ln\frac{y}{\ell}  + i\tau$,
\begin{eqnarray}
{\cal F}_B(\tau,y,\ell) &=& \left[ \frac{\cosh\alpha
-\displaystyle{\frac{y-\ell}{y+\ell}}
\sinh\alpha} 
 {\displaystyle \frac{\ell +
y}{\beta}\,\frac{\alpha}{\sinh\alpha}} \right]^{B/2}
  I_B(2\sqrt{Z(\tau,y,\ell)}), \cr
&& \cr
&& \cr
 Z(\tau,y,\ell) &=&
\frac{\ell + y}{\beta}\,
\frac{\alpha}{\sinh\alpha}\,
 \left(\cosh\alpha
%+ (1-2\zeta)
-\frac{y-\ell}{y+\ell}
\sinh\alpha\right); 
\label{eq:calFdef}
\end{eqnarray}
$I_B$ is the modified Bessel function of the first kind.

\vskip .75 cm

\subsection{The spectrum}
%%%%%%%%%%%%%%%%%%%%%%%%%

\vskip .5cm

On Fig.~12 below, we represent, on the left, the spectrum
as a function of the transverse momentum
(via $y$) for fixed $\ell$ and, on the right,  as a function
of the energy (via $\ell$) for fixed transverse momentum.

\bigskip

\vbox{
\begin{center}
\epsfig{file=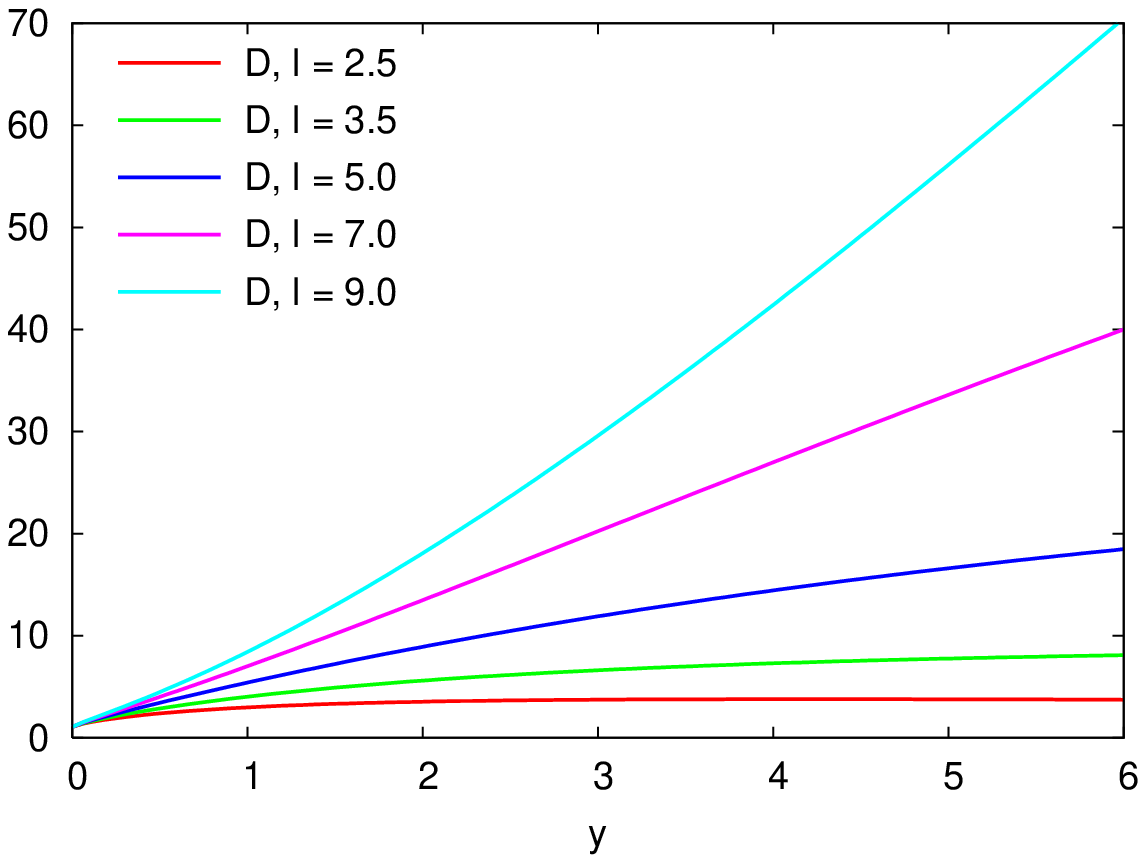, height=5truecm,width=7.5truecm}
\hfill
\epsfig{file=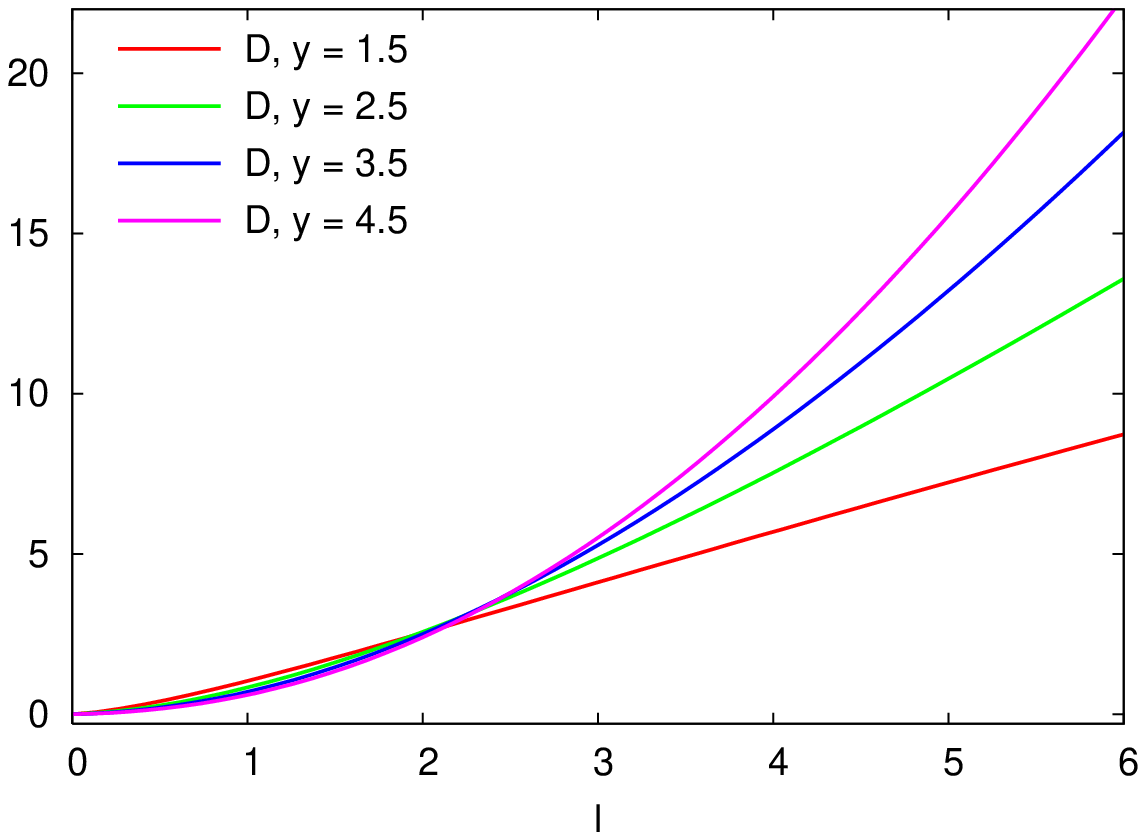, height=5truecm,width=7.5truecm}
\end{center}

\centerline{\em Fig.~12: spectrum $\tilde D(\ell,y)$ of emitted partons}

\centerline{\em as functions 
of transverse momentum (left) and energy (right)}

}

\bigskip
\bigskip

Fig.~13 shows enlargements of Fig.~12 for small values of $y$ and $\ell$
respectively; they ease the understanding of the curves for the derivatives
of the spectrum presented in subsection \ref{subsection:deriv}.

\bigskip
\bigskip

\vbox{
\begin{center}
\epsfig{file=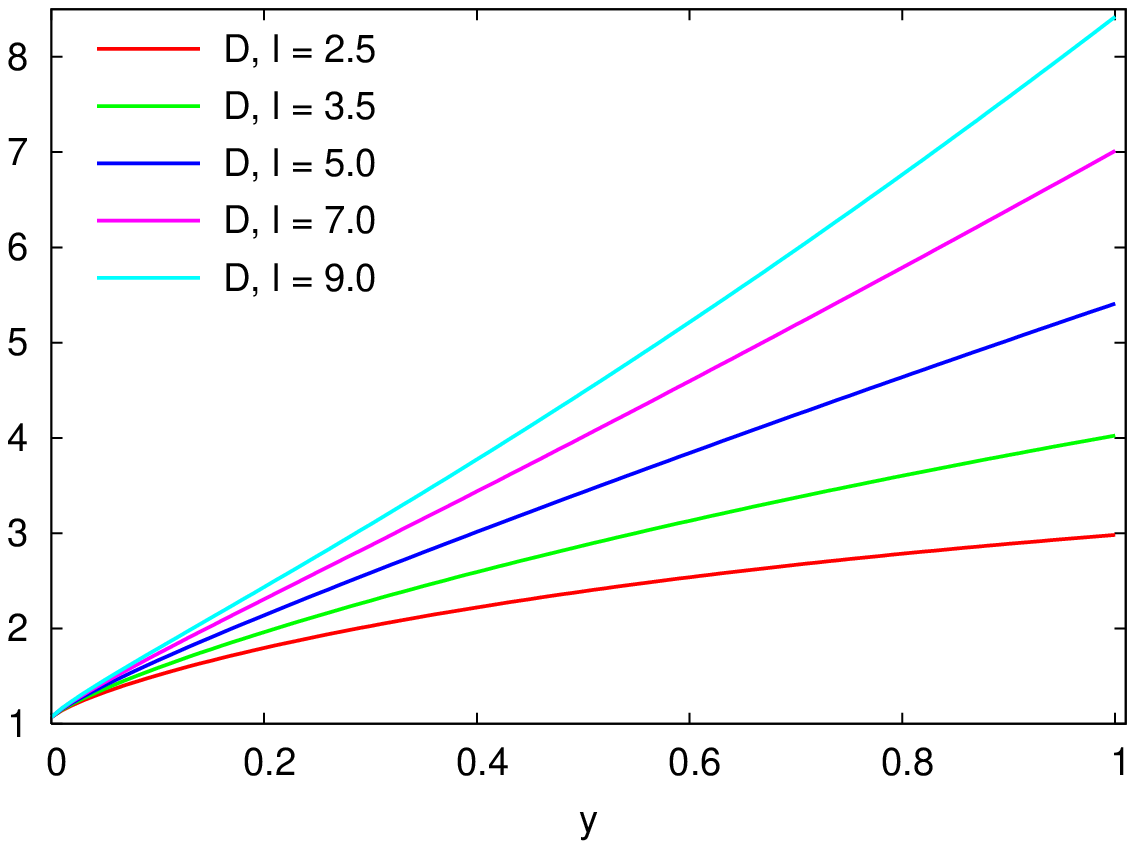, height=5truecm,width=7.5truecm}
\hfill
\epsfig{file=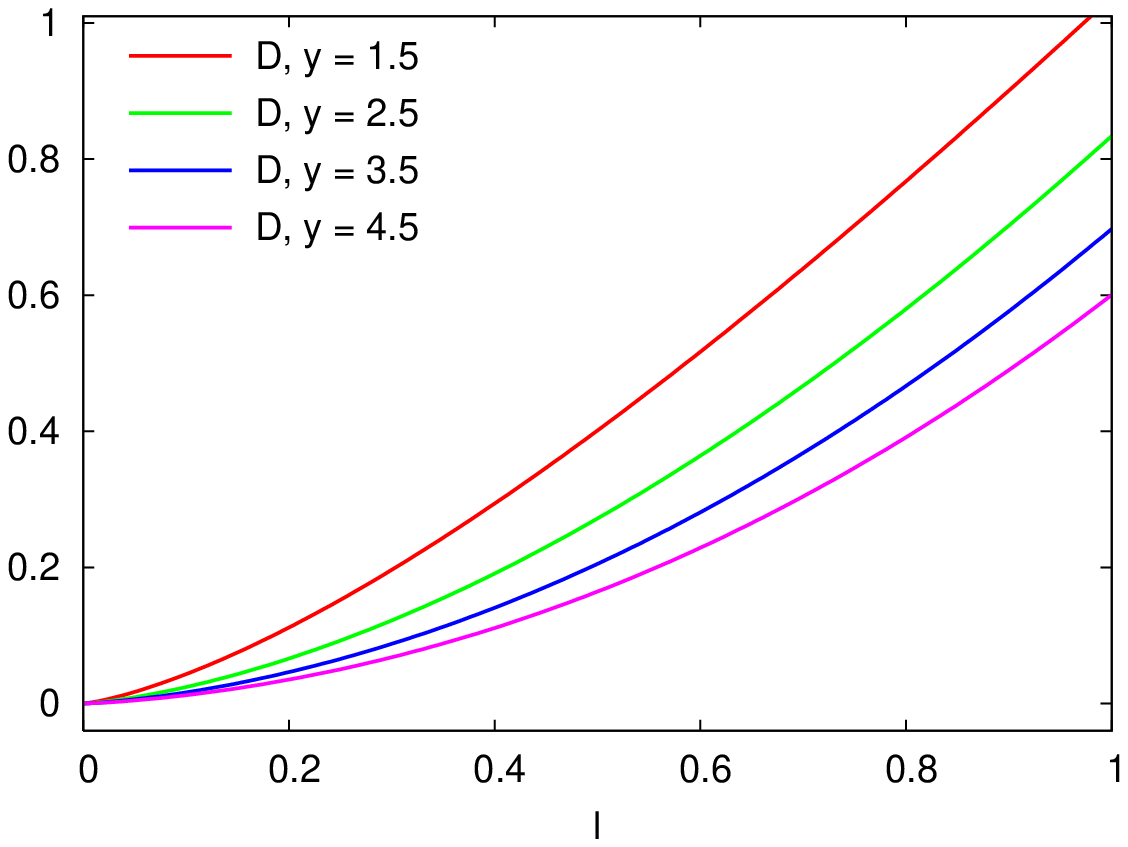, height=5truecm,width=7.5truecm}
\end{center}

\centerline{\em Fig.~13: spectrum $\tilde D(\ell,y)$ of emitted partons}

\centerline{\em as functions 
of transverse momentum (left) and energy (right): enlargement of Fig.~12}}

\bigskip

A comparison between MLLA and DLA calculations of the spectrum is done in
appendix \ref{subsection:DLAspec}.

\vskip .75 cm

\subsection{Derivatives of the spectrum}
\label{subsection:deriv}
%%%%%%%%%%%%%%%%%%%%%%%%%%%%%%%%%%%%%%%%%

\vskip .5cm

We evaluate below the derivatives of the spectrum w.r.t. $\ln k_{\perp}$
and $\ln(1/x)$.

We make use of the following property for the
confluent hypergeometric functions $\Phi$ ~\cite{SDP}: 
\begin{equation}\label{eq:derivconf}
\frac{d}{d\ell}\Phi\left(A+1,B+2,\omega\left(\ell+y\right)\right)
\equiv\frac{d}{dy}\Phi\left(A+1,B+2,\omega\left(\ell+y\right)\right)
=\omega\frac{A+1}{B+2}\Phi\left(A+2,B+3,\omega\left(\ell+y\right)\right).
\end{equation} 

$\bullet$\quad We first determine the derivative w.r.t. $\ell\equiv\ln(1/x)$.
Differentiating  (\ref{eq:confrep}) w.r.t. $\ell$, and
expanding (\ref{eq:derivconf}), one gets
\footnote{(\ref{eq:derivl}) and (\ref{eq:derivy}) have also been
checked by numerically differentiating (\ref{eq:ifD}).}
 \cite{Perez} 
\begin{equation}
\frac{d}{d\ell}\tilde  D_{g}\left(\ell,y\right)
 = 2\frac{\Gamma(B)}{\beta} \int_0^{\frac{\pi}2}\frac{d\tau}{\pi}\,
 e^{-B\alpha}
 \left[\frac1{\ell + y}\left(1+2e^{\alpha}
\sinh{\alpha}\right){\cal{F}}_B
+\frac1{\beta}e^{\alpha}{\cal{F}}_{B+1}\right];
\label{eq:derivl}
\end{equation}

$\bullet$\quad Differentiating w.r.t.
$y\equiv\ln\displaystyle{\frac{k_{\perp}}{Q_0}}$ yields
\begin{equation}
\frac{d}{dy}\tilde D_{g}\left(\ell,y\right)
= 2 \frac{\Gamma(B)}{\beta} \int_0^{\frac{\pi}2}
\frac{d\tau}{\pi}\,  e^{-B\alpha}
 \left[\frac1{\ell + y}
\left(1+2e^{\alpha}\sinh{\alpha}\right)
 {\cal{F}}_B
 +\frac1{\beta}
 e^{\alpha}{\cal{F}}_{B+1}\right.
\left.-\frac{2\sinh\alpha}{\ell +
y}{\cal{F}}_{B-1}\right].
\label{eq:derivy}
\end{equation}
In Fig.~14,  Fig.~15,  Fig.~16 and Fig.~17  below, we draw the curves for:

\smallskip

$\ast$\ $\displaystyle\frac{d\tilde D_g(\ell,y)}{dy}$ as a function
of $y$, for different values of $\ell$ fixed;

$\ast$\ $\displaystyle\frac{d\tilde D_g(\ell,y)}{dy}$ as a function
of $\ell$, for different values of $y$ fixed;

$\ast$\ $\displaystyle\frac{ d\tilde D_g(\ell,y)}{d\ell}$ as a
function of $\ell$ for different values of $y$ fixed;

$\ast$\ $\displaystyle\frac{ d\tilde D_g(\ell, y)}{d\ell}$
as a function of $y$ for different values of $\ell$ fixed.

\bigskip

In each case the right figure is an enlargement,
close to the origin of axes, of the left figure.

\vskip .75 cm

%\vbox{
\vbox{
\begin{center}
\epsfig{file=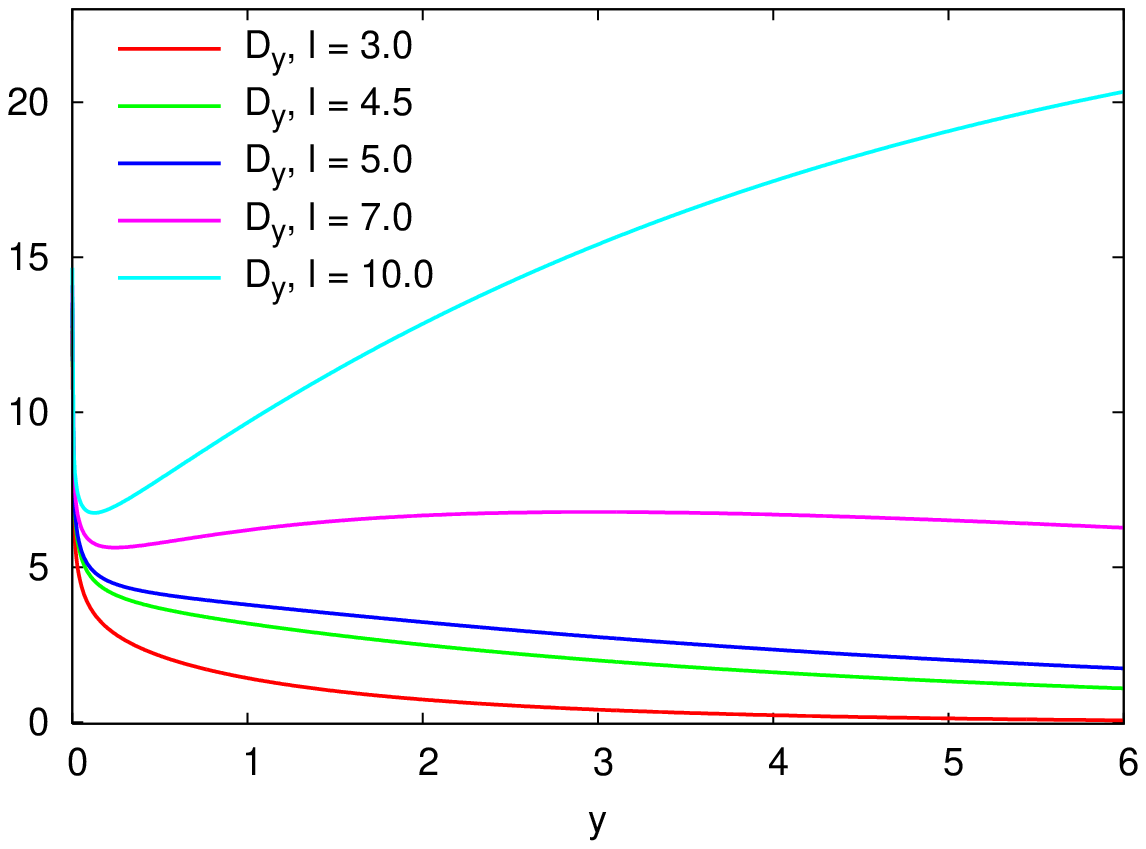, height=5truecm,width=7.5truecm}
\hfill
\epsfig{file=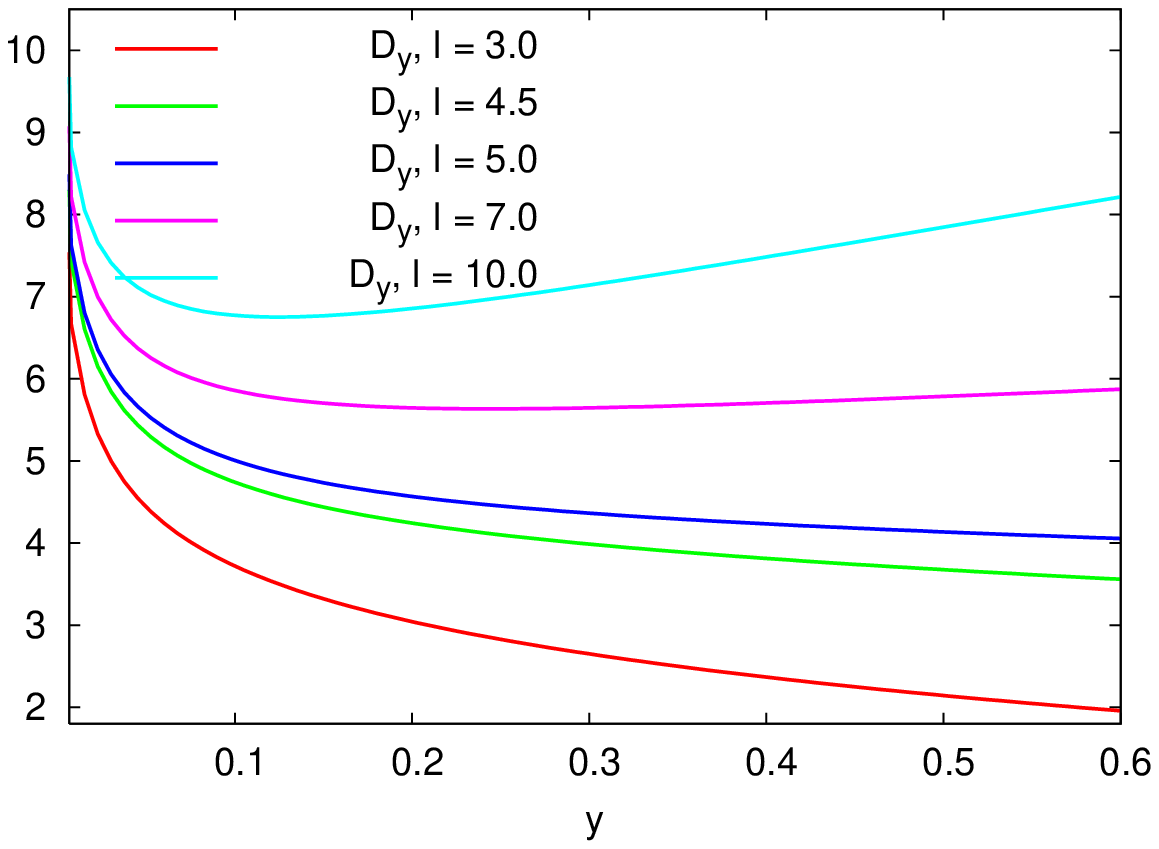, height=5truecm,width=7.5truecm}
\end{center}

\centerline{\em Fig.~14: $\frac{d\tilde D_g(\ell,y)}{dy}$ as a function of $y$
 for different values of $\ell$}

}

\medskip

\vbox{
\begin{center}
\epsfig{file=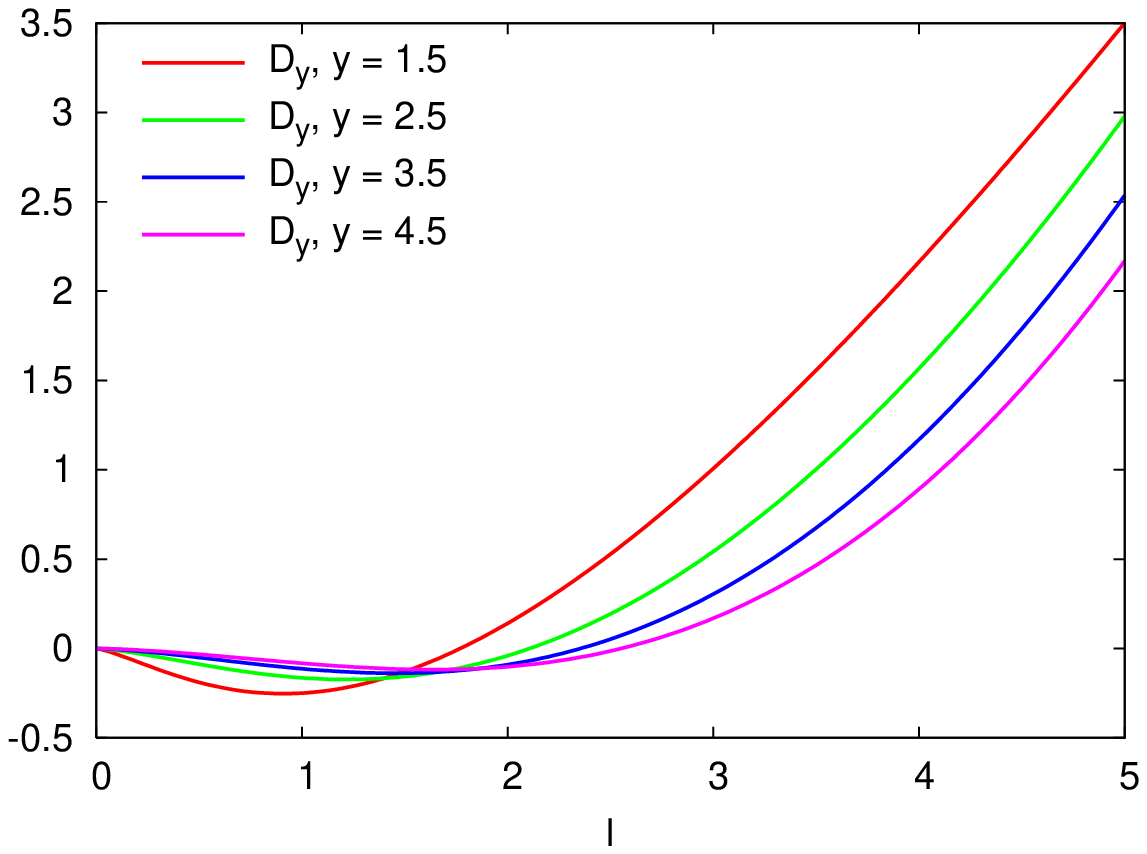,height=5truecm,width=7.5truecm}
\hfill
\epsfig{file=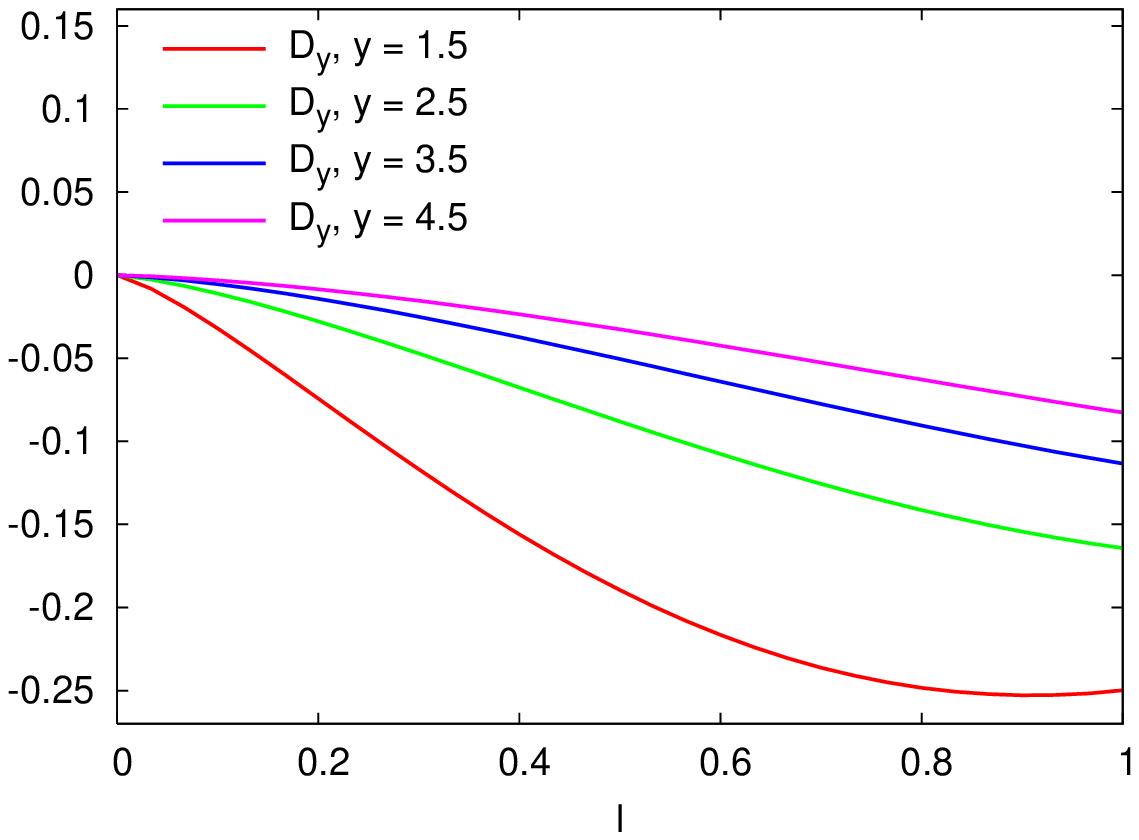, height=5truecm,width=7.5truecm}
\end{center}

\centerline{\em Fig.~15: $\frac{d\tilde D_g(\ell,y)}{dy}$ as a function of
$\ell$ for different values of $y$}

}

\medskip

\vbox{
\begin{center}
\epsfig{file=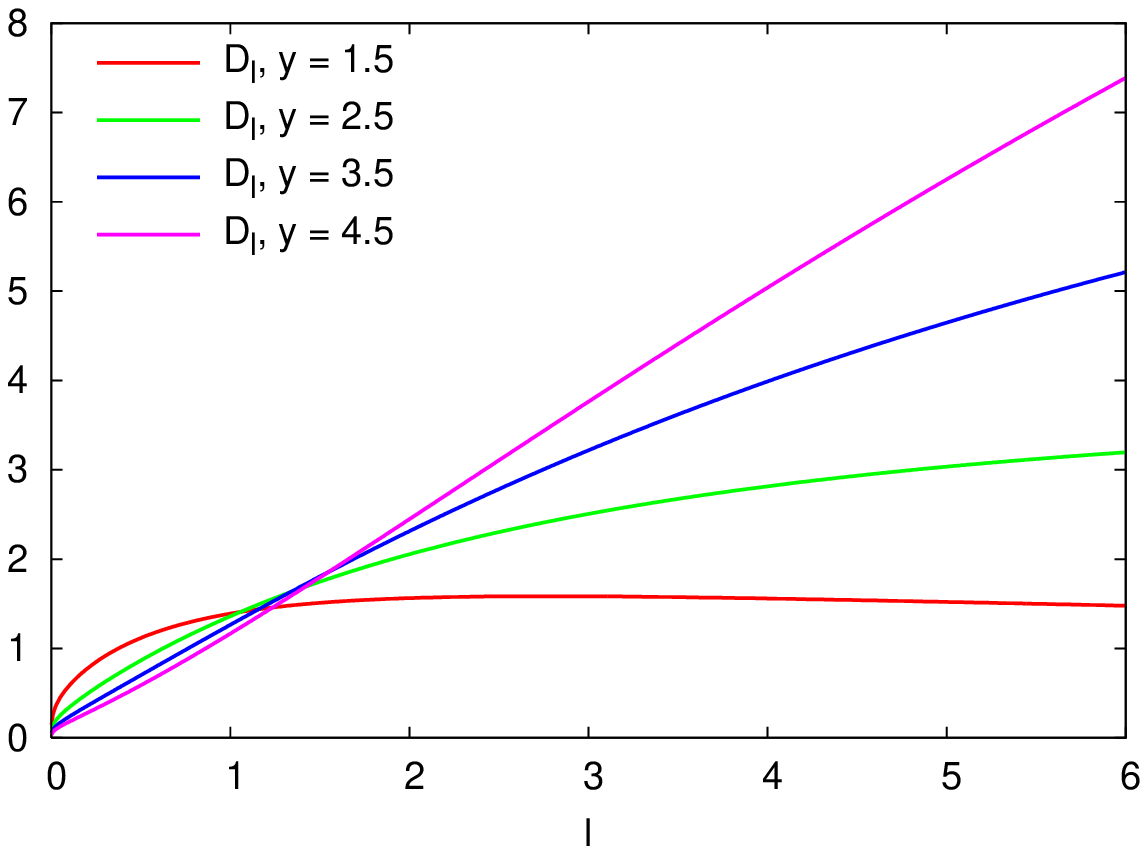, height=5truecm,width=7.5truecm}
\hfill
\epsfig{file=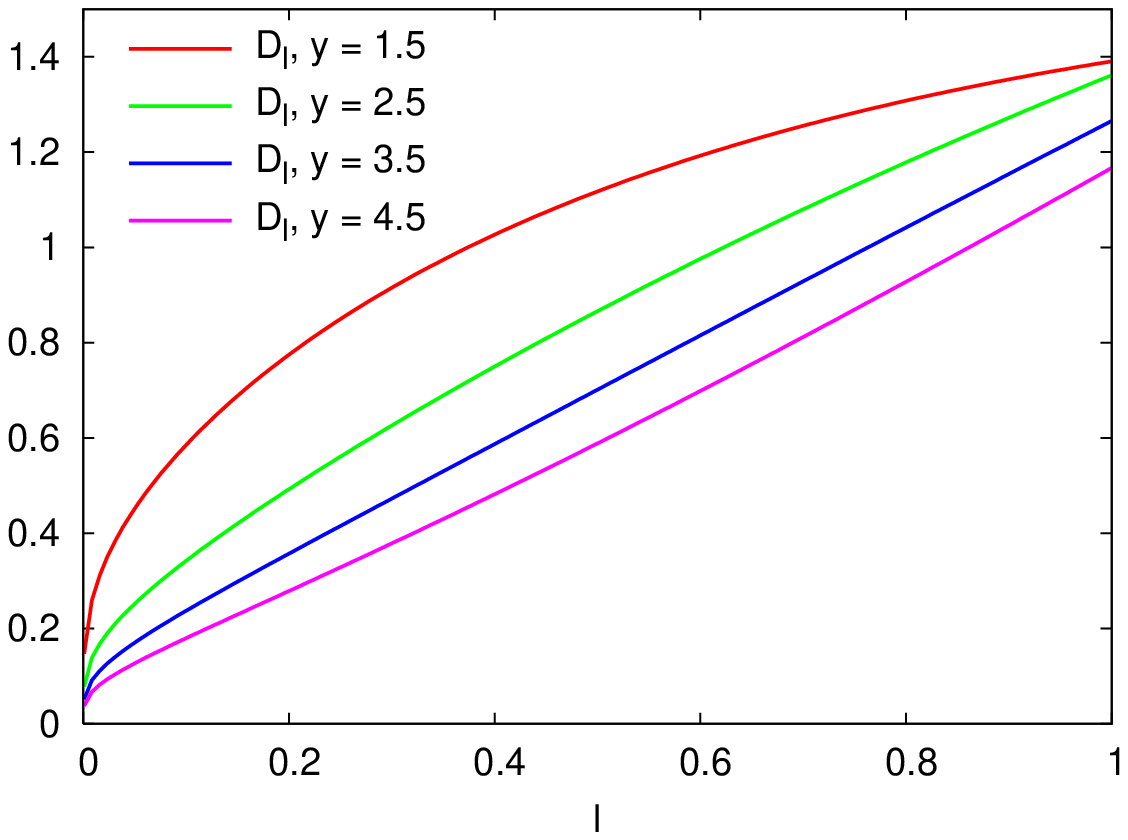, height=5truecm,width=7.5truecm}
\end{center}

\centerline{\em Fig.~16: $\frac{d\tilde D_g(\ell,y)}{d\ell}$ as a function of
$\ell$ for different values of $y$}
}

\medskip

\vbox{
\begin{center}
\epsfig{file=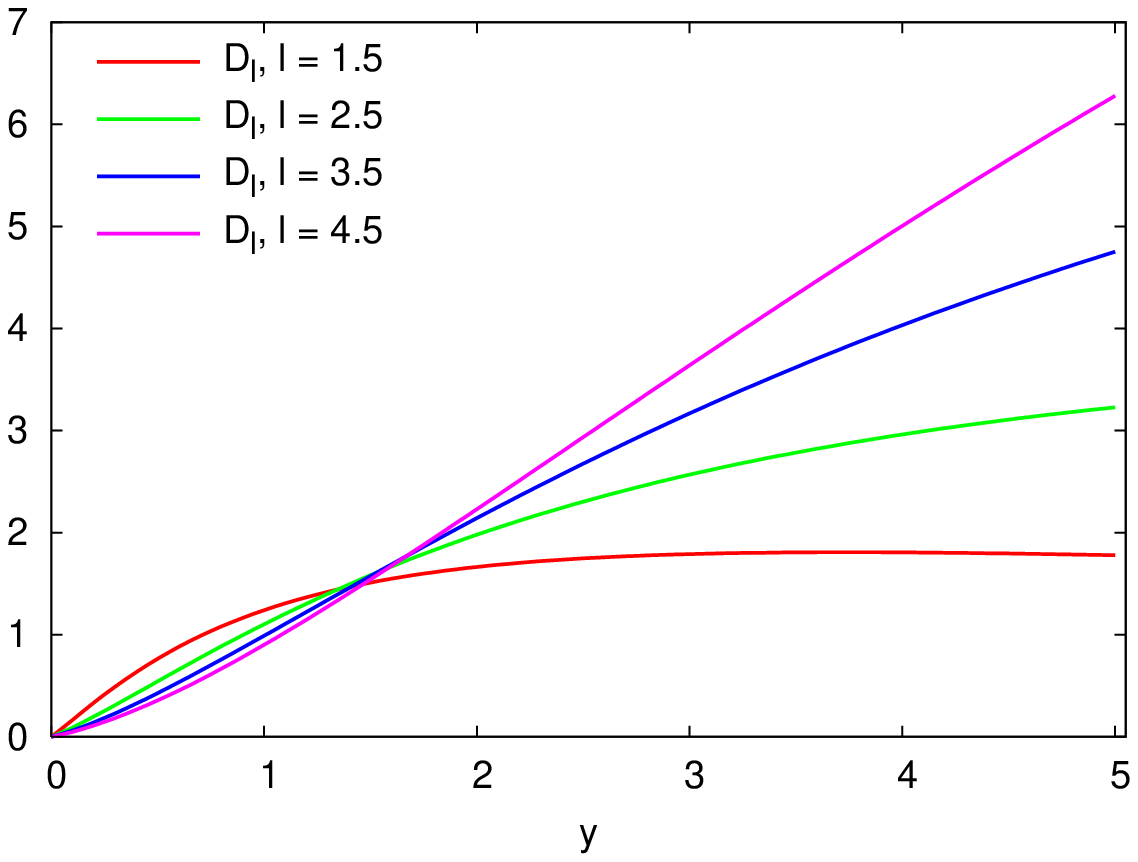, height=5truecm,width=7.5truecm}
\hfill
\epsfig{file=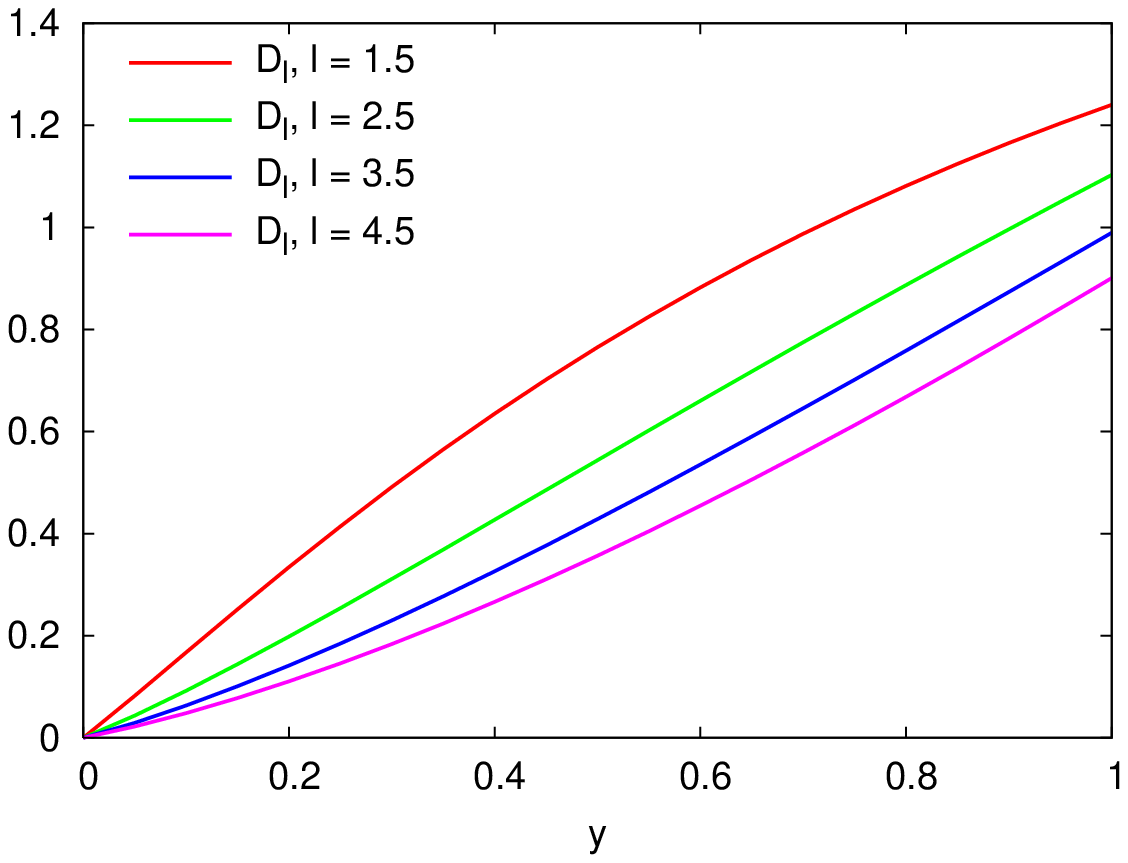, height=5truecm,width=7.5truecm}
\end{center}

\centerline{\em Fig.~17: $\frac{d\tilde D_g(\ell,y)}{d\ell}$ as a function of
$y$  for different values of $\ell$}
}
%}

\vskip .5cm

That $\displaystyle\frac{d\tilde D_g(\ell,y)}{dy}$ goes to infinity
when  $y \to 0$ is
in agreement with the analytic behavior in $\ln(\ell/y)$ of this
derivative.

\vskip .75 cm

%%%%%%%%%%%%%%%%%%%%%%%%%%%%%%%%%%%%%%%%%%%%%%%%%%%%%%%%%%%%%%%%%%%%%%%%%%%%
\section{LEADING CONTRIBUTIONS TO  $\boldsymbol{x_1 F_{A_0}^{h_1}(x_1, \Theta,
E, \Theta_0)}$ AT SMALL $\boldsymbol{x_1}$}
\label{section:leadingxF}
%%%%%%%%%%%%%%%%%%%%%%%%%%%%%%%%%%%%%%%%%%%%%%%%%%%%%%%%%%%%%%%%%%%%%%%%%%%%

\vskip .5cm

Using (\ref{eq:DgDq}), the leading terms of $x_1 F_{A_0}^{h_1}(x_1, \Theta,
E, \Theta_0)$ (\ref{eq:Fdev}) calculated at small $x_1$ read
\begin{eqnarray}
x_1 F_{g}^{h_1}\left(x_1,\Theta,E,\Theta_0\right)^0
&\approx&
\tilde D_g(\ell_1,y_1)\left(<u>^g_{g}
+ \frac{C_F}{N_c}<u>^q_{g}\right)
= \frac{<C>_g^0}{N_c}\;\tilde D_g(\ell_1,y_1),\cr
x_1 F_{q}^{h_1}\left(x_1,\Theta,E,\Theta_0\right)^0
&\approx&
\tilde D_g(\ell_1,y_1)\left(<u>^g_{q} 
+ \frac{C_F}{N_c}<u>^q_{q}\right)
=\frac{<C>_q^0}{N_c}\;\tilde D_g(\ell_1,y_1).\cr
&&
\label{eq:FA0}
\end{eqnarray}

The leading  $<C>_{g}^0$ and $<C>_{q}^0$ in (\ref{eq:C})
 for  a quark and a gluon jet are
 given respectively by (see \cite{EvEq}, chapt. 9
\footnote{The coefficient $\beta$, omitted in the exponents of eqs.
(9.12a), (9.12b), (9.12c) of \cite{EvEq} has been restored here. The factor
$4N_c$ is due to our normalization (see the beginning of section
\ref{section:descri}).}
)

\begin{eqnarray}
 <C>_{q}^0
&=&<C>_{\infty}-c_1\left(N_c-C_F\right)\left(
 \frac{\ln\left(E\Theta/\Lambda_{QCD}\right)}
 {\ln\left(E\Theta_0/\Lambda_{QCD}\right)}\right)^{(c_3/4N_c\beta)}\cr
&=&<C>_{\infty}-c_1\left(N_c-C_F\right)\left(
 \frac{Y_\Theta +\lambda}
 {Y_{\Theta_0} +\lambda}\right)^{(c_3/4N_c\beta)},
\label{eq:Cquark}
\end{eqnarray}
\begin{eqnarray}
 <C>_{g}^0
&=&<C>_{\infty}+c_2\left(N_c-C_F\right)\left(
 \frac{\ln\left(E\Theta/\Lambda_{QCD}\right)}
 {\ln\left(E\Theta_0/\Lambda_{QCD}\right)}\right)^{(c_3/4N_c\beta)}\cr
&=&<C>_{\infty}+c_2\left(N_c-C_F\right)\left(
 \frac{Y_{\Theta} +\lambda}
 {Y_{\Theta_0} +\lambda}\right)^{(c_3/4N_c\beta)},
\label{eq:Cgluon}
\end{eqnarray}
with
\begin{eqnarray}
<C>_{\infty} &=& c_1 N_c+c_2\, C_F,\cr
&& \cr
c_1=\displaystyle{\frac83}\displaystyle{\frac{C_F}{c_3}},\quad c_2 
&=&1-c_1=\displaystyle{\frac{2}{3}}\displaystyle{\frac{n_f}{c_3}},
\quad c_3=\displaystyle{\frac83}C_F+\displaystyle{\frac23}n_f;
\label{eq:c1c2c3}
\end{eqnarray}
in the r.h.s of (\ref{eq:Cquark}) (\ref{eq:Cgluon})
 we have used the definitions (\ref{eq:defY}) (\ref{eq:Y0}).
$<C>_\infty$ corresponds to the limit  $E\to \infty, \Theta \to 0$.
 
In practice, we take in this work
\begin{equation}
Q_0 \approx \Lambda_{QCD} \Leftrightarrow \lambda \approx 0,
\label{eq:lambdanul}
\end{equation}
which ensures in particular the consistency with the analytical
calculation of the MLLA spectrum (appendix \ref{section:exactsol}),
which can only be explicitly achieved in this limit.

\vskip .75 cm

%%%%%%%%%%%%%%%%%%%%%%%%%%%%%%%%%%%%%%%%%%%%%%%%%%%%%%%%%%%%%%%%%%%%%%%%%%%%
\section{ CALCULATION OF $\boldsymbol{\delta\!<C>_g}$ and $\boldsymbol{\delta\!<C>_q}$
OF SECTION \ref{section:lowEA}}
\label{section:udeltau}
%%%%%%%%%%%%%%%%%%%%%%%%%%%%%%%%%%%%%%%%%%%%%%%%%%%%%%%%%%%%%%%%%%%%%%%%%%%%

\vskip .5cm

\subsection{Explicit expressions for $\boldsymbol{<u>_{A_0}^A}$ and
$\boldsymbol{\delta\!<u>_{A_0}^A}$ defined in (\ref{eq:udef})}
\label{subsection:udu}
%%%%%%%%%%%%%%%%%%%%%%%%%%%%%%%%%%%%%%%%%%%%%%%%%%%%%%%%%%%%%%%%%%%%

\vskip .5cm

The expressions (\ref{eq:udef}) for $<u>^A_{A_0}$ and $\delta \!<u>^A_{A_0}$
are conveniently obtained from the Mellin-transformed DGLAP fragmentation
functions
\cite{EvEq}
\begin{equation}
        {\cal D}(j,\xi) = \int_0^1 du\, u^{j-1}  D(u,\xi), 
\label{eq:mellin}
\end{equation}
which, if one deals with $D_A^B(u,r^2,s^2)$, depends in reality on
the difference $\xi(r^2) - \xi(s^2)$:
\begin{equation}
\xi(Q^2) = \int_{\mu^2}^{Q^2} \;
\frac{dk^2}{k^2}\frac{\alpha_s(k^2)}{4\pi}, \quad
\xi(r^2) - \xi(s^2) \approx \frac{1}{4N_c\beta}
\ln\left(\frac{\ln(r^2/\Lambda_{QCD}^2)}{\ln(s^2/\Lambda_{QCD}^2)}\right).
\end{equation}
One has accordingly
\begin{equation}
<u>^A_{A_0} = {\cal D}_{A_0}^A (2,\xi(E\Theta_0)-\xi(E\Theta)),\quad
\delta\!<u>^A_{A_0} = \frac{d}{dj}{\cal D}_{A_0}^A(j,\xi(E\Theta_0)-\xi(E\Theta))\Big|_{j=2}.  \label{eq:udu}
\end{equation}
The DGLAP functions ${\cal D}(j,\xi)$ are expressed \cite{EvEq}
% chapt.~1)
in terms
of the anomalous dimensions $\nu_F(j)$, $\nu_G(j)$ and $\nu_\pm(j)$, the
$j$ dependence of which is  in particular known.

For the sake of completeness, we give below the expressions for the $<u>$'s
and $\delta<u>$'s.

\begin{eqnarray}
<u>^q_g &=& {\frac {9}{25}}\, \left(  \left( {\frac {{
Y_{\Theta_0}}+\lambda}{Y_\Theta+\lambda}}
 \right) ^{{\frac {50}{81}}}-1 \right)  \left( {\frac {{ Y_{\Theta_0}}+
\lambda}{Y_\Theta+\lambda}} \right) ^{-{\frac {50}{81}}},\cr
&&\cr
&&\cr
<u>^g_g &=&
1/25\, \left( 16\, \left( {\frac {{ Y_{\Theta_0}}+\lambda}{Y_\Theta+\lambda}}
 \right) ^{{\frac {50}{81}}}+9 \right)  \left( {\frac {{ Y_{\Theta_0}}+
\lambda}{Y_\Theta+\lambda}} \right) ^{-{\frac {50}{81}}},\cr
&&\cr
&&\cr
<u>^g_q &=&
{\frac {16}{25}}\, \left(  \left( {\frac {{
Y_{\Theta_0}}+\lambda}{Y_\Theta+\lambda}
} \right) ^{{\frac {50}{81}}}-1 \right)  \left( {\frac {{ Y_{\Theta_0}}+
\lambda}{Y_\Theta+\lambda}} \right) ^{-{\frac {50}{81}}},\cr
&&\cr
&&\cr
<u>^{sea}_q &=&
-1/25\, \left( -9\, \left( {\frac {{ Y_{\Theta_0}}+\lambda}{Y_\Theta+\lambda}}
 \right) ^{{\frac {50}{81}}}-16+25\, \left( {\frac {{ Y_{\Theta_0}}+\lambda}{
Y_\Theta+\lambda}} \right) ^{2/9} \right)  \left( {\frac {{
Y_{\Theta_0}}+\lambda}{Y_\Theta
+\lambda}} \right) ^{-{\frac {50}{81}}},\cr
&&\cr
&&\cr
<u>^{val} &=&
\left( {\frac {{ Y_{\Theta_0}}+\lambda}{Y_\Theta+\lambda}} \right) ^{-{\frac {32}{
81}}},\cr
&&\cr
&&\cr
<u>^{sea}_q + <u>^{val} &=& 1/25\, \left( 9\, \left( {\frac {{
Y_{\Theta_0}}+\lambda}{Y_\Theta+\lambda}}
 \right) ^{{\frac {50}{81}}}+16 \right)  \left( {\frac {{ Y_{\Theta_0}}+
\lambda}{Y_\Theta+\lambda}} \right) ^{-{\frac {50}{81}}};\cr
&&\cr
&&\cr
\delta <u>^q_g &=&
-{\frac {1}{337500}}\, \left( -43011\, \left( {\frac {{ Y_{\Theta_0}}+\lambda
}{Y_\Theta+\lambda}} \right) ^{{\frac {50}{81}}}+43011
-6804\,{\pi }^{2}
 \left( {\frac {{ Y_{\Theta_0}}+\lambda}{Y_\Theta+\lambda}} \right) ^{{\frac {50}{81
}}}
\right. \cr && \left.
+6804\,{\pi }^{2}-48600\,\ln  \left( {\frac {{
Y_{\Theta_0}}+\lambda}{Y_\Theta+
\lambda}} \right)  \left( {\frac {{ Y_{\Theta_0}}+\lambda}{Y_\Theta+\lambda}}
 \right) ^{{\frac {50}{81}}}
\right. \cr && \left.
+21600\,\ln  \left( {\frac {{ Y_{\Theta_0}}+
\lambda}{Y_\Theta+\lambda}} \right)  \left( {\frac {{
Y_{\Theta_0}}+\lambda}{Y_\Theta+
\lambda}} \right) ^{{\frac {50}{81}}}{\pi }^{2}+109525\,\ln  \left( {
\frac {{ Y_{\Theta_0}}+\lambda}{Y_\Theta+\lambda}} \right)
\right. \cr && \left.
 -17400\,\ln  \left( {
\frac {{ Y_{\Theta_0}}+\lambda}{Y_\Theta+\lambda}} \right) {\pi }^{2} \right) 
 \left( {\frac {{ Y_{\Theta_0}}+\lambda}{Y_\Theta+\lambda}} \right) ^{-{\frac {50}{
81}}},\cr
&&\cr
&&\cr
\delta <u>^g_g &=&
-{\frac {1}{337500}}\, \left( -11664\, \left( {\frac {{ Y_{\Theta_0}}+\lambda
}{Y_\Theta+\lambda}} \right) ^{{\frac {50}{81}}}+31104\,{\pi }^{2} \left( {
\frac {{ Y_{\Theta_0}}+\lambda}{Y_\Theta+\lambda}} \right) ^{{\frac {50}{81}}}
\right. \cr && \left.
-86400
\,\ln  \left( {\frac {{ Y_{\Theta_0}}+\lambda}{Y_\Theta+\lambda}} \right)  \left( {
\frac {{ Y_{\Theta_0}}+\lambda}{Y_\Theta+\lambda}} \right) ^{{\frac {50}{81}}}
\right. \cr && \left.
+38400
\,\ln  \left( {\frac {{ Y_{\Theta_0}}+\lambda}{Y_\Theta+\lambda}} \right)  \left( {
\frac {{ Y_{\Theta_0}}+\lambda}{Y_\Theta+\lambda}} \right) ^{{\frac {50}{81}}}{\pi }
^{2}
\right. \cr && \left.
+11664-31104\,{\pi }^{2}-109525\,\ln  \left( {\frac {{ Y_{\Theta_0}}+
\lambda}{Y_\Theta+\lambda}} \right)
\right. \cr && \left.
 +17400\,\ln  \left( {\frac {{ Y_{\Theta_0}}+
\lambda}{Y_\Theta+\lambda}} \right) {\pi }^{2} \right)  \left( {\frac {{ 
Y_{\Theta_0}}+\lambda}{Y_\Theta+\lambda}} \right) ^{-{\frac {50}{81}}},\cr
&&\cr
&&\cr
\delta <u>^g_q &=&
-{\frac {4}{759375}}\, \left( 48114\, \left( {\frac {{ Y_{\Theta_0}}+\lambda}
{Y_\Theta+\lambda}} \right) ^{{\frac {50}{81}}}-48114-6804\,{\pi }^{2}
 \left( {\frac {{ Y_{\Theta_0}}+\lambda}{Y_\Theta+\lambda}} \right) ^{{\frac {50}{81
}}}
\right. \cr && \left.
+6804\,{\pi }^{2}-48600\,\ln  \left( {\frac {{
Y_{\Theta_0}}+\lambda}{Y_\Theta+
\lambda}} \right)  \left( {\frac {{ Y_{\Theta_0}}+\lambda}{Y_\Theta+\lambda}}
 \right) ^{{\frac {50}{81}}}
\right. \cr && \left.
+21600\,\ln  \left( {\frac {{ Y_{\Theta_0}}+
\lambda}{Y_\Theta+\lambda}} \right)  \left( {\frac {{
Y_{\Theta_0}}+\lambda}{Y_\Theta+
\lambda}} \right) ^{{\frac {50}{81}}}{\pi }^{2}+109525\,\ln  \left( {
\frac {{ Y_{\Theta_0}}+\lambda}{Y_\Theta+\lambda}} \right)
\right. \cr && \left.
 -17400\,\ln  \left( {
\frac {{ Y_{\Theta_0}}+\lambda}{Y_\Theta+\lambda}} \right) {\pi }^{2} \right) 
 \left( {\frac {{ Y_{\Theta_0}}+\lambda}{Y_\Theta+\lambda}} \right) ^{-{\frac {50}{
81}}},\cr
&&\cr
&&\cr
\delta <u>^{sea}_q &=&
{\frac {2}{759375}}\, \left( -13122\, \left( {\frac {{ Y_{\Theta_0}}+\lambda}
{Y_\Theta+\lambda}} \right) ^{{\frac {50}{81}}}+34992\,{\pi }^{2} \left( {
\frac {{ Y_{\Theta_0}}+\lambda}{Y_\Theta+\lambda}} \right) ^{{\frac {50}{81}}}
\right. \cr && \left.
+54675
\,\ln  \left( {\frac {{ Y_{\Theta_0}}+\lambda}{Y_\Theta+\lambda}} \right)  \left( {
\frac {{ Y_{\Theta_0}}+\lambda}{Y_\Theta+\lambda}} \right) ^{{\frac {50}{81}}}
\right. \cr && \left.
-24300
\,\ln  \left( {\frac {{ Y_{\Theta_0}}+\lambda}{Y_\Theta+\lambda}} \right)  \left( {
\frac {{ Y_{\Theta_0}}+\lambda}{Y_\Theta+\lambda}} \right) ^{{\frac {50}{81}}}{\pi }
^{2}
\right. \cr && \left.
+13122-34992\,{\pi }^{2}+219050\,\ln  \left( {\frac {{ Y_{\Theta_0}}+
\lambda}{Y_\Theta+\lambda}} \right)
-34800\,\ln  \left( {\frac {{ Y_{\Theta_0}}+
\lambda}{Y_\Theta+\lambda}} \right) {\pi }^{2}
\right. \cr && \left.
-265625\,\ln  \left( {\frac {{
 Y_{\Theta_0}}+\lambda}{Y_\Theta+\lambda}} \right)  \left( {\frac {{
Y_{\Theta_0}}+\lambda}
{Y_\Theta+\lambda}} \right) ^{2/9}
\right. \cr && \left.
+37500\,\ln  \left( {\frac {{ Y_{\Theta_0}}+
\lambda}{Y_\Theta+\lambda}} \right)  \left( {\frac {{
Y_{\Theta_0}}+\lambda}{Y_\Theta+
\lambda}} \right) ^{2/9}{\pi }^{2} \right)  \left( {\frac {{
Y_{\Theta_0}}+
\lambda}{Y_\Theta+\lambda}} \right) ^{-{\frac {50}{81}}},\cr
&&\cr
&&\cr
\delta <u>^{val} &=&
-{\frac {2}{243}}\, \left( -85+12\,{\pi }^{2} \right) \ln  \left( {
\frac {{ Y_{\Theta_0}}+\lambda}{Y_\Theta+\lambda}} \right)  \left( {\frac {{
Y_{\Theta_0}}+
\lambda}{Y_\Theta+\lambda}} \right) ^{-{\frac {32}{81}}},\cr
&&\cr
&&\cr
\delta <u>^{val} + \delta <u>^{sea}_q &=&
-{\frac {2}{759375}}\, \left( 13122\, \left( {\frac {{ Y_{\Theta_0}}+\lambda}
{Y_\Theta+\lambda}} \right) ^{{\frac {50}{81}}}-34992\,{\pi }^{2} \left( {
\frac {{ Y_{\Theta_0}}+\lambda}{Y_\Theta+\lambda}} \right) ^{{\frac {50}{81}}}
\right. \cr && \left.
-54675
\,\ln  \left( {\frac {{ Y_{\Theta_0}}+\lambda}{Y_\Theta+\lambda}} \right)  \left( {
\frac {{ Y_{\Theta_0}}+\lambda}{Y_\Theta+\lambda}} \right) ^{{\frac {50}{81}}}
\right. \cr && \left.
+24300
\,\ln  \left( {\frac {{ Y_{\Theta_0}}+\lambda}{Y_\Theta+\lambda}} \right)  \left( {
\frac {{ Y_{\Theta_0}}+\lambda}{Y_\Theta+\lambda}} \right) ^{{\frac {50}{81}}}{\pi }
^{2}-13122+34992\,{\pi }^{2}
\right. \cr && \left.
-219050\,\ln  \left( {\frac {{ Y_{\Theta_0}}+
\lambda}{Y_\Theta+\lambda}} \right) +34800\,\ln  \left( {\frac {{ Y_{\Theta_0}}+
\lambda}{Y_\Theta+\lambda}} \right) {\pi }^{2} \right)  \left( {\frac {{ 
Y_{\Theta_0}}+\lambda}{Y_\Theta+\lambda}} \right) ^{-{\frac {50}{81}}}.\cr
&&
\label{eq:deltaus}
\end{eqnarray}

\bigskip

When $\Theta \to \Theta_0$, all $\delta\! <u>$'s vanish, ensuring that the
limits $\xi(E\Theta_0) -\xi(E\Theta) \to 0$ of the $(<C>_{A_0}^0 +
\delta\!<C>_{A_0})$'s
are the same as the ones of the $<C>_{A_0}^0$'s.

\vskip .75 cm

\subsection{$\boldsymbol{\delta\!<C>_q}$ and $\boldsymbol{\delta\!<C>_g}$}
%%%%%%%%%%%%%%%%%%%%%%%%%%%%%%%%%%%%%%%%%%%%%%%%%%%%%%%%%%%%%%%%%%%%%%%%%%

\vskip .5cm

They are given in (\ref{eq:deltaC}), and one uses
(\ref{eq:DgDq}) such that only $\psi_{g,\ell_1}$ (see (\ref{eq:psidef}))
appears.
Their full analytical expressions for the $\delta\!<C>$'s are 
too complicated to be easily written and manipulated.

Using the formul{\ae} of \ref{subsection:udu}, one gets the
approximate results
\begin{eqnarray}
\delta\!<C>_q &\approx& \left(
1.4676
- 1.4676\left(\frac{Y_{\Theta_0}+\lambda}{Y_\Theta+\lambda}\right)^{-\frac{50}{81}}
- 3.2510 \ln \left(\frac{Y_{\Theta_0}+\lambda}{Y_\Theta+\lambda}\right)\right.\cr
&& \left.  \hskip 2cm
+ 0.5461
\left(\frac{Y_{\Theta_0}+\lambda}{Y_\Theta+\lambda}\right)^{-\frac{50}{81}}
\ln \left(\frac{Y_{\Theta_0}+\lambda}{Y_\Theta+\lambda}\right)
\right)
\psi_{g,\ell_1}(\ell_1,y_1),
\label{eq:deltaCq}
\end{eqnarray}
and

\vbox{
\begin{eqnarray}
\delta\!<C>_g &\approx& \left(
-2.1898 
+ 2.1898 \left(\frac{Y_{\Theta_0}+\lambda}{Y_\Theta+\lambda}\right)^{-\frac{50}{81}}
- 3.2510 \ln \left(\frac{Y_{\Theta_0}+\lambda}{Y_\Theta+\lambda}\right)\right.\cr
&& \left.  \hskip 2cm
- 0.3072
\left(\frac{Y_{\Theta_0}+\lambda}{Y_\Theta+\lambda}\right)^{-\frac{50}{81}}
\ln \left(\frac{Y_{\Theta_0}+\lambda}{Y_\Theta+\lambda}\right)
\right)
\psi_{g,\ell_1}(\ell_1,y_1).
\label{eq:deltaCg}
\end{eqnarray}
}

The logarithmic derivative $\psi_{g,\ell_1}(\ell_1,y_1)$
(\ref{eq:psidef}) of the MLLA
spectrum $\tilde D_g(\ell_1,y_1)$ is obtained from (\ref{eq:ifD})
of appendix \ref{section:exactsol}.

\vskip .75 cm

%%%%%%%%%%%%%%%%%%%%%%%%%%%%%%%%%%%%%%%%%%%%%%%%%%%%%%%%%%%%%%
\section{AT LEP AND TEVATRON}
\label{section:LEP}
%%%%%%%%%%%%%%%%%%%%%%%%%%%%%%%%%%%%%%%%%%%%%%%%%%%%%%%%%%%%%%

\vskip .5cm

At LEP energy, the working conditions correspond to $Y_{\Theta_0}\approx
5.2$; they are not very different at the Tevatron where $Y_{\Theta_0}
\approx 5.6$. We first present the curves for LEP, then, after the
discussion concerning the size of the corrections and the domain of
validity of our calculations, we give our predictions for the Tevatron.

\vskip .5cm

\subsection{The average color current}
\label{subsection:accLEP}
%%%%%%%%%%%%%%%%%%%%%%%%%%%%%%%%%%%%%%%%

\bigskip

\vbox{
\begin{center}
\epsfig{file=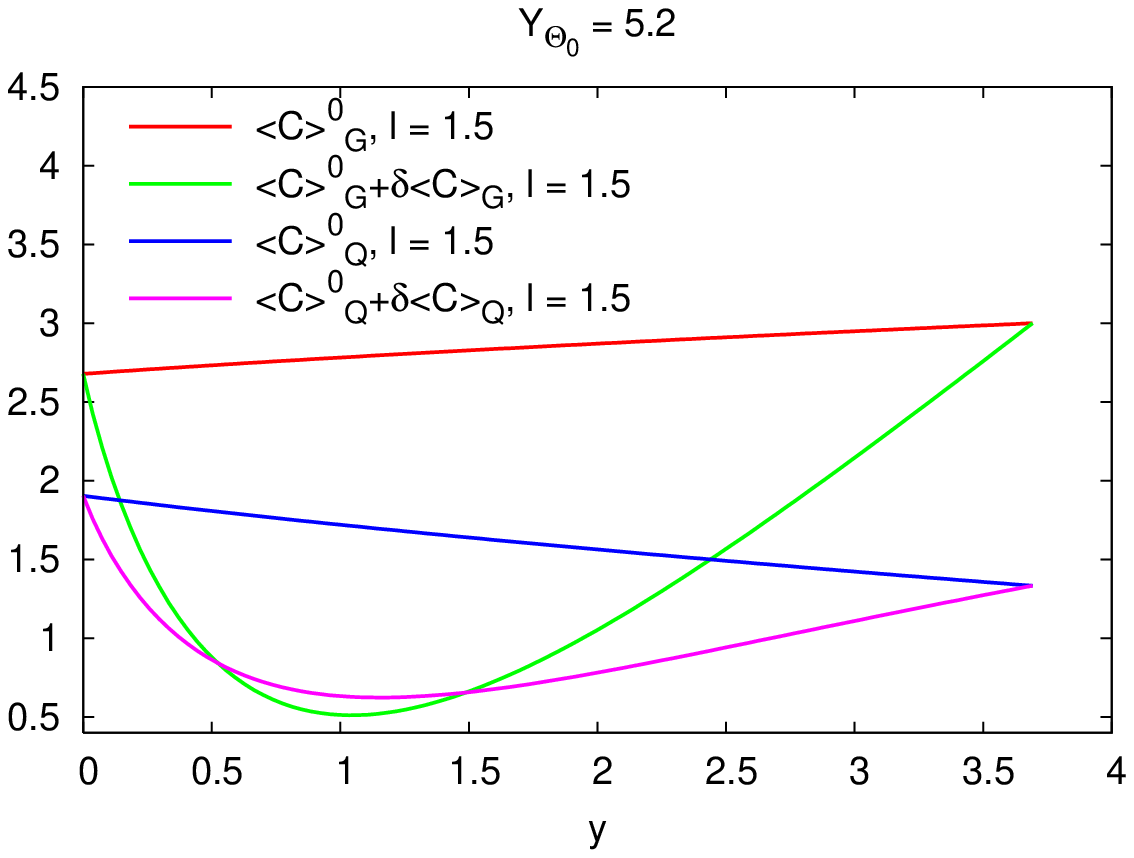, height=5truecm,width=7.5truecm}
\hfill
\epsfig{file=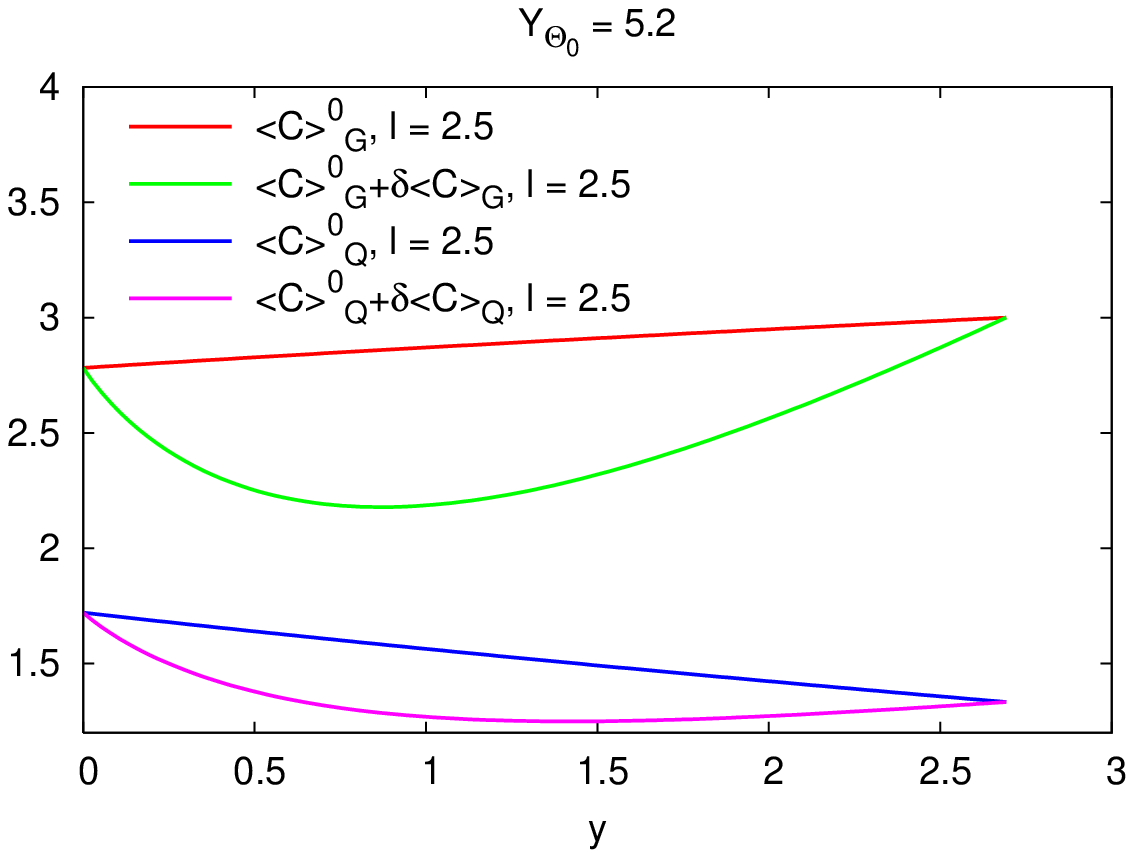, height=5truecm,width=7.5truecm}
\end{center}

\centerline{\em Fig.~18 $<C>_{A_0}^0$ and $<C>_{A_0}^0 + \delta\!<C>_{A_0}$
for quark and gluon jets, as functions of $y$,}

\centerline{\em for $Y_{\Theta_0}=5.2$, $\ell=1.5$ on the left and
$\ell=2.5$ on the right.}
}

\bigskip

Owing to the size of the (MLLA) corrections to the  $<C>$'s and their $y$
derivatives, we will keep to the lower bound $\ell_1 \geq 2.5$.

\vskip .5cm

\subsection{$\boldsymbol{\displaystyle\frac{d^2N}{d\ell_1\;d\ln k_\perp}}$
for a gluon jet}
\label{subsection:d2NgLEP}
%%%%%%%%%%%%%%%%%%%%%%%%%%%%%%%%%%%%%%%%%%%%%%%%%%%%%%%%%%%%%%%

\bigskip

We plot below $\frac{d^2N}{d\ell_1\;d\ln k_\perp}$ for the two values
$\ell=1.5$ and $\ell=2.5$.

\vbox{
\begin{center}
\epsfig{file=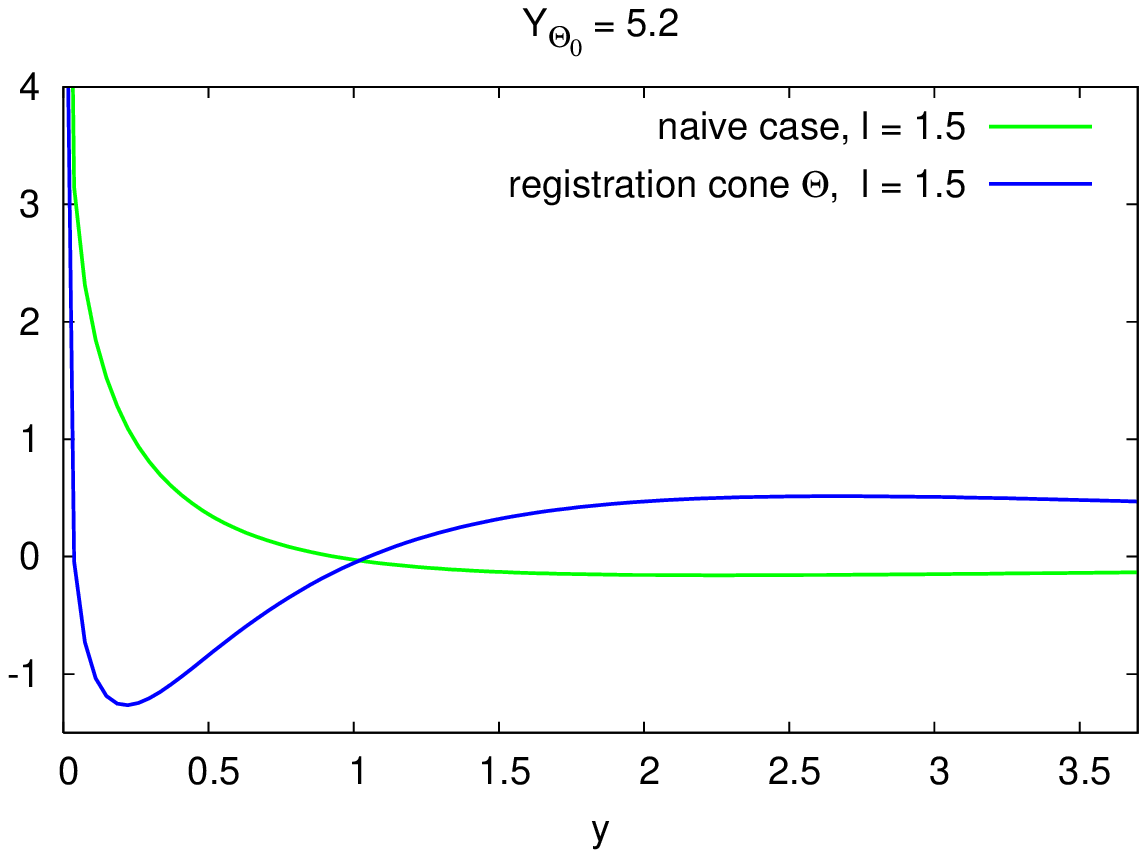, height=5truecm,width=7.5truecm}
\hfill
\epsfig{file=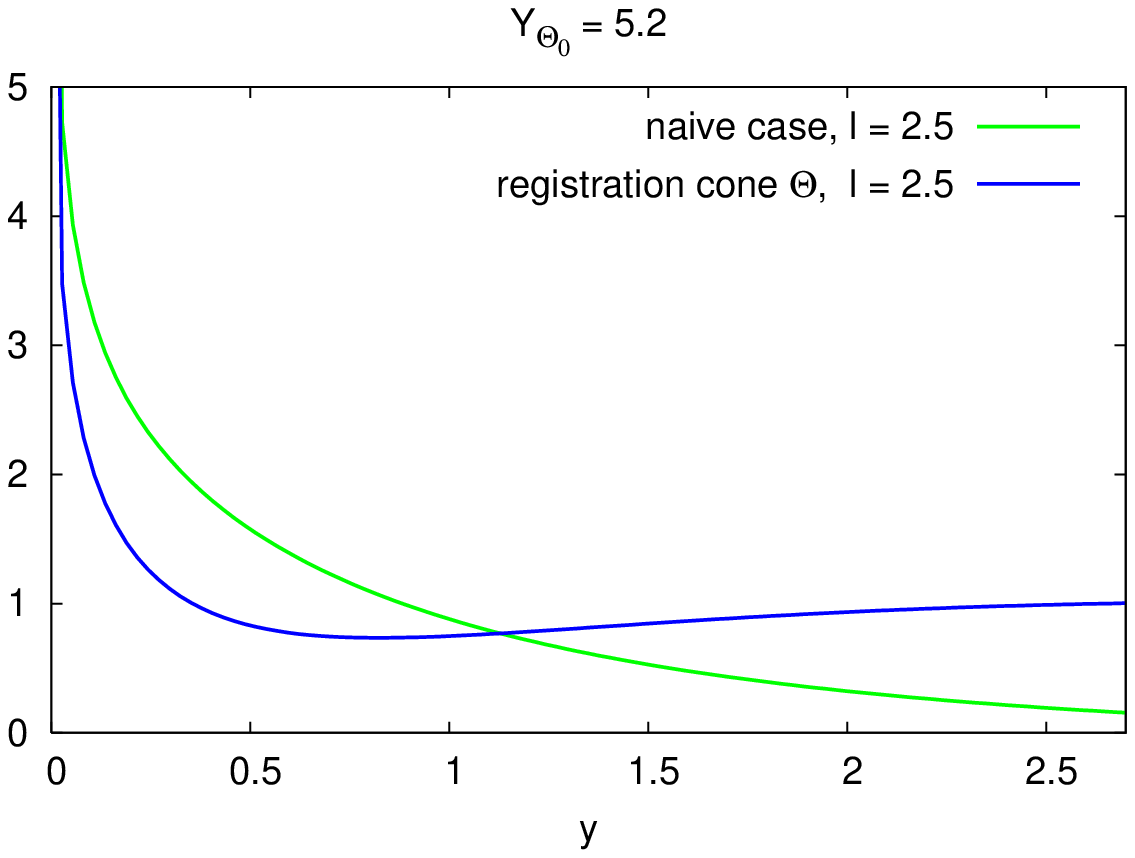, height=5truecm,width=7.5truecm}
\end{center}

\centerline{\em Fig.~19: $\frac{d^2N}{d\ell_1\;d\ln k_\perp}$ for a gluon jet
at fixed $\ell_1$,  MLLA and naive approach.}
}

\bigskip

The excessive size of the $\delta\!<C>$ corrections emphasized in
subsection \ref{subsection:accLEP} translates here into the loss  of
the positivity for $\frac{d^2N}{d\ell_1\;d\ln k_\perp}$ at $\ell=1.5$ for
$y<1$: our approximation is clearly not trustable there.

\vskip .5cm

\subsection{$\boldsymbol{\displaystyle\frac{d^2N}{d\ell_1\;d\ln k_\perp}}$
for a quark jet}
\label{subsection:d2NqLEP}
%%%%%%%%%%%%%%%%%%%%%%%%%%%%%%%%%%%%%%%%%%%%%%%%%%%%%%%%%%%%%%%

\bigskip

We consider the same two values of $\ell$ as above.

\vbox{
\begin{center}
\epsfig{file=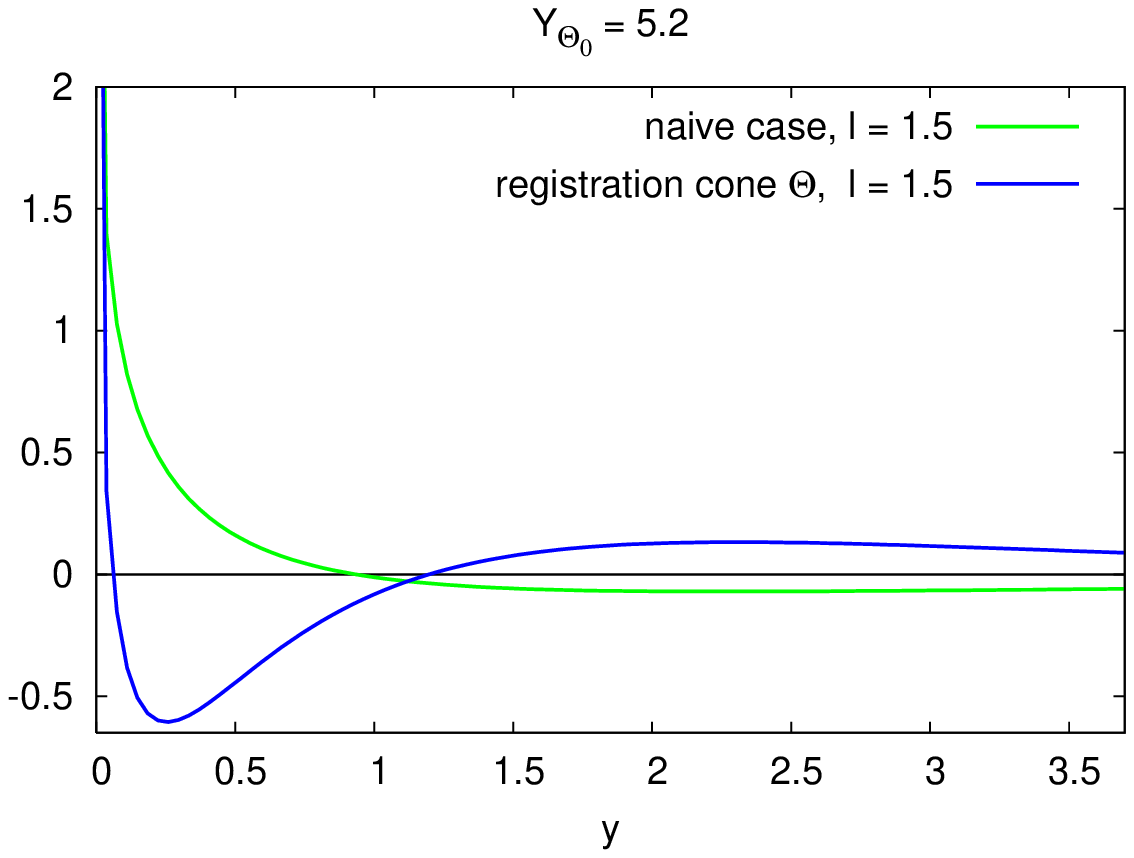, height=5truecm,width=7.5truecm}
\hfill
\epsfig{file=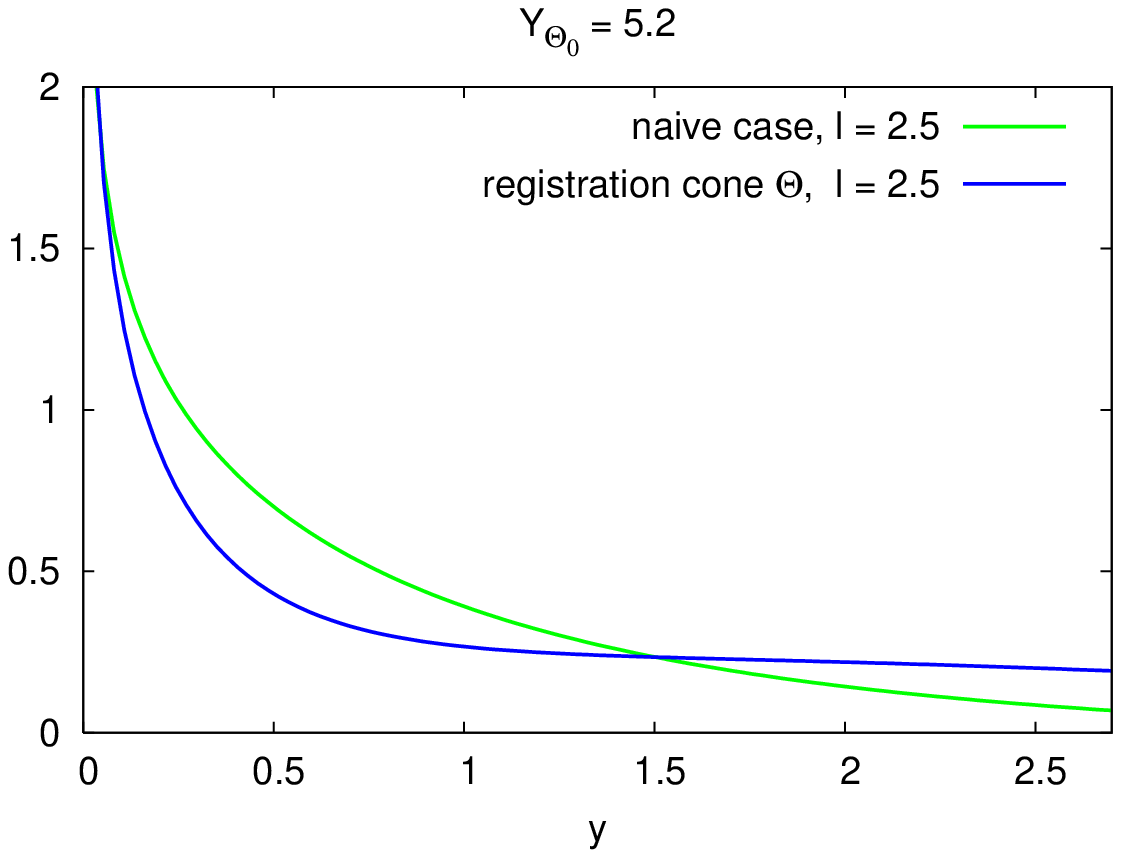, height=5truecm,width=7.5truecm}
\end{center}

\centerline{\em Fig.~20: $\frac{d^2N}{d\ell_1\;d\ln k_\perp}$ for a quark jet
at fixed $\ell_1$,  MLLA and naive approach.}
}

\bigskip

Like for the gluon jet, we encounter positivity problems at $\ell=1.5$ for
$y< 1.25$.

\vskip .5cm

\subsection{$\boldsymbol{\displaystyle\frac{dN}{d\ln k_\perp}}$ for a gluon jet}
\label{subsection:dNgLEP}
%%%%%%%%%%%%%%%%%%%%%%%%%%%%%%%%%%%%%%%%%%%%%%%%%%%%%%%%%%%%%%%%%%%%%%%%%%%%

\bigskip

We plot below $\frac{dN}{d\ln k_\perp}$ for a gluon jet obtained by the
``naive'' approach and including the jet evolution from $\Theta_0$ to
$\Theta$;  on the right is
an enlargement which shows how positivity is recovered when MLLA corrections
are included.

\vbox{
\begin{center}
\epsfig{file=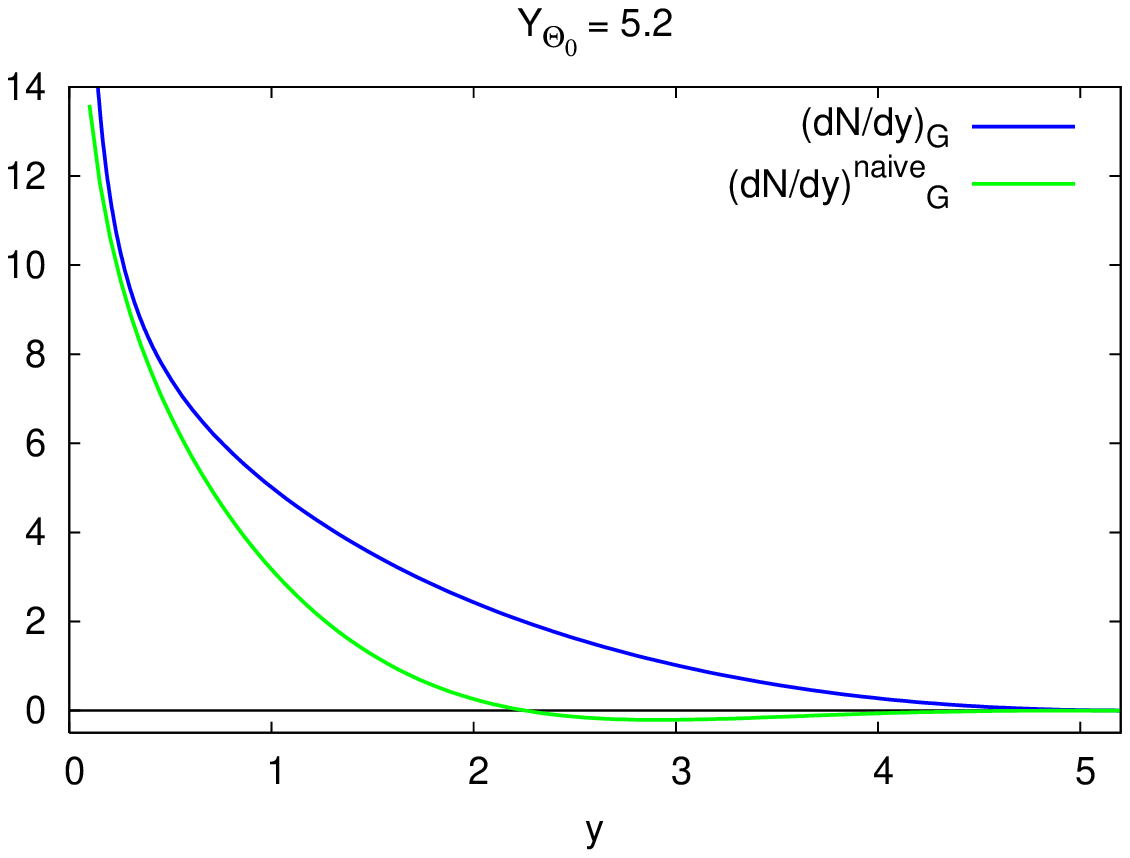, height=5truecm,width=7.5truecm}
\hfill
\epsfig{file=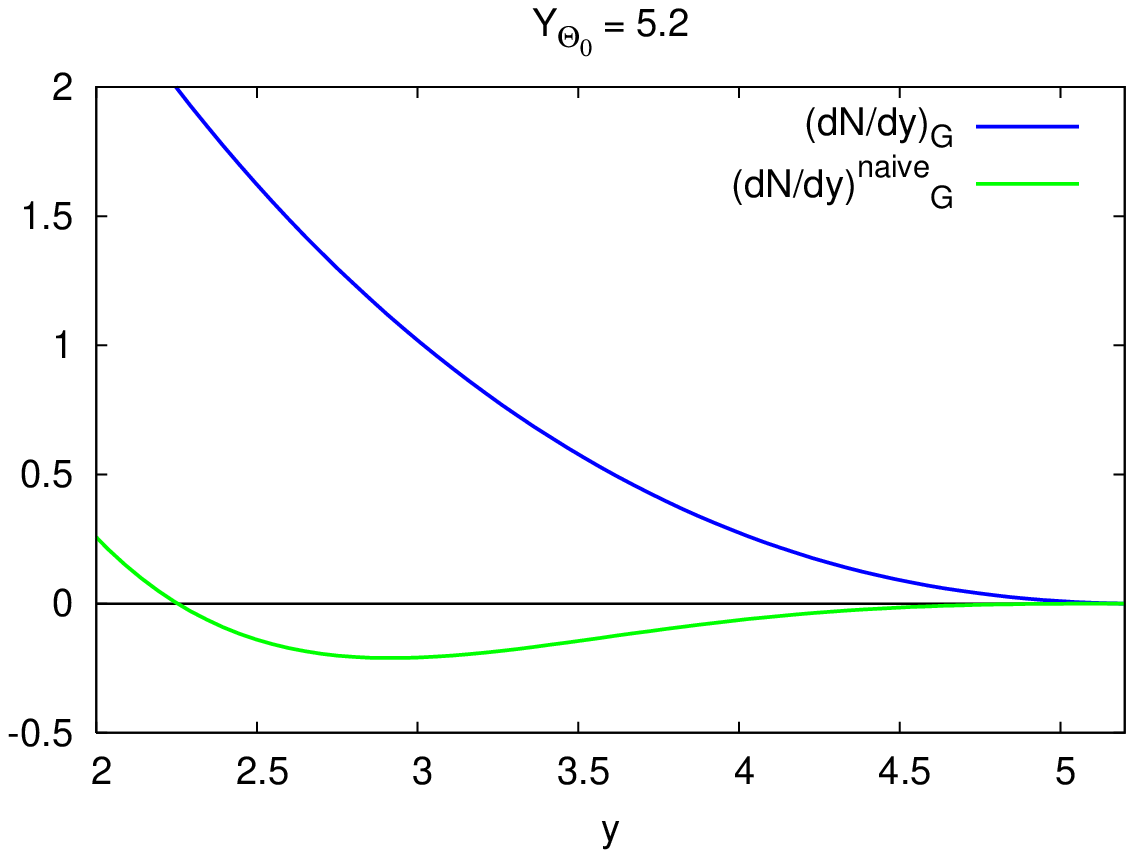, height=5truecm,width=7.5truecm}
\end{center}

\centerline{\em Fig.~21: $\frac{dN}{d\ln k_\perp}$ for a gluon jet,
MLLA and naive approach.}
}

\vskip .5cm

\subsection{$\boldsymbol{\displaystyle\frac{dN}{d\ln k_\perp}}$ for a quark jet}
\label{subsection:dNqLEP}
%%%%%%%%%%%%%%%%%%%%%%%%%%%%%%%%%%%%%%%%%%%%%%%%%%%%%%%%%%%%%%%%%%%%%%%%%%%%

\bigskip

We proceed like for a gluon jet. The curves below show the restoration of
positivity by MLLA corrections.

\vbox{
\begin{center}
\epsfig{file=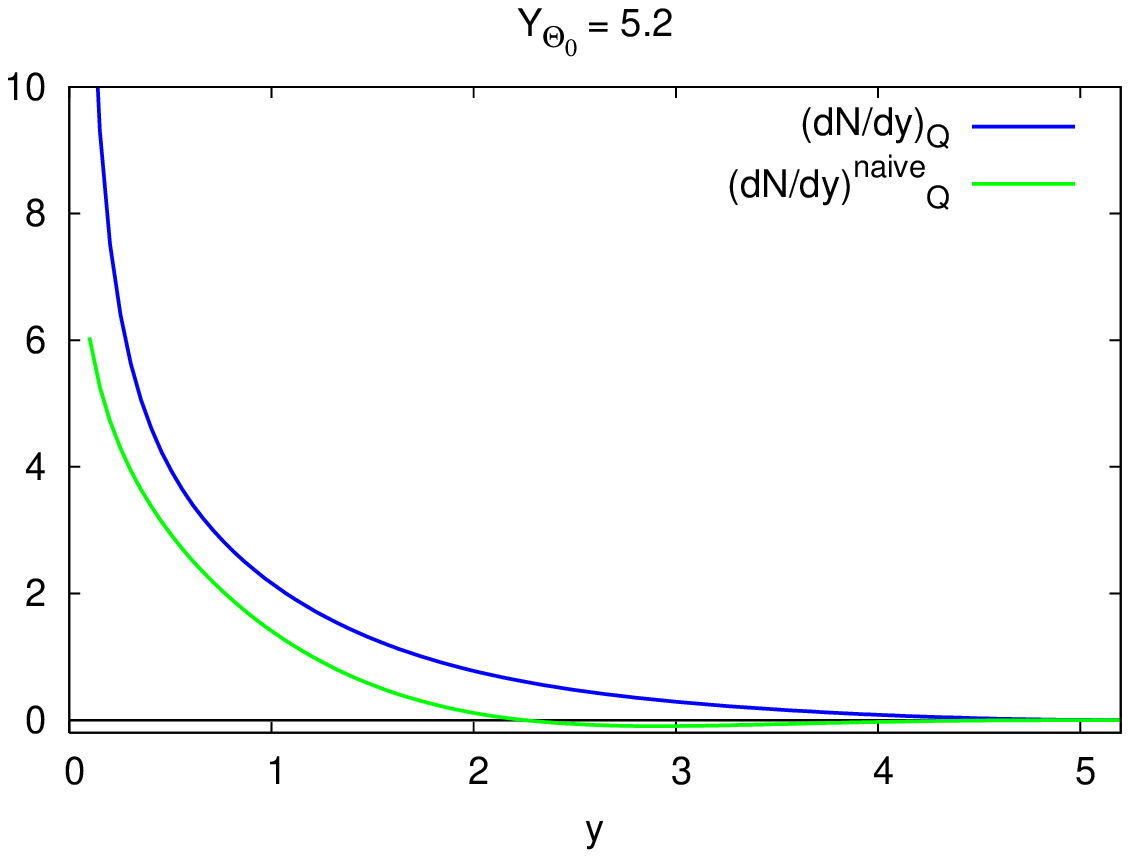, height=5truecm,width=7.5truecm}
\hfill
\epsfig{file=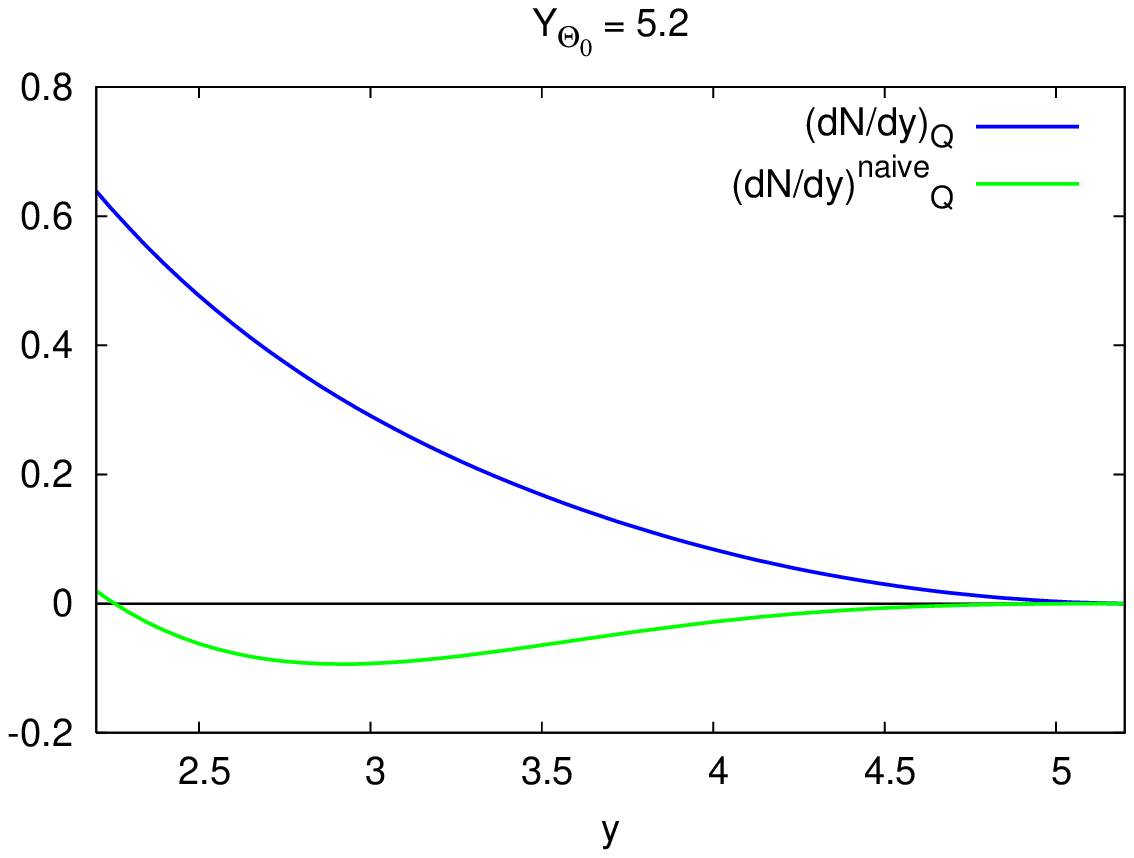, height=5truecm,width=7.5truecm}
\end{center}

\centerline{\em Fig.~22: $\frac{dN}{d\ln k_\perp}$ for a quark jet,
MLLA and naive approach.}
}

\vskip .5cm

That the upper bound of the $\ell_1$ domain of integration defining
$\frac{dN}{d\ln k_\perp}$ corresponds to a large enough $\ell_1 \geq 2.5$
requires that, for LEP, $y_1$ should be smaller that $5.2 - 2.5 = 2.7$;
combined with the necessity to stay in the perturbative regime, it yields
$1 \leq y_1 \leq 2.7$.

\subsection{Discussion and predictions for the Tevatron}
\label{subsection:discussion}
%%%%%%%%%%%%%%%%%%%%%%%%%%%%%%%%%%%%%%%%%%%%%%%%%%%%%%%%

The similar condition at Tevatron is $1 \leq y_1 \leq 5.6 - 2.5 = 3.1$;
like for LEP, it  does not extend to large values of $k_\perp$
because, there, the small $x$ approximation is no longer valid.
We give below the curves that we predict in this confidence
interval.

\vbox{
\begin{center}
\epsfig{file=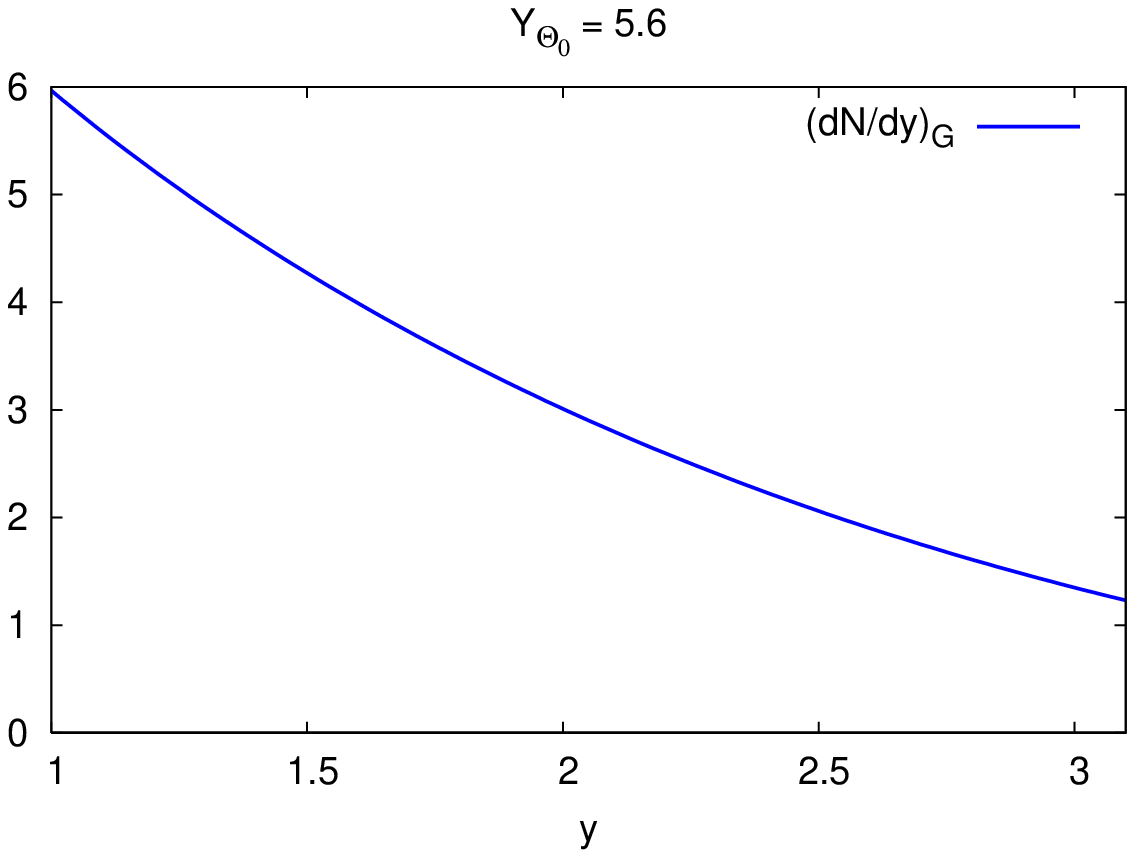, height=5truecm,width=7.5truecm}
\hfill
\epsfig{file=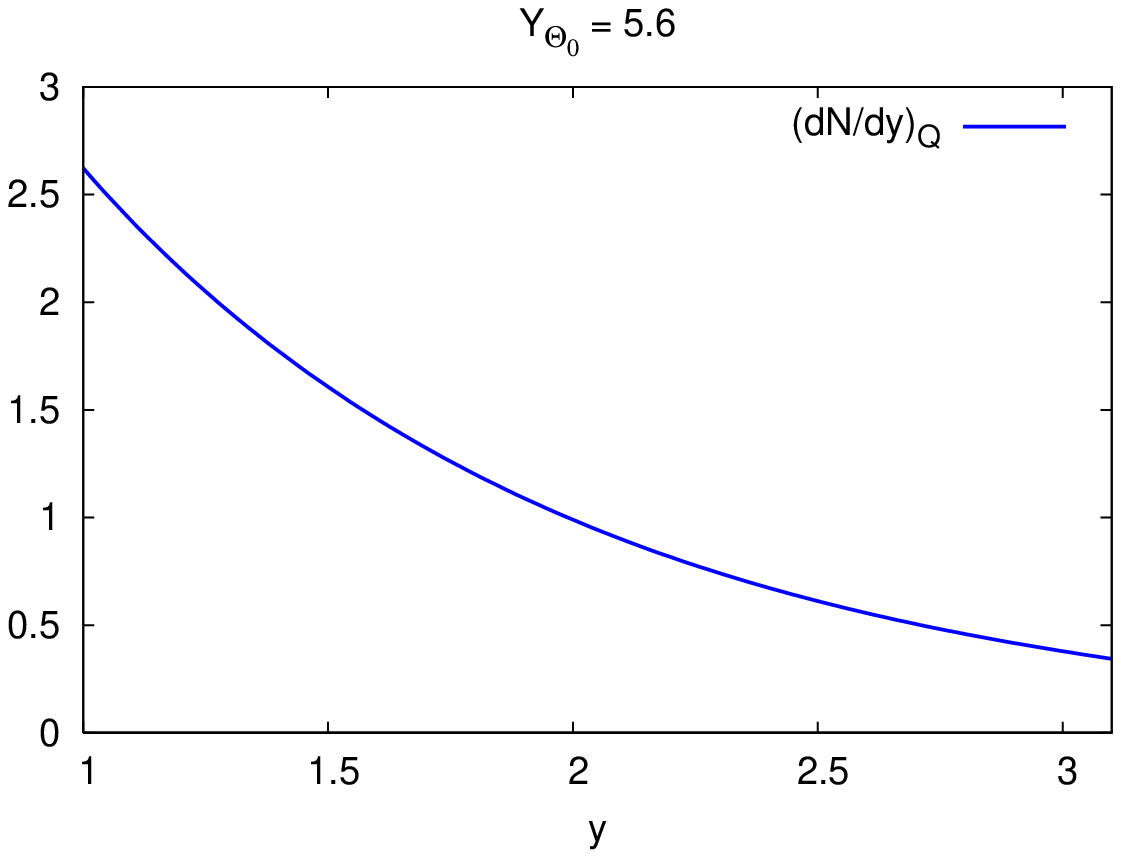, height=5truecm,width=7.5truecm}
\end{center}

\centerline{\em Fig.~23: $\frac{dN}{d\ln k_\perp}$ for a gluon (left) and
a quark (right) jets, MLLA predictions for the Tevatron.}
}

\bigskip

Since experimental results involve a mixture of gluon and quark jets, the
mixing parameter $\omega$ (subsection \ref{subsection:mixed}) has to be
introduced in the comparison with theoretical curves, together with the
phenomenological factor ${\cal K}^{ch}$ 
 normalizing partonic to charge hadrons distributions.

\vskip .75 cm

%%%%%%%%%%%%%%%%%%%%%%%%%%%%%%%%%%%%%%%%%%%%%%%%%%%%%%%%%%%%%%%%%%%%%%%%%%%%
\section{COMPARING DLA AND MLLA APPROXIMATIONS}
\label{section:DLA}
%%%%%%%%%%%%%%%%%%%%%%%%%%%%%%%%%%%%%%%%%%%%%%%%%%%%%%%%%%%%%%%%%%%%%%%%%%%%

\vskip .5cm

DLA \cite{DLA} \cite{DLA1} and MLLA approximations are very different
\cite{EvEq}; in particular,
the exact balance of energy (recoil effects of partons) is not accounted
for in DLA.

We compare  DLA and MLLA results for the two distributions of concern
in this work.
 Studying first their difference for the spectrum itself
eases the rest of the comparison.

We choose the two values $Y_{\Theta_0}=7.5$ and $Y_{\Theta_0}=15$.
While the first corresponds to the LHC working
conditions (see footnote \ref{footnote:LHC}), the second is purely academic
since, taking for example
$\Theta_0 \approx .5$ and $Q_0 \approx 250\; MeV$, it corresponds to an
energy of  $1635\; TeV$; it is however suitable, as we shall see in
subsection \ref{subsection:ktDLA} to disentangle the effects of coherence and
the ones of the divergence of $\alpha_s$ at low energy in the calculation
of the inclusive $k_\perp$ distribution.

\vskip .75 cm

\subsection{The spectrum}
\label{subsection:DLAspec}
%%%%%%%%%%%%%%%%%%%%%%%%%

\vskip .5cm

Fixing  $\alpha_s$  in DLA at the largest scale of the process, the
collision energy, enormously damps the corresponding spectrum (it does not
take into account the growing of $\alpha_s$ accompanying parton cascading),
which gives an unrealistic aspect to the comparison.

This is why, as far as the spectra are concerned, we shall compare their
MLLA evaluation with that obtained from the latter by 
taking to zero the coefficient $a$ given in (\ref{eq:adef}),
which also entails $B=0$; ${\cal F}_0(\tau,y,\ell)$ in
(\ref{eq:calFdef}) becomes $I_0(2\sqrt{Z(\tau,y.\ell)}$.
The infinite normalization that occurs in (\ref{eq:ifD}) because of
$\Gamma(B=0)$ we replace by a constant such that the two calculations can
be easily compared. 
This realizes a DLA  approximation (no accounting for recoil effects)
 ``with running $\alpha_s$''.

On Fig.~24 below are plotted the spectrum $\tilde D_g(\ell,y\equiv
Y_{\Theta_0}-\ell)$ for gluon jets in the MLLA and DLA ``with running
$\alpha_s$'' approximations.

\bigskip

\vbox{
\begin{center}
\epsfig{file=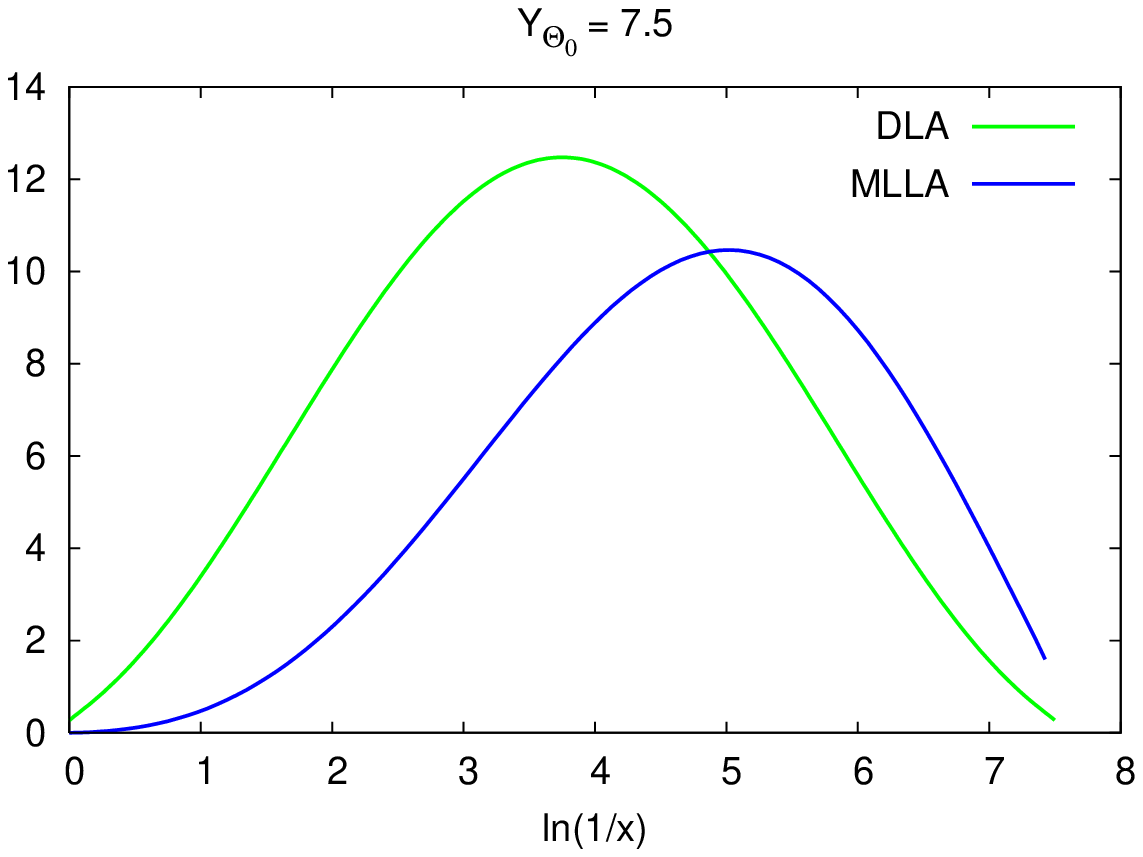, height=5truecm,width=7.5truecm}
\hfill
\epsfig{file=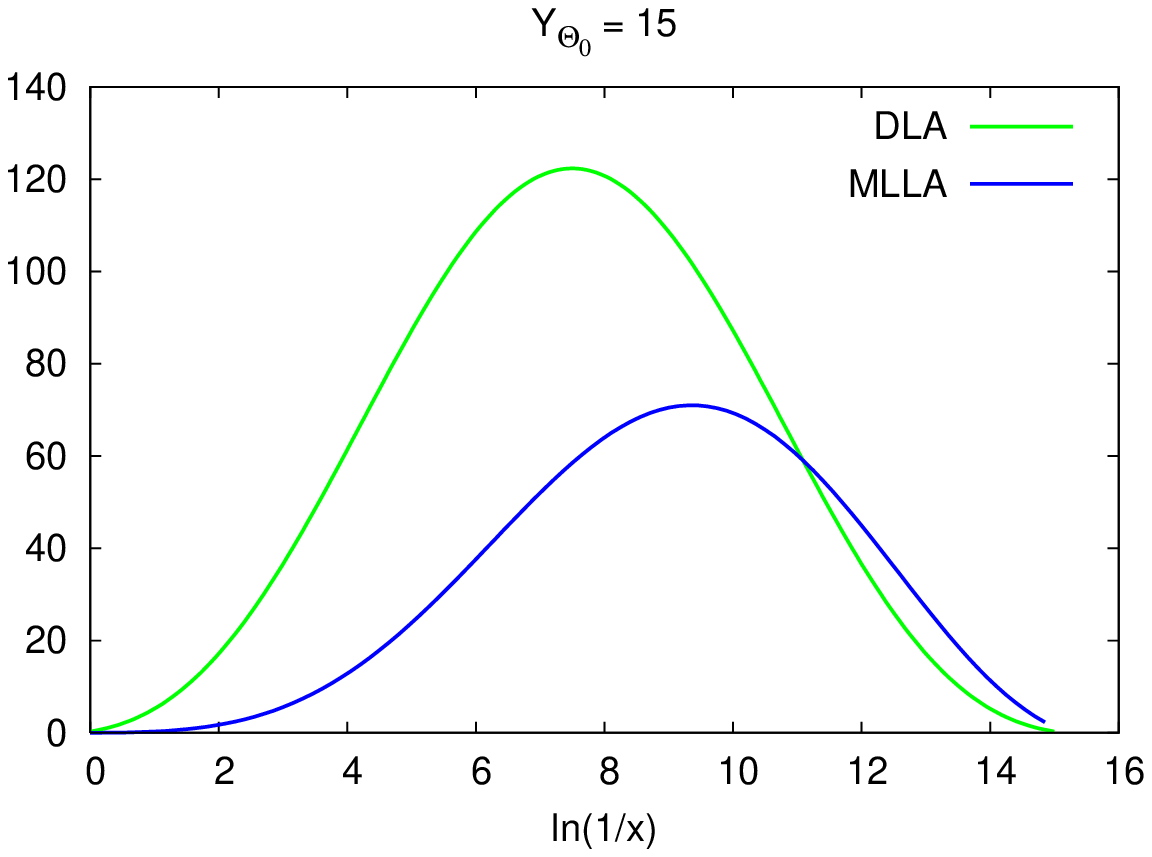, height=5truecm,width=7.5truecm}
\end{center}

\centerline{\em Fig.~24: the spectrum $\tilde D_g(\ell,Y_{\Theta_0}-\ell)$
for gluon jets;}
\centerline{\em  comparison between MLLA and DLA (``with running
$\alpha_s$'') calculations.}
}

\bigskip

The peak of the MLLA spectrum is seen, as expected,
 to occur at smaller values of the energy than that of DLA.

\vskip .5cm

\subsection{Double differential 1-particle inclusive distribution}
\label{subsection:doubleDLA}
%%%%%%%%%%%%%%%%%%%%%%%%%%%%%%%%%%%%%%%%%%%%%%%%%%%%%%%%%%%%%%%%%%%%%%%%%%%%

\vskip .5cm

The genuine MLLA calculations being already shown on Figs.~3 and 5, 
Fig.~25 displays, on the left, a ``modified''
MLLA calculation obtained by dividing by
$\alpha_s(k_\perp^2) \approx \frac{\pi}{2N_c \beta y}$
(see (\ref{eq:gamma0}) with $\lambda \to 0$);
subtracting in the MLLA calculations the dependence on $k_\perp$ due to the
running of $\alpha_s(k_\perp^2)$ allows a better comparison with DLA (with
fixed $\alpha_s$) by getting rid of the divergence when $k_\perp \to Q_0$.

 On the right are plotted the DLA results for gluon jets, in which
$\alpha_s$ has been fixed at the collision energy (it is thus very small).
Since their normalizations are now different, only the {\em shapes} of the two
types of curves must be compared;
we indeed observe that the DLA growing  of
$\frac{d^2N}{d\ell_1\,d\ln k_\perp}$ with $k_\perp$
(or $y_1$) also occurs in the ``modified'' MLLA curves.

\vskip .7cm 

\vbox{
\begin{center}
\epsfig{file=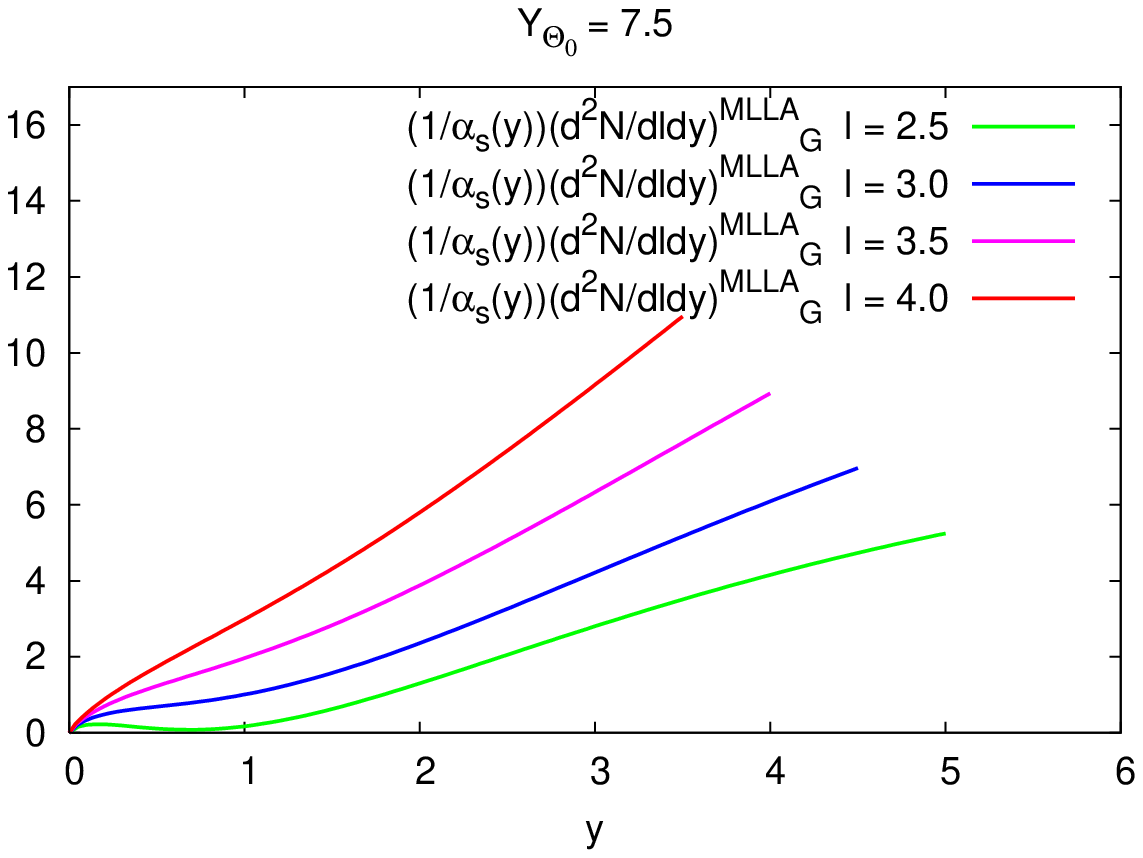, height=5truecm,width=7.5truecm}
\hfill
\epsfig{file=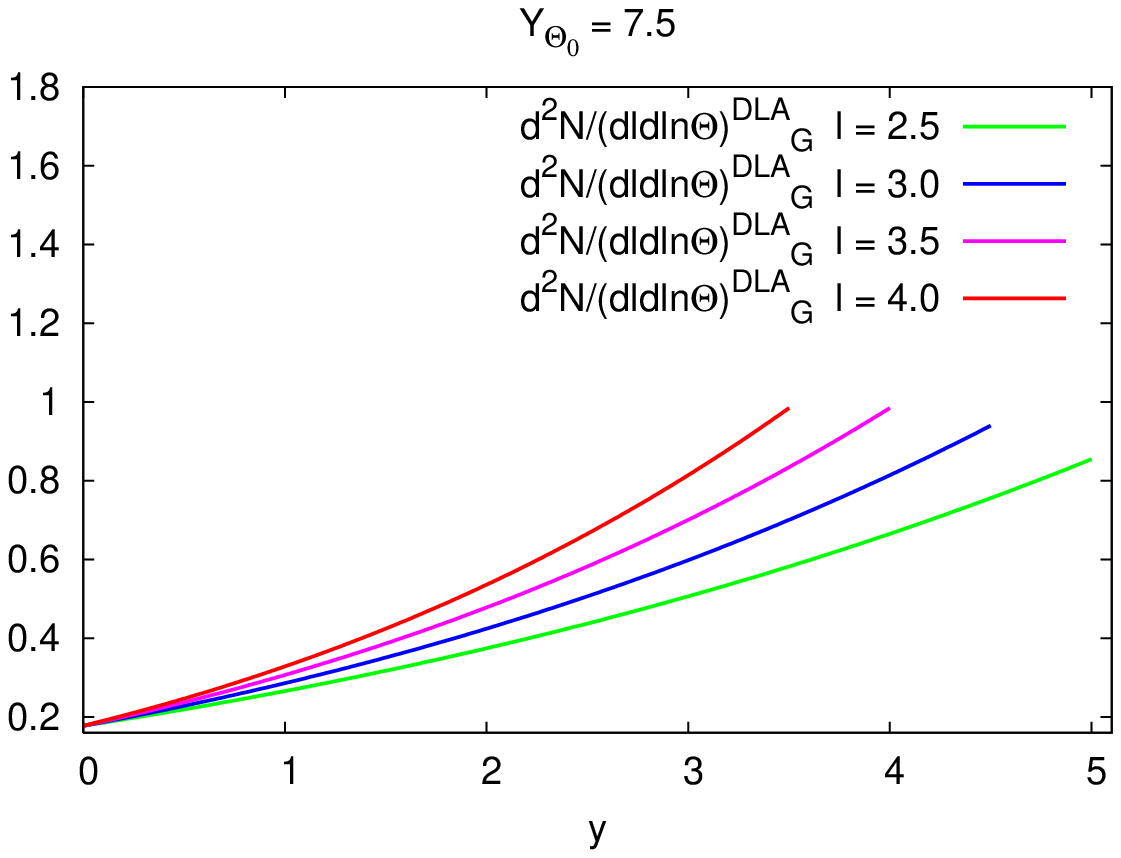, height=5truecm,width=7.5truecm}
\end{center}

\centerline{\em Fig.~25: comparison between MLLA (after dividing by
$\alpha_s(k_\perp^2)$, on the left)}
\centerline{\em  and DLA calculation with $\alpha_s$ fixed (on the right)
of $\frac{d^2N}{dy\,d\ln k_\perp}$ for gluon jets.}
}

\bigskip

The DLA distribution for quark jets is obtained from that of gluon jets
by multiplication by the factor $C_F/N_c$; it it thus also a
growing function of $y_1$.

The MLLA distribution for quark jets, which is, unlike that for gluon jets,
a decreasing function of $y_1$ (see Fig.~6), becomes, like the latter, 
growing, after the dependence on $\alpha_s(k_\perp^2)$ has been
factored out: one  finds the same behavior as in DLA.

\vskip .75 cm

\subsection{Inclusive $\boldsymbol{k_\perp}$ distribution}
\label{subsection:ktDLA}
%%%%%%%%%%%%%%%%%%%%%%%%%%%%%%%%%%%%%%%%%%%%%%%%%%%%%%%%%%%%%%%%%%%%%%%%%%%%

\vskip .5cm

On Fig.~26  we have plotted, at $Y_{\Theta_0}= 7.5$:

- the MLLA calculation of $\frac{dN}{d\ln k_\perp}$
divided by $\alpha_s(k_\perp^2)$,
such that the divergence due to the running of $\alpha_s$
has been factored out, leaving  unperturbed the damping  due to
coherence effects;

- the DLA calculation of  $\frac{dN}{d\ln k_\perp}$ with $\alpha_s$ fixed
  at the collision energy.

Like in \ref{subsection:doubleDLA}, because of the division by $\alpha_s$,
the two curves are  not normalized alike, such that only
their {\em shapes} should be compared.

The comparison of the DLA curve (at fixed $\alpha_s$)
 with the genuine MLLA calculation displayed in Fig.~7  (left)
 shows how different are the outputs of the two
approximations; while at large $k_\perp$ they are both decreasing, at small
$k_\perp$  the running of $\alpha_s$  makes the sole MLLA
distribution diverge when $k_\perp \to Q_0$ (non-perturbative
domain).

\vskip .5cm

\vbox{
\begin{center}
\epsfig{file=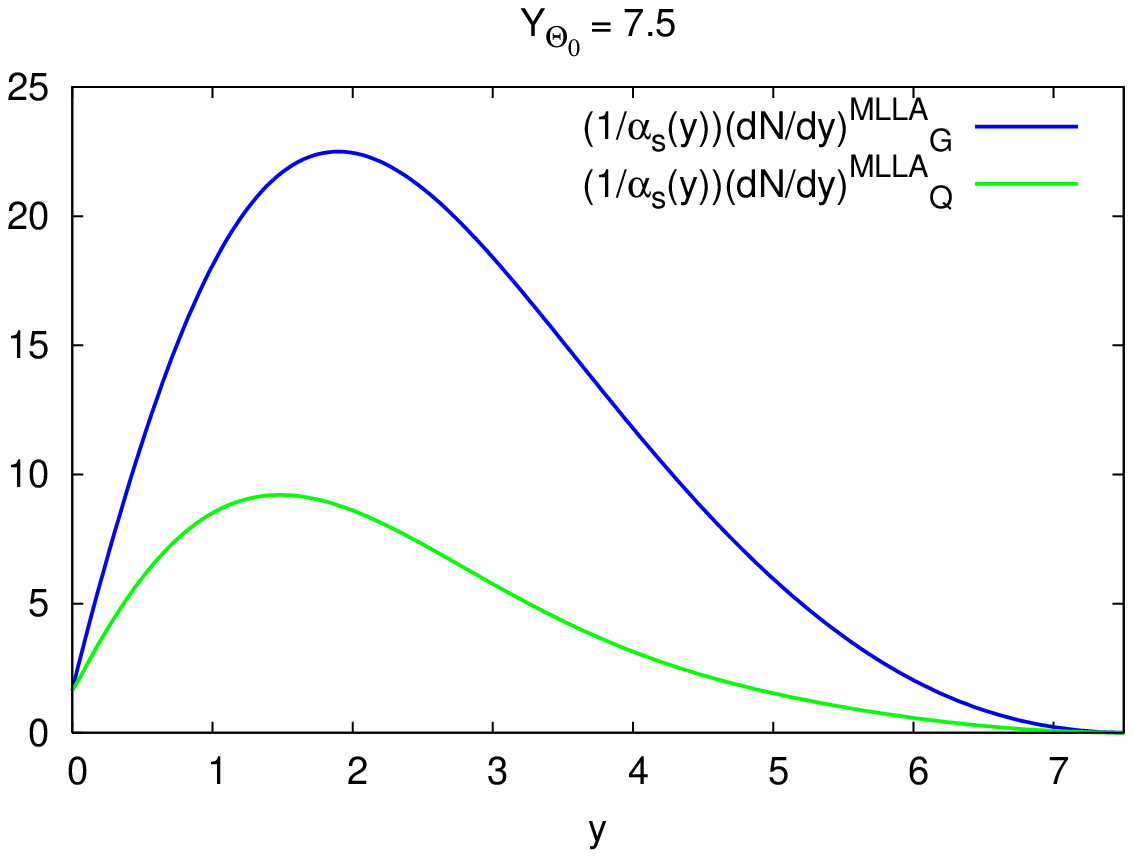, height=5truecm,width=7.5truecm}
\hfill
\epsfig{file=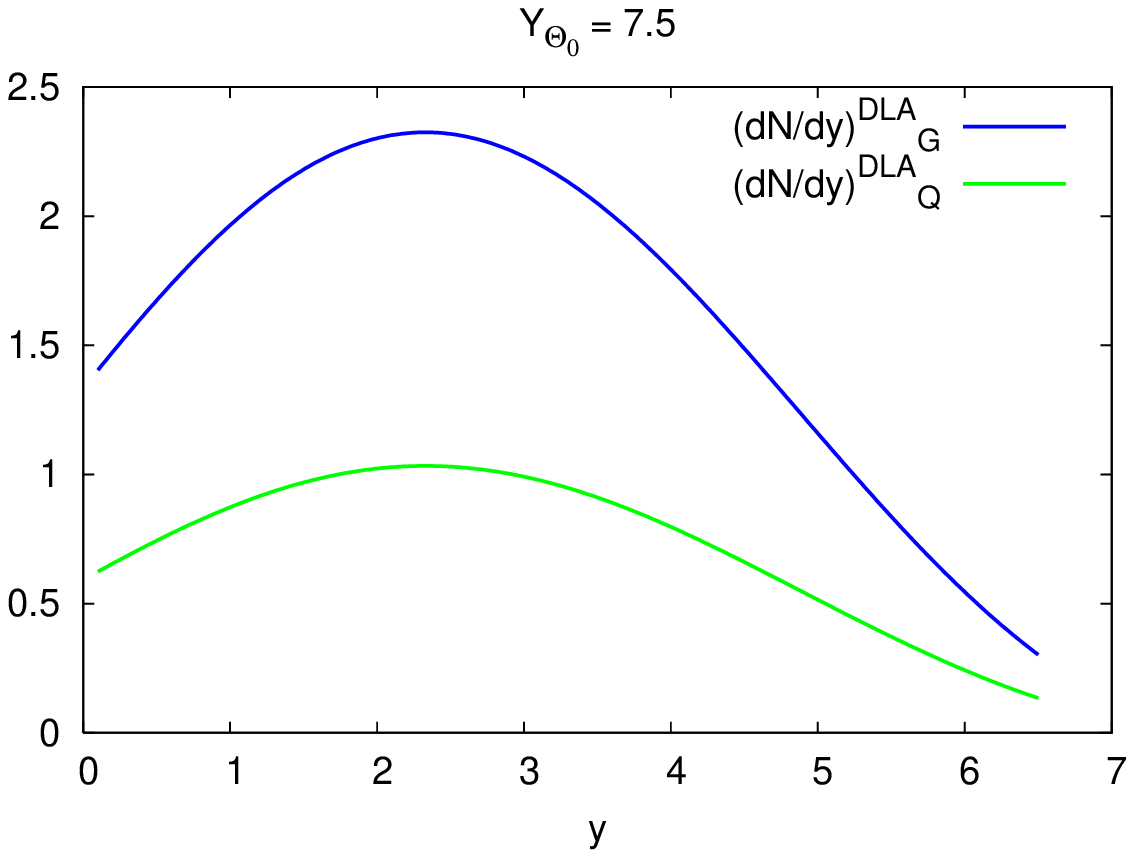, height=5truecm,width=7.5truecm}
\end{center}

\centerline{\em Fig.~26: $Y_{\Theta_0}=7.5$:
comparing  MLLA and DLA calculations
of $\frac{dN}{d\ln k_\perp}$ (see also Fig.~7);}
\centerline{\em from left to right: 
$\frac{1}{\alpha_s(k_T^2)}$MLLA
and DLA ($\alpha_s$ fixed).}
}

\vskip .5cm

In the extremely high domain of energy $Y_{\Theta_0}=15$ used for Fig.~27,
the two competing phenomena occurring
 at small $y_1$ can then be neatly distinguished.

The first plot, showing MLLA results,  cleanly separates
coherence effects from the running of  $\alpha_s$;
in the second figure we have plotted the  MLLA calculation
divided by $\alpha_s(k_\perp^2)$: damping at small $y_1$ due to coherence
effects appears now unspoiled;
finally, DLA calculations  clearly exhibit, too, the damping due to coherence
\footnote{The DLA points corresponding to $y_1=0$ can be analytically
determined to be $4N_c/n_f$ (gluon jet) and $4C_F/n_f$ (quark jet); they
are independent of the energy $Y_{\Theta_0}$.}
.

The large difference of magnitude  observed between the first
(genuine MLLA) and the last (DLA) plots occurs because DLA calculations
have been performed with $\alpha_s$ fixed at the very high collision energy.

Like in \ref{subsection:doubleDLA}, because of the division by $\alpha_s$,
the second curve is not normalized like the two others, such that only
its {\em shape} should be compared with theirs.

\begin{center}
\epsfig{file=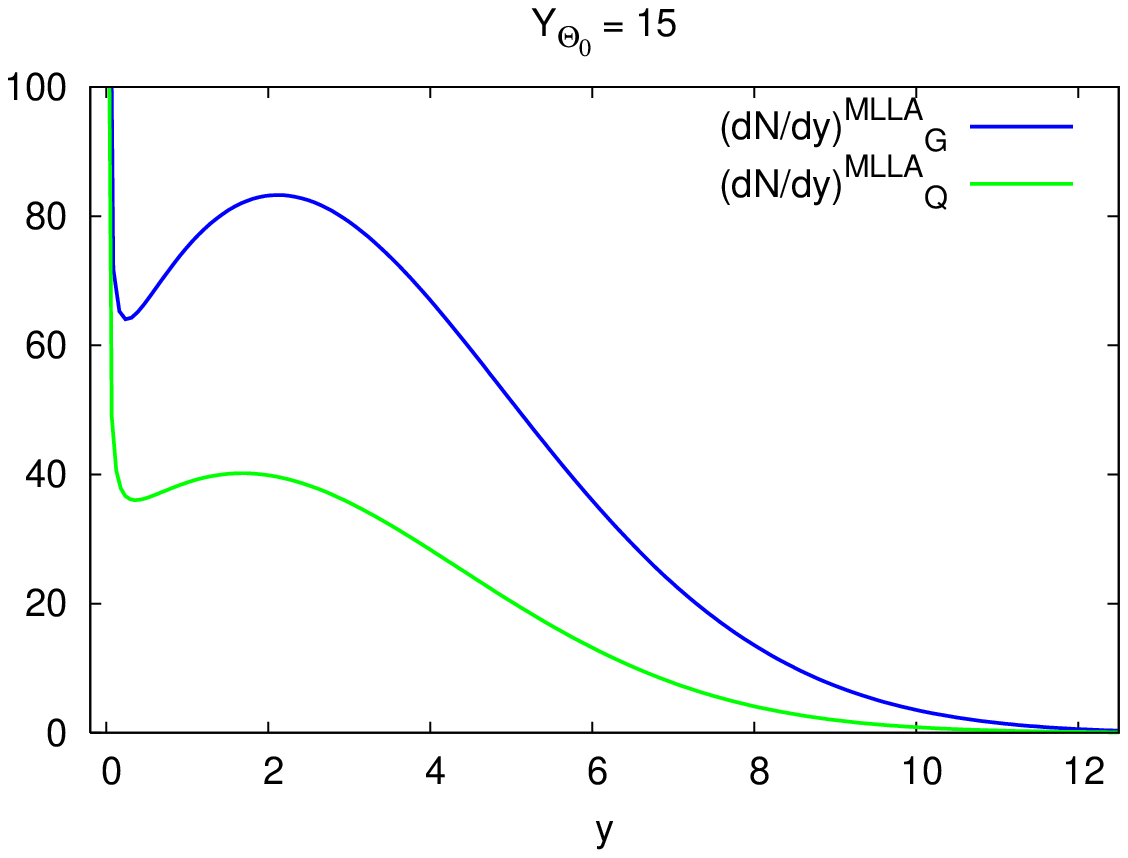, height=5truecm,width=7.5truecm}
\hfill
\epsfig{file=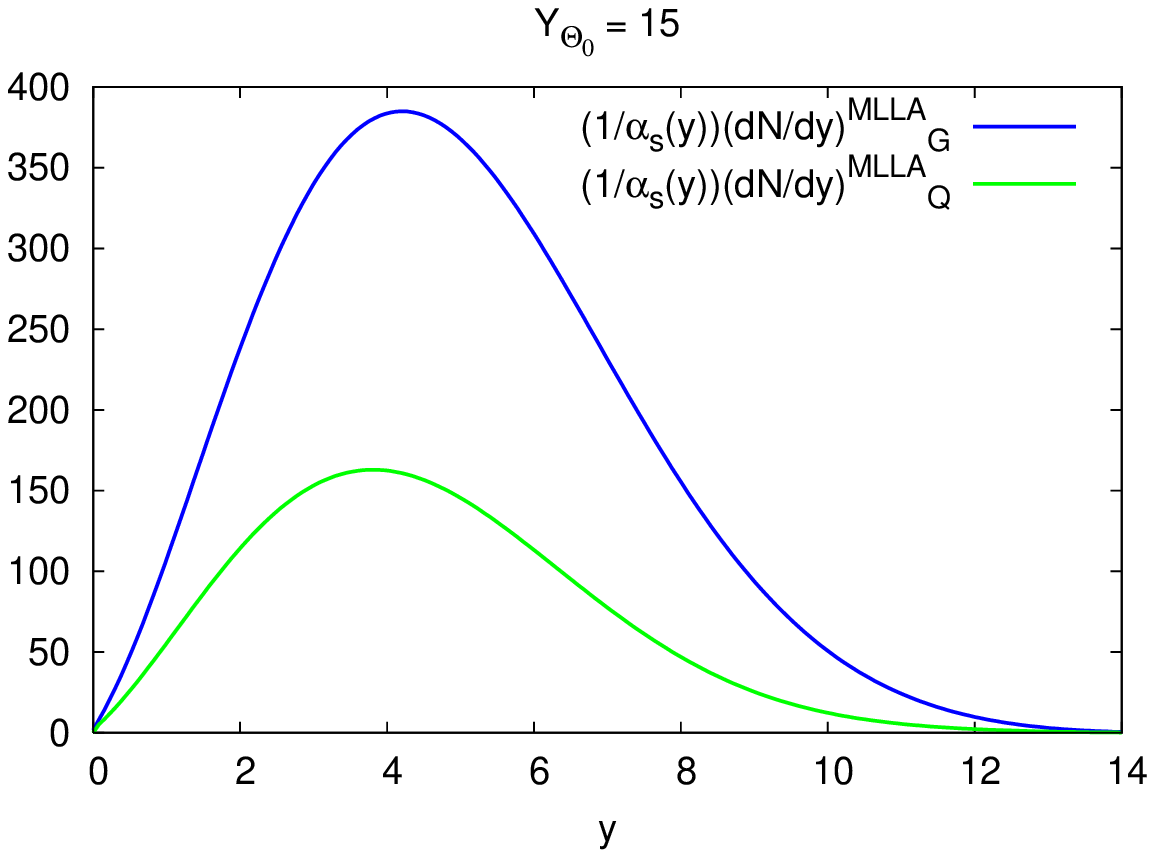, height=5truecm,width=7.5truecm}
\end{center}

\vbox{
\begin{center}
\epsfig{file=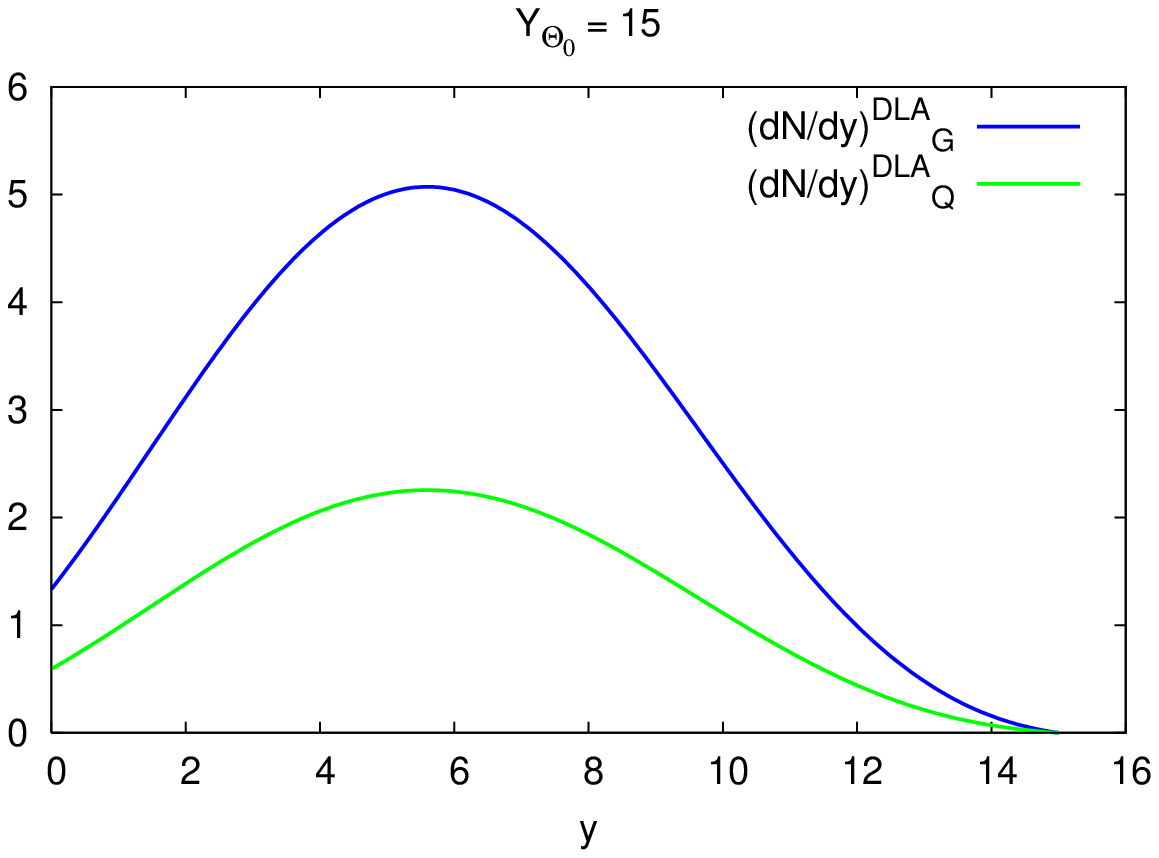, height=5truecm,width=7.5truecm}
\end{center}

\centerline{\em Fig.~27: $Y_{\Theta_0} = 15$: 
comparing  MLLA and DLA calculations
of $\frac{dN}{d\ln k_\perp}$;}
\centerline{\em from left to right: MLLA,
$\frac{1}{\alpha_s(k_T^2)}$MLLA and DLA ($\alpha_s$ fixed).}
}

\bigskip

%%%%%%%%%%%%%%%%%%%%%%%%%%%%%%%%%%%%%%%%%%%%%%%%%%%%%%%%%%%%%%%%%%%%%%%%%%%%
%%%%%%%%%%%%%%%%%%%%%%%%%%%%%%%%%%%%%%%%%%%%%%%%%%%%%%%%%%%%%%%%%%%%%%%%%%%%
\newpage

{\bf\Large Figure captions}

\vskip .75 cm

Fig.~1: the process under consideration: two hadrons $h_1$ and $h_2$ inside
one jet;

\smallskip

Fig.~2: $<C>_{A_0}^0$ and $<C>_{A_0}^0 + \delta<C>_{A_0}$ for quark and gluon
jets, as functions of $y$,
for $Y_{\Theta_0}=7.5$, $\ell=2.5$  and $\ell=3.5$;

\smallskip

Fig.~3: $\frac{d^2N}{d\ell_1\;d\ln k_\perp}$ for a gluon jet,
$Y_{\Theta_0} =7.5$ and $Y_{\Theta_0}=10$;

\smallskip

Fig.~4: $\frac{d^2N}{d\ell_1\;d\ln k_\perp}$ at fixed $\ell_1$ for a gluon jet,
 comparison between MLLA and the naive approach;

\smallskip

Fig.~5: $\frac{d^2N}{d\ell_1\;d\ln k_\perp}$ for a quark jet,
 $Y_{\Theta_0} =7.5$ and $Y_{\Theta_0}=10$;

\smallskip

Fig.~6: $\frac{d^2N}{d\ell_1\;d\ln k_\perp}$ at fixed $\ell_1$ for a quark jet,
 comparison between MLLA and the naive approach;

\smallskip

Fig.~7: inclusive $k_\perp$ distribution $\frac{d{N}} {d\ln k_\perp}$
for a gluon jet,  $Y_{\Theta_0} =7.5$ and $Y_{\Theta_0}=10$,
MLLA and naive approach, both for $\ell_{min}=0$;

\smallskip

Fig.~8: enlargements of Fig.~6 at large $k_\perp$;

\smallskip

Fig.~9: inclusive $k_\perp$ distribution  $\frac{d{N}} {d\ln k_\perp}$
for a quark jet,  $Y_{\Theta_0} =7.5$ and
$Y_{\Theta_0}=10$,
MLLA and naive approach, both for $\ell_{min}=0$;

\smallskip

Fig.~10: enlargements of Fig.~8 at large $k_\perp$;

\smallskip

Fig.~11: role of the upper limit of integration over $x_1$
in the inclusive $k_\perp$ distribution $\frac{dN}{d\ln\,k_\perp}$
for gluon (left) and quark (right) jet;

\smallskip

Fig.~12: spectrum $\tilde D_g(\ell,y)$ of emitted partons as functions 
of transverse momentum (left) and energy (right);

\smallskip

Fig.~13:  enlargements of Fig.~11 close to the origin;

\smallskip

Fig.~14:  $\frac{d\tilde D_g(\ell,y)}{dy}$ as a function of $y$
for different values of $\ell$;

\smallskip

Fig.~15: $\frac{d\tilde D_g(\ell,y)}{dy}$ as a function of $\ell$
for different values of $y$;

\smallskip

Fig.~16: $\frac{d\tilde D_g(\ell,y)}{d\ell}$ as a function of $\ell$
for different values of $y$;

\smallskip

Fig.~17: $\frac{d\tilde D_g(\ell,y)}{d\ell}$ as a function of $y$
for different values of $\ell$;

\smallskip

Fig.~18: $<C>_{A_0}^0$ and $<C>_{A_0}^0 + \delta\!<C>_{A_0}$
for quark and gluon jets, as functions of $y$,
for $Y_{\Theta_0}=5.2$, $\ell=1.5$ and $\ell=2.5$;

\smallskip

Fig.~19: $\frac{d^2N}{d\ell_1\,d\ln k_\perp}$ for a gluon jet
for $Y_{\Theta_0}=5.2$ at fixed $\ell_1$,  MLLA and naive approach;

\smallskip

Fig.~20: $\frac{d^2N}{d\ell_1\,d\ln k_\perp}$ for a quark jet
for $Y_{\Theta_0}=5.2$ at fixed $\ell_1$,  MLLA and naive approach;

\smallskip

Fig.~21: $\frac{dN}{d\ln k_\perp}$ for a gluon jet
for $Y_{\Theta_0}=5.2$,  MLLA and naive approach;

\smallskip

Fig.~22: $\frac{dN}{d\ln k_\perp}$ for a quark jet
for $Y_{\Theta_0}=5.2$,  MLLA and naive approach;

\smallskip

Fig.~23: $\frac{dN}{d\ln k_\perp}$ for a gluon and a quark
jets, MLLA predictions for the Tevatron.

\smallskip

Fig.~24: the spectrum $\tilde D_g(\ell,Y_{\Theta_0}-\ell)$
for gluon jets;  comparison between MLLA and DLA (``with running
$\alpha_s$'') calculations;

\smallskip

Fig.~25:  comparison between MLLA (after dividing by
$\alpha_s(k_\perp^2)$, on the left)
and DLA calculation with $\alpha_s$ fixed (on the right)
of $\frac{d^2N}{dy\,d\ln k_\perp}$ for gluon jets;

\smallskip

Fig.~26: $Y_{\Theta_0}=7.5$: comparing  MLLA and DLA calculations
of $\frac{dN}{d\ln k_\perp}$ (see also Fig.~6); from left to right: 
$\frac{1}{\alpha_s(k_T^2)}$MLLA and DLA ($\alpha_s$ fixed);

\smallskip

Fig.~27: $Y_{\Theta_0} = 15$: comparing  MLLA and DLA calculations
of $\frac{dN}{d\ln k_\perp}$; from left to right: MLLA,
$\frac{1}{\alpha_s(k_T^2)}$MLLA and DLA ($\alpha_s$ fixed).

%%%%%%%%%%%%%%%%%%%%%%%%%%%%%%%%%%%%%%%%%%%%%%%%%%%%%%%%%%%%%%%%%%%%%%%%%%%%%%
%%%%%%%%%%%%%%%%%%%%%%%%%%%%%%%%%%%%%%%%%%%%%%%%%%%%%%%%%%%%%%%%%%%%%%%%%%%%%%

%%%%%%%%%%%%%%%%%%%%%%%%%%%%%%%%%%%%%%%%%%%%%%%%%%%%%%%%%%%%%%%%%%%%%%%%%%%%%%

\end{document}